\documentclass[a4paper,12pt]{spieman}  % 12pt font required by SPIE
\usepackage{amsmath,amsfonts,amssymb}
\usepackage{graphicx}
\usepackage{setspace}
\usepackage{tocloft}
\usepackage{ulem}
\usepackage{xcolor}
\usepackage{makecell}
\usepackage{subcaption}
\usepackage{lineno}
\usepackage{hyperref}
\usepackage{comment}
\usepackage{booktabs}
\usepackage{gensymb}

\title{Development of ProtoPol: a medium resolution echelle spectro-polarimeter for PRL telescopes, Mt Abu, India - Part I :  the design, development and laboratory characterization}

\author[a,*]{Mudit K. Srivastava}
\author[a,b,*]{Arijit Maiti}
\author[a, $\dagger$]{Vipin Kumar}
\author[a]{Bhaveshkumar Mistry}
\author[a]{Ankita Patel}
\author[a, $\dagger \dagger$]{Vaibhav Dixit}
\author[a]{Kevikumar A. Lad}

\affil[a]{Astronomy and Astrophysics Division, Physical Research Laboratory, Ahmedabad, 380009,
India}
\affil[b]{Indian Institute of Technology, Gandhinagar, 382335, India}
\affil[$\dagger$]{\textit{Current affiliation}: I. Physikalisches Institut, Universit\"at zu K\"oln, Z\"ulpicher Stra\ss e 77, 50937, K\"oln, Germany}
\affil[$\dagger \dagger$]{\textit{Current affiliation}: Advanced Engineering Group, Azista Industries, Applewoods Township, Ahmedabad, 380058, Gujarat, India}

\cftpagenumbersoff{figure}
\cftpagenumbersoff{table} 
\begin{document} 

\maketitle

\begin{abstract}
ProtoPol is a medium-resolution echelle spectro-polarimeter developed for Physical Research Laboratory (PRL) 1.2m and 2.5m telescopes, Mt. Abu, India. Though initially conceived to evaluate the development methodology of the echelle spectro-polarimeter, it was subsequently elevated to the level of a full-fledged back-end instrument for PRL telescopes. ProtoPol is developed on the traditional concept of using a half-wave plate with Wollaston prism to achieve the separation of two mutually orthogonal polarized beams. These separated beams are modulated and directed into an echelle spectrometer which is employs an echelle grating and two plane reflection gratings as the cross-dispersers. Therefore, the cross-dispersed spectra for two orthogonal polarized beams are recorded in multiple orders on a CCD detector. ProtoPol is designed to operate in the visible and near IR spectral range, 4000 - 9600 $\AA$, with a spectral resolution ($\delta$$\lambda$) around 0.4-0.75$\AA$.  The uniqueness of ProtoPol lies in its design which has entirely been developed with commercially available off-the-shelf optical and opto-mechanical components. This feature makes ProtoPol a noteworthy development as it offers a cost-effective way to develop spectro-polarimeters with such resolutions for small-aperture (2-3m) telescopes around the world, in a much shorter development period. ProtoPol has been successfully developed and commissioned on PRL 1.2m and 2.5m telescopes since December 2023, and a variety of observations have been carried out for instrument characterization, performance verification, and scientific purposes.  This is the first of the two-part research articles series, wherein we present the design and development methodology of ProtoPol, along with its laboratory characterization and performance.

\end{abstract}

\keywords{Telescope, Instrumentation, Spectro-polarimeter, Echelle spectrograph, Optical Design}

{\noindent \footnotesize\textbf{*}Corresponding author(s),  \linkable{mudit@prl.res.in}, \linkable{arijitmaiti@prl.res.in}}

\newcommand\blfootnote[1]{%
  \begingroup
  \renewcommand\thefootnote{}%
  \footnote{#1}%
  \addtocounter{footnote}{-1}%
  \endgroup
}

\begin{spacing}{2} 

%%%%%%%%%%%%%%%%%%%%%%%%%%%%%%%%%%%%%%%%%%%%%%%%%%%%%%%%%%%%%%%%%%

\section{Introduction}\label{sec-Intro}
 
Physical Research Laboratory (PRL) has been operating a 1.2m, $f/13$ optical/near-infrared (NIR) telescope at Gurushikhar, Mt. Abu since the mid-1990s \cite{srivastava2021design}. Recently (2022-2023), a new 2.5m, $f/8$ optical/near-infrared (NIR) telescope \cite{Pirnay2018} has also been set up at the same site. Over the past decades, the 1.2m telescope has been extremely successful in delivering important contributions to important astrophysical research fields like novae, supernovae, planetary nebulae, star formation, exoplanet detection, AGNS, cometary science, etc \cite{rajpurohit2018exploring, Srivastava2016, Kaur2017, Chakraborty2018, Joshi2017}. The new 2.5m telescope is only expected to further enhance the prospects of research in these observational research fields in several new arenas.  A suite of instruments has been/are being developed for the newer 2.5m telescope. As first-light instruments, a wide-field camera and a high-resolution spectrograph ($\sim$100000) called PARAS-2 \cite{Chakraborty2018b} were included. Another instrument—the Near-Infrared Spectrometer and Polarimeter (NISP), for near-infrared (NIR) observations, is also in development \cite{rai2020optical, sarkar2020electronics}, offering the capabilities of imaging, low-resolution spectroscopy (R$\sim$2000-3000), and imaging polarimetry in near-infrared bands. A Faint Object Spectrograph and Camera (FOSC)-style instrument \cite{srivastava2021design, buzzoni1984eso, andersen1995new, omar2017scientific} was initially planned, providing low-resolution spectroscopy and seeing-limited imaging capability in visible wavelengths; however, to address other key scientific needs (to be discussed in later sections), this was upgraded to an extended version of FOSC, incorporating the functionalities of an intermediate-resolution spectro-polarimeter alongside the standard FOSC capabilities. The M-FOSC-EP (Mt. Abu Faint Object Spectrograph and Camera-Echelle Polarimeter) instrument was, thus, conceived to cater to these needs. It is a two-channel multimode instrument currently being developed for PRL 2.5m telescope at Mt. Abu, India \cite{kumar2022designs}. One of the channels, named low-resolution arm (LRA), would provide the capabilities of seeing-limited imaging and low-resolution (R$\sim$500-800) spectroscopy, while the other is the intermediate-resolution (R$\sim$15000) channel, providing the capabilities of intermediate resolution spectro-polarimetry/ spectroscopy in visible and near IR wavelengths (3900-9900 $\AA$). The preliminary design details of M-FOSC-EP are presented in \cite{kumar2022designs}.
\par
A successful pathfinder instrument for the FOSC mode was previously developed for PRL 1.2m telescope for a variety of science goals that could be achieved from low-resolution spectroscopy and imaging over a moderate field of view (few arc-minutes a side) \cite{srivastava2021design, Rajpurohit2020, kumar2022optical}. Similar science goals with improved magnitude limits were thus envisioned from the LRA mode of M-FOSC-EP on PRL 2.5m telescope, with lower resolutions. Readers are referred to \cite{srivastava2021design} for details. However, it was the additional science benefits of a larger aperture of telescope that prompted us to conceive an instrument beyond the traditional FOSC regime. While imaging polarimetry had been a significant observing program of the observatory since long \cite{ganesh2020empol, aarthy2019nicspol},  the newer 2.5m telescope opens up the window for spectro-polarimetry as well. 
\par 
Spectropolarimetry, being a photon-hungry technique, had predominantly been employed on large aperture telescopes ($>3$m diameter), e.g. HARPSpol (High Accuracy Radial velocity Planet Searcher, R$\sim$120000) at the ESO 3.6m Telescope in La Silla, Chile \cite{piskunov2011harpspol}; ESPaDOnS (Echelle SpectroPolarimetric Device for the Observation of Stars, R$\sim$120000) at the 3.6m Canada-France-Hawaii Telescope on Mauna Kea \cite{donati2003espadons}; SARG (R$\sim$160000) at the 3.55m telescope on the Roque de Los Muchachos, La Palma, Canary Islands, Spain \cite{gratton2001sarg}; and HRS (High-Resolution Spectrograph, R$\sim$16000-67000) at the 11m Southern African Large Telescope (SALT) \cite{bramall2010salt} etc. While such large facilities have been highly successful in advancing the field of spectro-polarimetric studies, it is often difficult to plan a dedicated spectro-polarimetric science program or high cadence observations on such facilities due to over-subscription of the observing time. In contrast, small to medium aperture telescopes (2-3m diameter) equipped with lower-resolution spectropolarimeters (R$\sim$6000-15000) present a valuable alternative, given the relative ease of securing observing time on such telescopes. A few existing instruments on similar-class telescopes worldwide serve this purpose, including PEPSI (Potsdam Echelle Polarimetric and Spectroscopic Instrument) (R$\sim$130000) at 2×8.4m Large Binocular Telescope (LBT) in southern Arizona \cite{strassmeier2008pepsi}, SPIRou NIR spectro-polarimeter (R$\sim$70000) at 3.6m Canada–France–Hawaii Telescope (CFHT) in USA \cite{donati2020spirou}, BOES-BOAO Echelle Spectrograph,(R$\sim$60000) at the 1.8m Bohyunsan Optical Astronomy Observatory in Korea \cite{kim2007boes}; MuSiCoS (MUlti SIte COntinuous Spectroscopy, R$\sim$35000) at the 2m Bernard Lyot Telescope in Pic du Midi, France \cite{baudrand1992musicos}; VESPolA (Very Precise Echelle SpectroPolarimeter, R$\sim$8000) at the 1.3m Araki Telescope in the Koyama Astronomical Observatory of Kyoto Sangyo University, Kyoto, Japan \cite{arasaki2015very}; and LIPS (LIne Polarimeter and Spectrograph, R$\sim$7000) at the 2.2m University of Hawaii Telescope \cite{ikeda2003development}. Therefore, having a spectro-polarimeter on PRL 2.5m telescope would be of immense importance.
\par 
There have been several interesting astronomical targets—such as symbiotic stars, Herbig stars, M-dwarfs, AGNs, novae, supernovae etc. that would significantly benefit from higher spectral resolution (R $\sim$ 10,000–15,000) and spectro-polarimetric capabilities. Such key scientific objectives include inspecting spectral line variability to investigate the structure and dynamics of symbiotic systems, using spectro-polarimetry to explore the surroundings of obscured objects (e.g., novae, supernovae, Herbig stars, etc.) by studying the polarization profile across various emission lines, and conducting spectroscopic analyses of M-dwarfs etc. All such goals demand intermediate-resolution spectroscopy and spectro-polarimetry, complementing the standard functions of a FOSC instrument.
\par 
While the pathfinder of LRA (the FOSC mode) had already been successfully developed for PRL 1.2m telescope \cite{srivastava2021design}, a prototype spectro-polarimeter was also planned to be developed, as a precursor, to evaluate the development methodology of the spectro-polarimeter arm of M-FOSC-EP. Though initially conceived only as a laboratory prototype, it was later elevated to the status of a full-fledged back-end instrument, named ProtoPol, for both the PRL telescopes, owing to a successful design utilizing only commercially available off-the-shelf (COTS) components \cite{kumar2022designs}. This aspect of ProtoPol is its most unique feature, which makes its development a noteworthy achievement for the global astronomy community. ProtoPol has been designed completely with COTS optical and opto-mechanical components, ensuring cost efficiency and a faster development period. Similar to M-FOSC-EP design, ProtoPol is also developed on the concept of echelle and cross-disperser (CD) gratings to perform spectro-polarimetry in the wavelength range of $\sim$4000–9600 $\AA$ with comparatively lower resolution (0.4-0.75 $\AA$). Considering such wavelength range and spectral resolution, the instrument provides a valuable platform for diverse spectro-polarimetry studies, including analyzing polarization signatures in emission lines from symbiotic stars, novae, and supernovae, planetary nebulae, Classical Be stars, Herbig Ae/Be stars, investigating dust polarization around cool evolved stars, etc. In spectroscopy mode, its medium-resolution capability makes it well-suited for studying M-dwarfs, relatively brighter active galactic nuclei (AGNs), and other astrophysical targets etc.
\par 
ProtoPol has been successfully developed and commissioned on PRL first on the PRL 1.2m telescope in December 2023 and then on PRL 2.5m telescope in February 2024. Following a series of instrumental characterization observations, it has since been used for science observations of a variety of astrophysical targets. These observations successfully confirmed the designed performance of ProtoPol, viz., throughput, spectral range \& resolution, polarization accuracy, etc. This success, then, presents ProtoPol design and development methodology as one of the most suitable ones for the development of spectro-polarimeters for small/medium aperture optical telescopes around the world.  Here, we present the complete design of ProtoPol, development methodology, laboratory, and on-sky characterization showing its performance, and first science results in the two-part research articles. In this first part, we shall describe the optical and opto-mechanical designs of ProtoPol, subsequent assembly-integration-testing (AIT) in the laboratory, followed by its successful on-sky commissioning and preliminary results. The second part would cover the development of ProtoPol's data-analysis pipeline, its complete on-sky characterization and performance verification, as well as first science results. In the subsequent sections, we shall first describe the potential science cases, which are being/can be explored with ProtoPol, and then its design, development, AIT, and commissioning aspects.

%%%%%%%%%%%%%%%%%%%%%%%%%%%%%%%%%%%%%%%%%%%%%%%%%%%%%%%%%%%%%%%%%
%%%%%%%%%%%%%%%%%%%%%%%%%%%%%%%%%%%%%%%%%%%%%%%%%%%%%%%%%%%%%%%%%

\section{Science programs to be explored with ProtoPol} \label{sec-SciencePrograms}

As explained in section~\ref{sec-Intro} above, ProtoPol was initially conceived only as the laboratory prototype, which was later elevated as a full-fledged instrument. Therefore, unlike its parent M-FOSC-EP, which is being designed to achieve certain science objectives \cite{kumar2022designs, srivastava2024development}, ProtoPol's science targets were selected considering its expected performance as evaluated from the design. Its medium resolution (R$\sim$7000–8000) was found most suitable to study scattering-induced polarization in various astrophysical situations wherein the spectral features are reasonably broad, e.g., Raman scattered features in symbiotic stars \cite{schmid1994raman, harries1996raman} etc. Other suitable targets where the derived spectral resolution could have been useful are transients (e.g. novae, supernovae, etc.), M Dwarfs, classical Be stars, etc., which have been well studied with PRL 1.2m telescope \cite{Rajpurohit2020, kumar2022optical}. Their key scientific objectives include analyzing spectral line profile variations to study the structure and dynamics of the underlying stellar systems, spectro-polarimetry to probe the surroundings of obscured objects (e.g., novae, supernovae, symbiotic stars etc), spectro-polarimetry of emission lines as a tool to probe the inner disc region of objects like Classical Be stars and Herbig stars which are otherwise not resolvable, tracing the continuum polarization of AGB/post-AGB stars, and conducting detailed spectroscopic studies of M dwarfs. These investigations highlight the need for intermediate-resolution spectroscopy to achieve these science goals. Some of these potential research avenues have been discussed below:
\par
\textbf{Spectro-polarimetry of symbiotic stars:} Symbiotic stars are interacting binary systems consisting of an evolved red giant and a hot radiating source, typically a white dwarf, with orbital periods of several hundred days to longer \cite{kenyon1986symbiotic, munari2019symbiotic}. One unique feature of the Symbiotic star spectra is the presence of the emission features at 6825$\AA$ and 7082$\AA$, which were later identified as features produced by the Raman scattering of far UV O VI doublet lines 1032$\AA$  and 1038$\AA$ originating near the hot component by the neutral hydrogen wind of the cool star \cite{nussbaumer1989raman, schmid1989identification}. As those features originate from a scattering event, they exhibit strong polarization, as studied by \cite{schmid1994raman, harries1996raman}. ProtoPol, with its designed resolution, would be able to resolve the polarimetric profile of the Raman scattered lines in a lot more detail. 
\par
Several symbiotic stars show broad H$\alpha$ wings \cite{lee2000raman}. This broad wing could result from hydrodynamical motion or inelastic Raman scattering of Lyman$\beta$ by neutral hydrogen. To decouple between these two scenarios, spectro-polarimetry of the H$\alpha$ profile of such stars provides a powerful diagnostic tool. If the broadening of the H$\alpha$ wing is indeed due to Raman scattering phenomena, some polarimetric signature is expected to be detected at the wings \cite{yoo2002polarization, lee2018h}. This kind of spectro-polarimetric study has been done by \cite{ikeda2004polarized} using LIPS spectro-polarimeter \cite{ikeda2003development} with a resolution of 7000. ProtoPol, with a similar resolution, should also be capable of conducting similar studies. 
\par 
\textbf{Spectro-polarimetry of Herbig Ae/Be stars:} Herbig Ae/Be stars are optically visible pre-main-sequence stars that are the more massive counterparts of T Tauri stars \cite{de1994new, mottram2007difference}. Due to their intermediate masses, they play an essential role in bridging the gap between the formation mechanisms of low-mass stars and high-mass stars, whose formation theories pose a lot of challenges. Herbig Ae/Be stars are one of the few higher mass pre-main-sequence stars that are visible at optical wavelengths. Although clear-cut evidence of the presence of circumstellar gas and dust in such systems has been reported by \cite{waelkens1997comet}, there is no general consensus about the geometry of such systems. Spectro-polarimetry across emission lines of spectra remains the most powerful tool to probe the inner circumstellar discs around the stars on scales of a few stellar radii. H$\alpha$ spectro-polarimetry of Herbig stars has been done by \cite{vink2002probing} with a velocity resolution of 35 kms$^{-1}$ at H$\alpha$. ProtoPol with a similar velocity resolution (42 kms$^{-1}$) can also be used to conduct similar spectro-polarimetric studies of H$\alpha$ and other Hydrogen Balmer lines.
\par 
\textbf{Spectroscopic studies of M-Dwarfs:} M dwarfs are the lowest-mass stars, occupying the bottom of the main sequence in the Hertzsprung-Russell (HR) diagram. They amount to roughly 70\% of the stars in our Galaxy, while contributing around 40\% of its total stellar mass \cite{henry2024character}. Due to their small size and low mass, they are ideal targets for detecting planets within the habitable zone, offering valuable insights into exoplanet formation. The study of M dwarfs relies on determining key stellar parameters such as effective temperature, surface gravity, and metallicity. While low-resolution spectroscopy combined with synthetic models can estimate effective temperature and surface gravity \cite{Rajpurohit2020}, measuring metallicity demands medium-to-high-resolution spectroscopy. High-resolution spectra with broad wavelength coverage also help investigate dust and cloud formation at cooler temperatures, shedding light on atmospheric processes in these stars. Recently, \cite{rajpurohit2018exploring} conducted a similar study on 292 M dwarfs to derive their stellar parameters. ProtPol, with a resolution of 7000–8000, will enable such detailed investigations in the future.

%%%%%%%%%%%%%%%%%%%%%%%%%%%%%%%%%%%%%%%%%%%%%%%%%%%%%%%%%%%%%
%%%%%%%%%%%%%%%%%%%%%%%%%%%%%%%%%%%%%%%%%%%%%%%%%%%%%%%%%%%%%

\section{Optical and Opto-Mechanical Designs of ProtoPol}
\label{sec-DesignFull}

\par
\subsection{PRL 1.2m and 2.5m Telescopes} 
\label{subsec-PRLTel}
\par 
The older 1.2m telescope features a 1.2m primary mirror and a 0.3m secondary mirror in equatorial mount configuration. It delivers an $f/13$ beam at the telescope focal plane \cite{Deshpande1995} and its focal plane position can be adjusted via motorized movement of the secondary mirror along the optical axis. The plate scale of the telescope is 76 $\mu$m per arc-second and an aberration-free field of view of approximately 10 arc-minutes in diameter \cite{Banerjee1997}. The weight limit of the telescope focal plane instrumentation is 125 kg, while the available space below the primary mirror cell is restricted to about 1m in diameter and 0.8m in depth. Consequently, any backend instrument for this telescope must adhere to these weight and size constraints.
\par
The newer 2.5m telescope has recently become operational in 2022-2023. It is an alt-azimuth-mounted telescope which follows a Ritchey-Chrétien optical design \cite{Pirnay2018}, featuring a hyperbolic concave primary mirror (2.5m, $f/2$) paired with a hyperbolic convex secondary mirror (0.779m, $f/2.4$). The telescope delivers an $f/8$ beam at the telescope focal plane, with a scientific field of view (FOV) of 25 arc-minutes in diameter ({\href{https://www.prl.res.in/~miro/}{https://www.prl.res.in/~miro/}}). The telescope incorporates an active optics system for mirror support that uses the feedback of a wavefront sensor, which in turn uses an off-axis guide star for wavefront sensing. The primary mirror’s shape is maintained across different orientations using multiple axial and radial actuators, while the secondary mirror is adjusted via a hexapod for precise positioning and fine alignments. The telescope supports the mounting of three instruments simultaneously: one at the main Cassegrain focus and two at side ports. The maximum permissible instrument weights are 341 kg for the main port and 150 kg for each side port.

%%%%%%%%%%%%%%%%%%%%%%%%%%%%%%%%%%%%%%%%%%%%%%%%%%%
%%%%%%%%%%%%%%%%%%%%%%%%%%%%%%%%%%%%%%%%%%%%%%%%%%%%%

%%%%%%%%%%%%%%%%%%%%%%%%%%%%%%%%%%%%%%%%%%%%%%%%%%%%%

\begin{table*}
	\centering
	\caption{Design Parameters of ProtoPol:}
	\begin{tabular}{p{5cm}p{5cm}}
		\hline
		\hline
		&          \\
		\textbf{Parameters} &	\textbf{Values} \\
		&          \\
		\hline
		\hline 
		&           \\
		Optimized Wavelength Range &	4000 – 9600 $\AA$ \\
        Telescopes    &  1.2m, $f/13$ with plate-scale of 76 $\mu$m per arc-second \\
                      &  2.5m, $f/8$ with plate-scale of 97 $\mu$m per arc-second \\
		Slit mapping	& $\sim$3.6 pixels per arc-second on 2.5m Telescope \\
        
		Image Quality Requirement  &	$\sim80\%$ Encircled Energy Diameter is to be within 1.5 pixels \\
		Magnification of the ProtoPol optical Chain &	$\times$0.67 (Polarimeter Unit), $\times$0.61 (Spectrometer Unit), Total $\times$0.41 \\
		Camera optics	& Canon EF 200mm f/2L IS USM ($f/2$ to $f/32$)\\
		CCD Detector	& 1K $\times$ 1K ANDOR M934 CCD with 13$\mu$m a side pixel size \\
		%Filters &	Astronomy standard Bessell’s BVRI filters. Narrow band H-$\alpha$ filter \\		Spectral Coverage of Gratings &	$\sim$4500-8500$\AA$ using three different Gratings \\
		Spectral Resolution $(\delta\lambda)$ &	$\sim$0.40-0.75$\AA$ for 1 arc-second ($\sim97\mu m$) slit width on 2.5m PRL Telescope \\
		Pupil Diameter & $\sim6.78$mm in Polarimeter and  $\sim$61 mm in Spectrometer \\
		Grating’s Specifications & One Echelle (54.49 lp/mm, 46 degree blazed angle) and two Plane reflection gratings with 150 lp/mm and 300 lp/mm and blazed at 8000$\AA$ and 4220$\AA$, respectively, working as cross-disperser gratings. \\ 
		
		\hline		
		\hline
		
	\end{tabular}
	\label{table-para}	
\end{table*}
%%%%%%%%%%%%%%%%%%%%%%%%%%%%%%%%%%%%%%%%%%%%%

\subsection{Optical Design of ProtoPol} 
\label{subsec-OpticalDesign}
\par
\begin{figure}[ht]
	\centering
	\includegraphics[width=\textwidth]{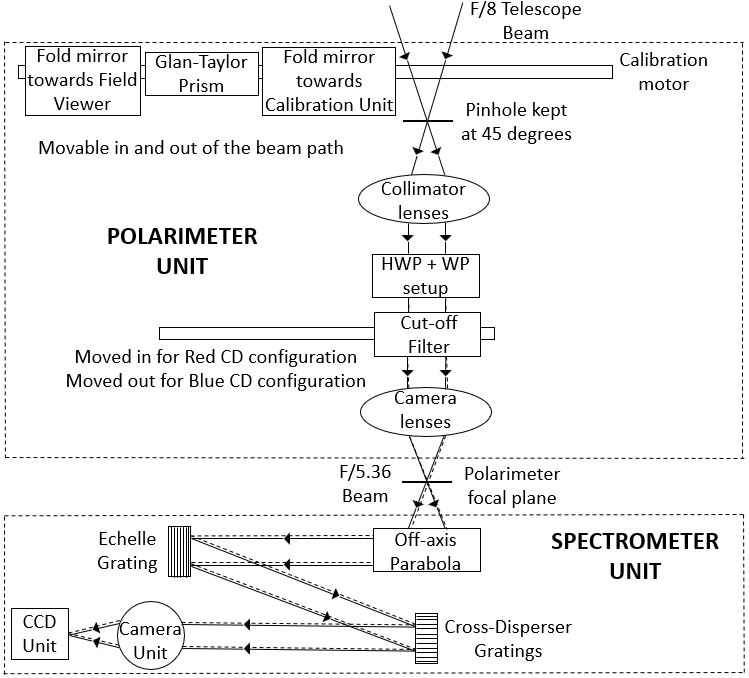}
	\caption{Schematics of ProtoPol system design}\label{BlockDiagram}
\end{figure}

The preliminary optical and opto-mechanical designs of ProtoPol are presented in \cite{kumar2022designs, srivastava2024development}. Here we will describe the layout of the instrumentation system of ProtoPol, the optical designs, and the derived performance of each of the sub-systems for 2.5m PRL telescope. The instrument (Figure~\ref{BlockDiagram}) maps the telescope’s focal plane onto a 150-micron pinhole kept at an orientation of 45$\degree$ (thus making a projection of 106-micron $\sim$1.1 arc-seconds on 2.5m telescope's focal plane), serving as the slit. ProtoPol's optical chain provides the magnification of $\times$0.41, which, when coupled with other effects such as grating's anamorphic magnification, slit-tilt caused by echelle grating (see section \ref{subsec-SlitTilt}), etc., would provide a sampling of $\sim$3.6 pixels per arc-second on 2.5m Telescope. The incoming light is then split into two orthogonally polarized beams (o-ray and e-ray) by the polarimeter module, which comprises four off-the-shelf achromatic doublet lenses, a Wollaston prism, and an achromatic half-waveplate (HWP). All these optical components were selected from the catalog of commercial optics manufacturers as required by the instrument's design. These separated o- and e- beams are fed to the spectrometer section, which features a commercially available off-the-shelf off-axis parabolic mirror as a collimator, an echelle grating, and interchangeable CD gratings: one optimized for the blue wavelength range (4000–6000 $\AA$) and another for the red wavelength range (5600–9600 $\AA$). The dispersed spectra, spanning multiple echelle orders, are imaged by a commercially available off-the-shelf Canon 200mm focal length camera lens coupled to an ANDOR  off-the-shelf 1K × 1K CCD detector, which records the final cross-dispersed o-ray and e-ray spectra for various echelle orders. The block diagram of the ProtoPol is given in Figure~\ref{BlockDiagram}. Key technical specifications are listed in Table~\ref{table-para}. 

%================================
%================================

\par

%================================
\begin{figure}[ht]
	\centering
	\includegraphics[width=\textwidth]{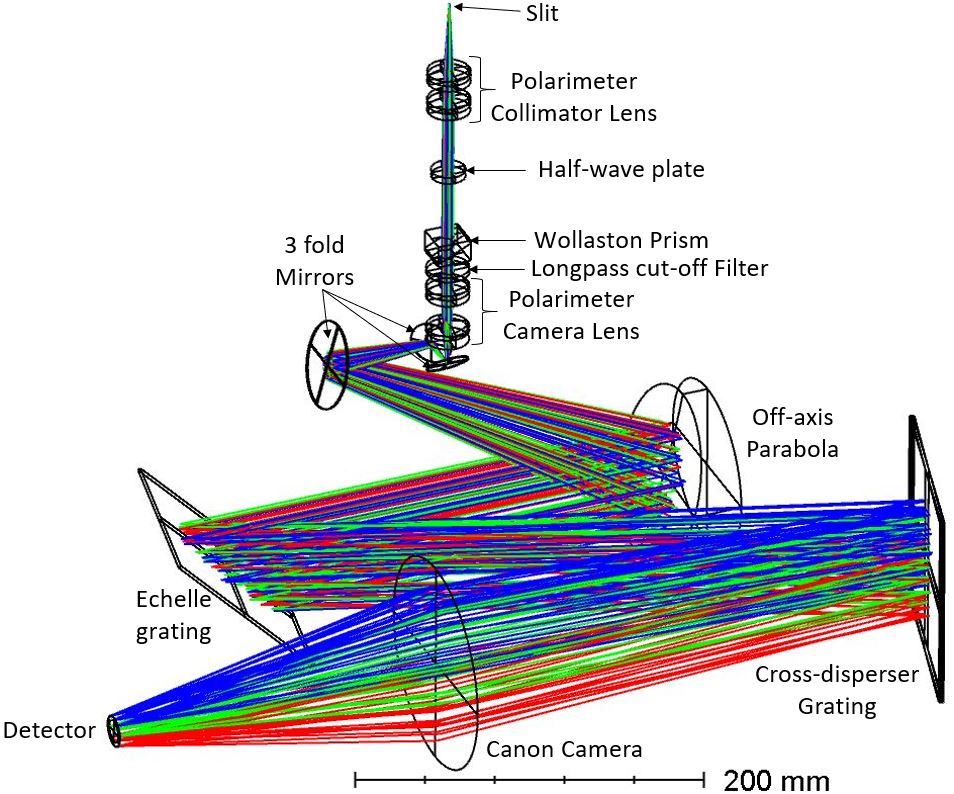}
	\caption{Optical layout of ProtoPol. The Canon camera lens is represented by an ideal lens surface of focal length 200mm in the layout.}\label{OpticsLayouts}
\end{figure}
%================================

ProtoPol is also equipped with a calibration unit to simulate the telescope pupil using calibration beams from a Uranium-Argon (U-Ar) spectral lamp (for wavelength calibration) and a halogen lamp (for tracing the spectral orders). A Glan-Taylor (GT) prism (to generate 100\% polarized light) is also mounted before the focal plane to check for instrumental depolarization. The prism is mounted on a linear translational stage, which is also equipped with two reflecting fold mirrors. Fold mirror-1 redirects the telescope’s incoming beam towards a field viewer camera, equipped with a Bessel’s V filter in the beam path. It is used for sky field identification for target selection and matching. Fold mirror-2 blocks the telescope beam while directing light from the calibration unit lamps into the instrument's optical chain. The half-wave plate (HWP) is mounted into an in-house designed and developed stepper motor-driven rotation stage, which can be positioned at the required angles for spectro-polarimetric measurements. The instrument is controlled entirely via a fully in-house developed instrument control system and operating software, while the detector communication is handled via the camera manufacturer’s user interface. ProtoPol is compatible with both the PRL 1.2m and 2.5m telescopes, with only a change in the mechanical interface adaptations and replacing the pupil mask inside the calibration unit for switching between systems. The optical design of these sub-systems of ProtoPol, viz. polarimeter unit, spectrometer unit, and calibration Unit, and the design performances are discussed below. The optical design software - \href{https://www.ansys.com/en-in/products/optics/ansys-zemax-opticstudio}{ZEMAX} is used for various design and evaluation purposes.

%%%%%%%%%%%%%%%%%%%%%%%%%%%%%%%%%%%%%%%%%%%%%%%%%%%%%%%%%%%%%%%

\subsubsection{Polarimeter: Optical design \& Performance}
\label{subsec-Polarimeter}

\begin{figure}[ht]
	\centering
	\fbox{\includegraphics[width=\textwidth]{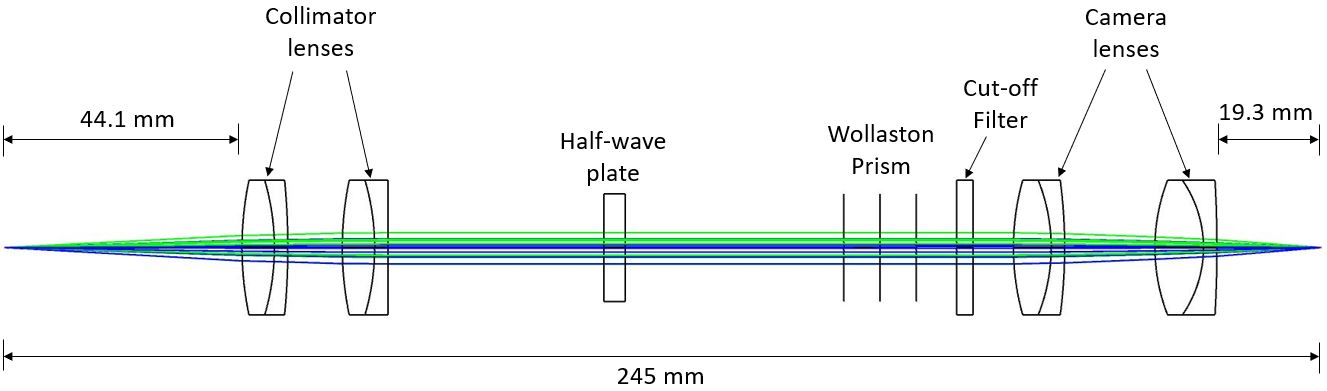}}
	\caption{\label{fig:protopol_polarimeter} 2-D layout of the optical design of Polarimeter section}
\end{figure}

The optics chain of the polarimeter optics is designed with an $f/8$ collimator and an $f/5.4$ camera, providing a net magnification of 0.67. The pupil diameter produced by the collimator is $\sim$6.8mm within the polarimeter optical chain, with the system having a field of view of 1$\times$1 arc-minute$^2$. A COTS quartz Wollaston prism is incorporated within the polarimeter section as a polarizing beam splitter, while a rotatable COTS half-wave plate made of PMMA material functions as a polychromatic modulator. The Wollaston prism splits the incoming beam into ordinary (o-ray) and extraordinary (e-ray) components, which are focused by the polarimeter camera optics onto the object plane of the echelle spectrometer for performing simultaneous spectroscopy. The focused o- and e- beams are separated by $\sim$324$\mu$m at this plane. Both the collimator and camera optics are designed with two COTS achromatic doublet lenses each. A long-pass filter (cut-off wavelength: 5250$\AA$) is placed immediately after the Wollaston prism to avoid the second-order contamination while observing with Red CD. The filter is mounted on a small linear translational stage to retract and insert in the path of the beam as and when required. The 2D optical layout of the polarimeter, as modeled in ZEMAX, is shown in Figure~\ref{fig:protopol_polarimeter}, while the design specifications and list of COTS components are provided in Tables~\ref{proto_Polarimeter_Spec} and ~\ref{proto_polarimeter_components}.

%%%%%%%%%%%%%%%%%%%%%%%%%%%
%%%%%%%%%%%%%%%%%%%%%%%%%%%

%=====================================
\begin{table*} 
\caption{Design specifications of the polarimeter section} \label{proto_Polarimeter_Spec}
\centering
\setlength{\tabcolsep}{5pt}
\renewcommand{\arraystretch}{1.2}
	\begin{tabular}{cc cc }
           \hline
		  \hline

\textbf{Parameters} & \textbf{Values} \\
\hline
\hline
            Input f/\# at telescope focal plane & 8 \\ 
			Output f/\# at Polarimeter focal plane & 5.4 \\
			Collimator focal length & 54.2 mm \\ 
			Polarimeter camera focal length & 36.3 mm \\ 
			Magnification & 0.67 \\ 
			Pupil diameter & 6.8 mm \\ 
			Half wave plate specification & \textit{Material:} PMMA, \textit{Retardance:} 180 degree \\ 
			Polarizer & \textit{Type:} Wollaston prism, \textit{Material:} Quartz\\ 
			Wollaston Prism Cut angle  & 25.5 degrees \\
			Wollaston Prism Deviation angle  & 0.5 degrees \\ 
			\begin{tabular}[c]{@{}l@{}}Separation between o- and e-rays\\ at polarimeter focal plane\end{tabular} & $\sim$ 324$\mu$m \\

		\hline
		\hline
	\end{tabular}
\end{table*}

%=====================================
\begin{table*} 
\caption{Details of the components of the polarimeter section }\label{proto_polarimeter_components}
\centering
\setlength{\tabcolsep}{5pt}
\renewcommand{\arraystretch}{1.2}
	\begin{tabular}{cc cc cc cc}
           \hline
		  \hline

\textbf{Component} & \textbf{Specific} & \textbf{Specification} & \textbf{Part number}\\
                   & \textbf{element}  &                        & \textbf{\& Manufacturer}\\
\hline
\hline
           & 1.Doublet Lens & 1.f=100mm & 1. \#49-360$^a$ \\ 
Collimator & 2.Doublet Lens & 2.f=100mm & 2. \#49-360 \\ 
           &           &           & (Edmund Optics) \\ 
\\
Half Wave plate & Super Achromatic & 3900-9200 $\AA$ & APSAW-5-Wide$^b$ \\ 
                &     Retarder     &            &    20 mm \\ 
\\
Polarizer & Quartz  & 2300-28000 $\AA$ & PWQ 30.20$^c$  \\ 
          & Wollaston Prism & Deviation: 0.5 degrees &  (Bernhard Halle Nachfl) \\ 
\\
Cut-off filter & Long-pass filter & 5250$\AA$  & 84-744$^d$ \\
               &                 &     Cut-off wavelength  & (Edmund Optics) \\ 
\\
       & 1. Doublet Lens & 1. f=75mm & 1. \#49-358$^e$\\ 
Camera & 2. Doublet Lens & 2. f=50mm & 2. \#49-356$^f$\\ 
       &            &           & (Edmund Optics)\\ 
\\
Slit & Pinhole & Diameter: & P150MB$^g$ (Thorlabs) \\ 
     &                  & 150 micron  & \\         
		\hline
		\hline
	\end{tabular}

    \begin{itemize}
        \item a:\href{https://www.edmundoptics.in/p/25mm-dia-x-100mm-fl-vis-nir-coated-achromatic-lens/9763/}{https://www.edmundoptics.in/p/25mm-dia-x-100mm-fl-vis-nir-coated-achromatic-lens/9763/}
        \item b:\href{http://astropribor.com/super-achromatic-quarter-and-half-waveplate/}{http://astropribor.com/super-achromatic-quarter-and-half-waveplate/}
        \item c:\href{https://www.b-halle.de/produkte/polarisatoren/wollaston-polarisatoren.html}{https://www.b-halle.de/produkte/polarisatoren/wollaston-polarisatoren.html}
        \item d:\href{https://www.edmundoptics.in/p/525nm-25mm-dia-high-performance-longpass-filter/27813/}{https://www.edmundoptics.in/p/525nm-25mm-dia-high-performance-longpass-filter/27813/}
        \item e:\href{https://www.edmundoptics.in/p/25mm-dia-x-75mm-fl-vis-nir-coated-achromatic-lens/9759/}{https://www.edmundoptics.in/p/25mm-dia-x-75mm-fl-vis-nir-coated-achromatic-lens/9759/}
        \item f:\href{https://www.edmundoptics.in/p/25mm-dia-x-50mm-fl-vis-nir-coated-achromatic-lens/9755/}{https://www.edmundoptics.in/p/25mm-dia-x-50mm-fl-vis-nir-coated-achromatic-lens/9755/}
        \item g:\href{https://www.thorlabs.com/thorproduct.cfm?partnumber=P150MB}{https://www.thorlabs.com/thorproduct.cfm?partnumber=P150MB}
    \end{itemize}

\end{table*}
%===================================================

%===============================================
\begin{table}
	\centering
	\caption{Lens Parameters for the ProtoPol polarimeter Optical Chain: This table contains ZEMAX Optical Design Data}
		\begin{tabular}{cccccc} 
	%\begin{tabular}{p{3.0cm}lcccp{1.5cm}}
		\hline
		\hline	
		Element &  Surface & Glass & Catalog & Radius of  & Thickness (mm)  \\
		&          &       &         &     Curvature (mm)            &   (Distance to   \\
		&          &       &         &                     &   next Surface) \\
		
		\hline    
        \hline 
		&          &       &         &                     &   \\    
		Col-D-1$^{a}$ & First  & N-BK7  & SCHOTT  &  61.47 (CX)$^{b}$      & 6.0    \\
		Col-D-1 & Second & N-SF5  & SCHOTT  &  -44.64 (CX-CC)$^{b}$  & 2.5    \\
		Col-D-1 & Third  &          &        &  -129.94 (CX)$^{b}$      & 10.1   \\
		Col-D-2 & First  & N-BK7  & SCHOTT  &  61.47 (CX)     & 6.0    \\
		Col-D-2 & Second & N-SF5  & SCHOTT  &  -44.64 (CX-CC)  & 2.5    \\
		Col-D-2 & Third  &          &        &  -129.94 (CX)      & 116$^{c}$   \\
		Cam-D-1$^{a}$     & First  &  N-BK7  & SCHOTT  &  46.44 (CX)      & 7.0    \\
		Cam-D-1     & Second &  N-SF5  & SCHOTT  & -33.77 (CX-CC) & 2.5    \\
		Cam-D-1     & Third  &          &        &  -95.94 (CX)   & 16.75  \\
		Cam-D-2     & First  & N-BAF10  & SCHOTT & 34.53 (CX)      & 9.0    \\
		Cam-D-2    & Second & N-SF10  & SCHOTT &  -21.98 (CX-CC)  & 2.5    \\
		Cam-D-2     & Third  &          &        &  -214.63 (CX)   & 19.3     \\
	
		\hline
		\hline
		\vspace{-0.2cm}
	\end{tabular}
	\label{table-OpticsData}
	\begin{list}{}{}
        \item a: Col-D-1: Collimator-Doublet-1, Col-D-2: Collimator-Doublet-2, Cam-D-1: Camera-Doublet-1, Camera-Doublet-2.
		\item b: CC: Concave Surface. CX: Convex Surface. The intermediate surface of a doublet lens is described as CC-CX or CX-CC.
		\item c: The 116.0mm (= 40.0+4.0+40.5+6.75+6.75+18 mm) is the distance between the last surface of the collimator lens to the first surface of the camera lens. 40.0mm is the distance between the collimator and the Half-wave plate (HWP), 4.0mm is the thickness of the HWP, 40.5mm is the distance from the HWP to Wollaston prism (WP), (6.75mm + 6.75mm) is the thickness of the WP, and 18mm is the distance of the first surface of the camera lens from WP. The cut-off filter can be inserted in this space.

	\end{list}
\end{table}
%==========================================================

\par
\begin{figure}[]
	\centering
	\fbox{\includegraphics[width=\textwidth]{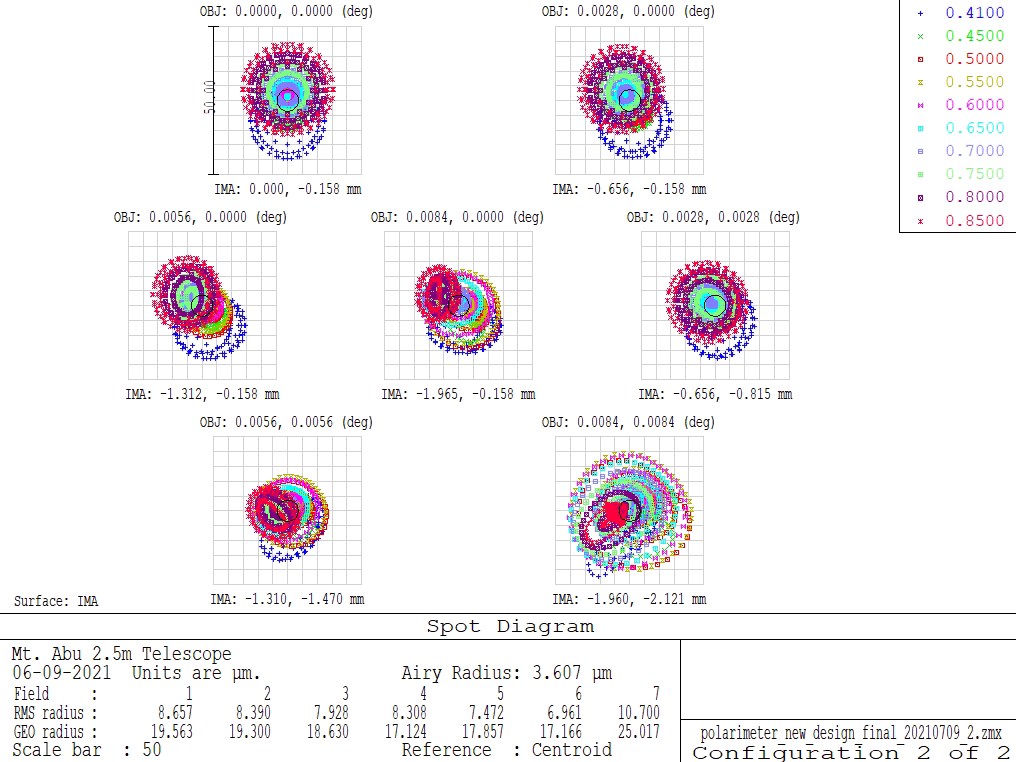}}
	\caption{\label{proto_pol-per-o-ray}Spot diagram for the polarimeter section of ProtoPol at its focal plane for o-ray configuration (0.41-0.85 $\mu$m). The box size is 50$\mu$m ($\sim$0.77 arc-seconds).}
\end{figure}
\begin{figure}[]
	\centering
	\includegraphics[width=\textwidth]{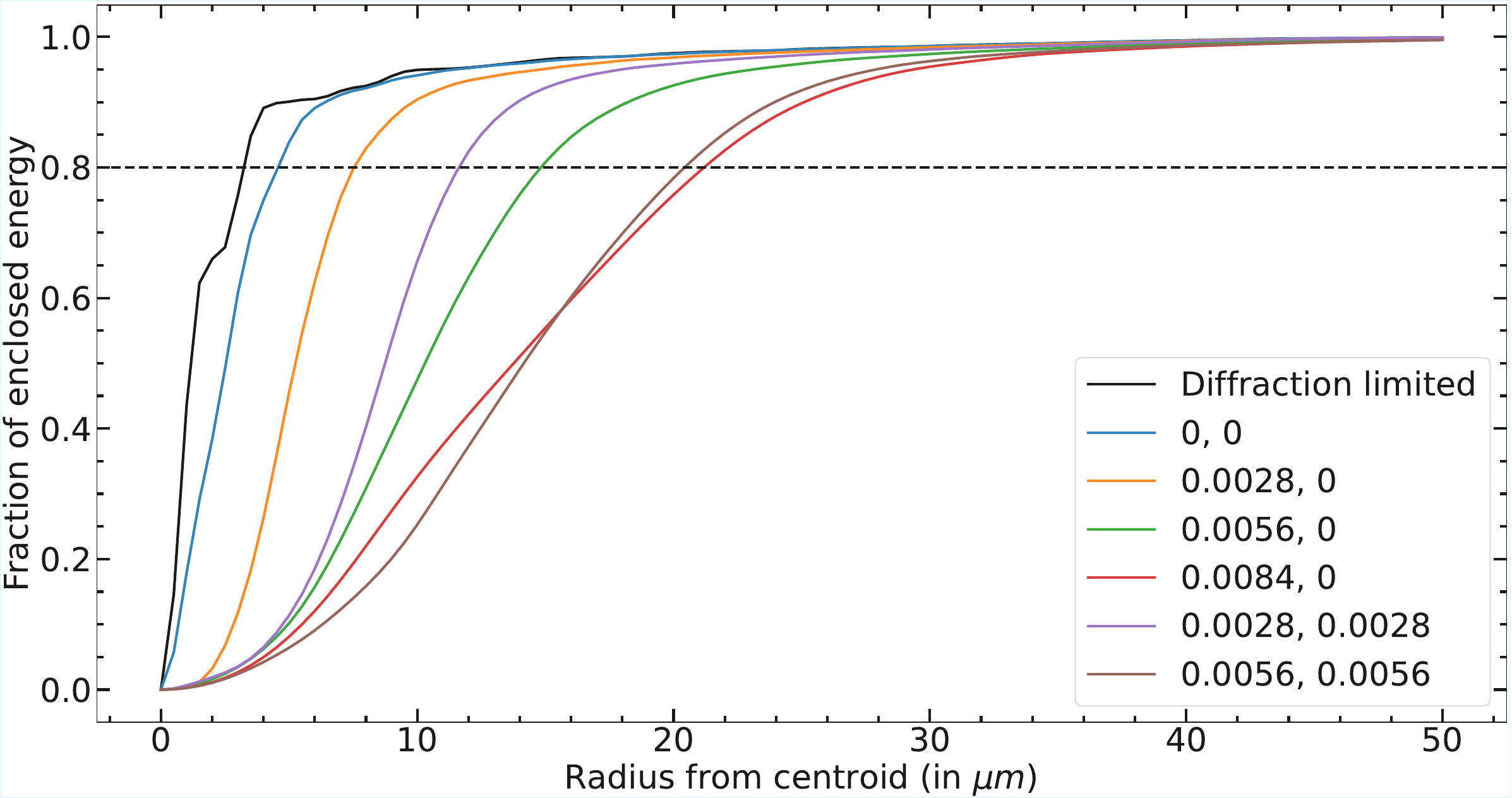}
	\caption{\label{proto_pol-E80-o-ray}Encircled energy diagram for o-ray configuration at the polarimeter focal plane (0.6 $\mu$m) for 1$\times$1 arc-minute$^2$ field. All field points are mentioned in degrees.}
\end{figure}

A pinhole of diameter 150$\mu$m is used as a slit (at 45$\degree$) at the object plane of the collimator, which coincides with the focal plane of the telescope. The slit size (45-degree projection of 150$\mu$m slit), projected at the output of polarimeter, i.e., on the re-mapped focal plane, would be 71$\mu$m (due to a magnification factor of 0.67), with the Encircled Energy (EE80) diameters within 29$\mu$m (0.45 arc-seconds) across various field points. Nearly identical performance is observed between the o-ray and e-ray designs. The spot diagram corresponding to the o-ray is presented in Figure~\ref{proto_pol-per-o-ray}, while the corresponding encircled energy distribution is illustrated in Figure~\ref{proto_pol-E80-o-ray}. The imaging performance of the polarimeter for different fields is shown in Table~\ref{tab:proto_polarimeter_per} in terms of EE80, RMS, and geometrical spot diameters.

%===========================
%================================

%=====================================
\begin{table*} 
\caption{EE80, RMS and geometrical diameters at the output of the polarimeter for 1$\times$1 arc-minute$^2$ fields (for 0.41-0.85$\mu$m).}\label{tab:proto_polarimeter_per}
\centering
\setlength{\tabcolsep}{5pt}
\renewcommand{\arraystretch}{1.2}
	\begin{tabular}{cc cc cc cc}
           \hline
		  \hline
		\textbf{Field} & \textbf{RMS} & \textbf{EE80} & \textbf{Geometrical} \\
		\textbf{(arc-minutes, arc-minutes)} & \textbf{diameters ($\boldsymbol{\mu}$m)} & \textbf{diameters ($\boldsymbol{\mu}$m)} & \textbf{diameters ($\boldsymbol{\mu}$m)} \\ 
        \hline
        \hline
		(0, 0) & 17.3 & 28.6 & 39.1 \\ 
		(0.17, 0) & 16.8 & 28.0 & 38.6 \\ 
		(0.34, 0) & 15.8 & 26.3 & 37.3 \\ 
		(0.50, 0) & 16.6 & 24.8 & 34.3 \\ 
		(0.17, 0.17) & 15.0 & 25.6 & 35.7 \\ 
		(0.34, 0.34) & 14.0 & 22.0 & 34.4 \\
		(0.50, 0.50) & 21.4 & 26.2 & 50.0 \\
		\hline
		\hline
	\end{tabular}
\end{table*}
%===================================

The tolerance analysis of the polarimeter unit was done in ZEMAX, where the various tolerance parameters were estimated as follows: 1) tolerances from the optics manufacturer's perspective, which includes the radius of curvature, optics thickness, decenter/tilt of the various optical surfaces, refractive index, abbe number, etc. 2) from the perspective of opto-mechanical assembly, including decenter/tilt with respect to optic axis, ambient temperature and pressure, etc. The sensitivity analysis was performed with ZEMAX to determine the required error limits to be imposed upon each parameter, keeping the RMS spot diameter as the optimization condition. Monte Carlo simulations were performed to simulate the random combinations of such errors to be simultaneously incorporated in the given optical system. After generating a large number of such random optical systems, the statistical performance of the system is evaluated to conclude the success criteria of the given optical system. The tolerance values of the various parameters are given in Table~\ref{ProtoPol_Polarimeter_tolerance_table}. The results (RMS diameters) of 10000 Monte Carlo simulations are mentioned in Table~\ref{ProtoPol_Polarimeter_tolerance_MC}. The RMS spot diameters for all wavelengths were calculated to be $< 35\: \mu m (\sim 0.5\: \text{arc-seconds)}$ with 98$\%$ confidence limit in worst-case scenario.

%=====================================
\begin{table*} 
\caption{Tolerance parameters for the polarimeter and spectrometer sections}\label{ProtoPol_Polarimeter_tolerance_table}
\centering
\setlength{\tabcolsep}{5pt}
\renewcommand{\arraystretch}{1.2}
	\begin{tabular}{cc cc }
           \hline
		  \hline

\textbf{Criterion} & \textbf{Tolerance} \\
\hline
\hline
\textbf{Surface} &  \\
            Thickness (mm) & $\pm$ 0.2 \\ 
			Radius (fr) & 3 \\
			Decenter X,Y (mm) & $\pm$ 0.015 \\ 
			Tilt X,Y (deg) & $\pm$ 0.017 (1 arc-minute) \\ 
			Irregularity (fr) & 0.5 \\ 
			Refractive Index & $\pm$ 0.0001 \\ 
			Abbe Number ($\%$) & $\pm$ 0.5 \\ 
\textbf{Element} &  \\
			Decenter X,Y (mm) & $\pm$ 0.05 \\ 
			Tilt X,Y (deg) & $\pm$ 0.05 (3 arc-minutes) \\
			Temperature (degrees)  &  $\pm$ 10 \\ 
			Pressure (atm) &  $\pm$ 0.01 \\

		\hline
		\hline
	\end{tabular}

    \begin{list}{}{}
        \item a: The ambient temperature and pressure were assumed to be 15$^\circ$ and 0.82 atm, respectively. The ambient pressure value was estimated from the atmospheric pressure at the altitude of Mt. Abu observatory with respect to sea-level pressure.
	\end{list}
    
\end{table*}

%=====================================

%=====================================
\begin{table*} 
\caption{Tolerance analysis of polarimeter section for both o- and e-ray in the 450-850 nm wavelength range with 10000 Monte Carlo runs for each configuration. The distance of the polarimeter focal plane from the last camera lens surface was used as a compensator ($\pm 1mm$).} \label{ProtoPol_Polarimeter_tolerance_MC}
\centering
\setlength{\tabcolsep}{5pt}
\renewcommand{\arraystretch}{1.2}
	\begin{tabular}{c ccccc ccccc}
           \hline
		  \hline
          
        %\toprule
        \multicolumn{1}{c}{\textbf{Wavelength}} & \multicolumn{5}{c}{\textbf{O-ray RMS diameter ($\boldsymbol{\mu m}$)}} & \multicolumn{5}{c}{\textbf{E-ray RMS diameter ($\boldsymbol{\mu m}$)}} \\
        \cmidrule(lr){2-6} \cmidrule(lr){7-11}
       (\textbf{nm}) & Nominal & 98 $\%$ & 90 $\%$ & Mean & Sigma & Nominal & 98 $\%$ & 90 $\%$ & Mean & Sigma \\
        %\midrule
        \hline
		  \hline
        450 & 8.28 & 11.32 & 10.16 & 8.48 & 1.20 & 8.30 & 11.42 & 10.24 & 8.54 & 1.20 \\
        550 & 6.04 & 8.64 & 7.58 & 6.60 & 0.76 & 6.06 & 8.76 & 7.70 & 6.68 & 0.78 \\
        650 & 9.38 & 13.86 & 12.26 & 9.68 & 1.82 & 9.40 & 13.98 & 12.36 & 9.72 & 1.84 \\
        750 & 17.46 & 23.22 & 21.16 & 17.46 & 2.76 & 17.46 & 23.48 & 21.40 & 17.54 & 2.88 \\
        850 & 27.08 & 34.02 & 31.58 & 27.08 & 3.38 & 27.08 & 34.26 & 31.80 & 27.10 & 3.50 \\
        %\bottomrule
\hline
\hline
	\end{tabular}
    
\end{table*}

%=====================================
%=====================================

%%%%%%%%%%%%%%%%%%%%%%%%%%%%%%%%%%%%%%%%%%%%%%%%%%%%%%%%%%%%%%%%%%%%%%%%%%
%%%%%%%%%%%%%%%%%%%%%%%%%%%%%%%%%%%%%%%%%%%%%%%%%%%%%%%%%%%%%%%%%%%%%%%%%%

\subsection{Echelle Spectrometer: Optical design \& performance}

The o-ray and e-ray from the polarimeter section are fed into the spectrometer section of ProtoPol. A COTS off-axis parabola-mirror operates as the collimator, while a COTS Canon camera lens system (f/2L IS USM)  is employed as the camera optics. A magnification factor of $\times$0.61 is provided to the slit width by the spectrometer unit. As a result, $\sim$71$\mu$m at the input of the spectrometer would fall over 3.3 pixels at the detector (with pixel size of 13$\mu$m a side). A pupil diameter of $\sim$61 mm is formed at the echelle grating (54.49 lp/mm, 46 degrees blaze angle) by the spectrometer collimator. The dispersed beam from the echelle is subsequently cross-dispersed by two grating CDs. Two plane ruled reflection gratings - one with 150 lp/mm blazed at 8000 $\AA$ and the other with 300 lp/mm blazed at 4220 $\AA$ - are implemented to cover the wavelength ranges of $\sim$3810-5860 $\AA$ (Blue CD) and 5800-9770 $\AA$ (Red CD), respectively. Finally, the cross-dispersed multi-order spectra are recorded by the camera optics of the spectrometer unit onto a COTS 1K $\times$ 1K {ANDOR iKon M934} CCD detector featuring 13$\mu$m a side pixels.
\par 
The aforementioned design configuration provides a separation of 15 pixels ($\sim$4.9 arc-seconds) between the dispersed spectra of o-ray and e-ray within any given order. For consecutive orders, the minimum separation between o-rays is 28 pixels at the lowest wavelengths, with a progressive increase in separation at longer wavelengths. The baseline specifications of the echelle spectrometer module are provided in Table~\ref{proto_EchelleSpec-Specs}, while the corresponding COTS components are listed in Table~\ref{proto_spectrometer_component}.

%%%%%%%%%%%%%%%%%%%%%%%%%%%%%%%%%%%%%%%%%%%%%%%%%%
\begin{table*} 
\caption{Specification of Echelle spectrometer section of ProtoPol.}\label{proto_EchelleSpec-Specs}
\centering
\setlength{\tabcolsep}{5pt}
\renewcommand{\arraystretch}{1.2}
	\begin{tabular}{cc cc }
           \hline
		  \hline

\textbf{Parameters} & \textbf{Values} \\
\hline
\hline
            Pupil size                                       & $\sim$ 61mm                             \\ 
			Echelle spectrometer input f/\#                  & 5.4                                \\ 
			Echelle spectrometer output f/\#                 & 3.3                              \\ 
			Magnification                                    & 0.612                             \\ 
			Collimator focal length                          & 326.7mm                            \\ 
			Camera focal length                              & 200mm                            \\ 
			1 arc-second at the spectrometer object/slit plane          & $\sim$ 65$\mu$m \\ 
			Minimum separation between two order             & $\sim$ 28 pixels                         \\ 
			Separation between o \& e-ray at input           & 324 $\mu$m ($\sim$15 pixel on detector)\\ 
			Anamorphic magnification                         & $\sim$ 0.95-1.02					\\ 
			Slit sampling     & 3.6 pixels\\ 
            \hline
			Echelle Specifications                            &                                  
            \\ 
            \hline
			Require over-sized size (available size) in mm$^{2}$         & 102 X 102 \\ 
			Alpha and Beta angles for Echelle                & 46 degree              \\ 
			Out of plane gamma angle for Echelle                          & 10.5 degree                        \\ 
            \hline
			Cross Disperser Specifications                                  &                                 
            \\
            \hline
			\textbf{1. Blue cross disperser}                          &                                  \\ 
			Require over-sized size (available size) in mm$^{2}$         & 120 X 120 \\
			Alpha and Beta angles                            & 6.26 and 14.74 degree              \\ 
			Order range (Wavelength range)                   & 45-67 (3810$\AA$-5860$\AA$)               \\ 
			\textbf{2. Red cross disperser}                           &                                  \\
			Require over-sized size (available size) in mm$^{2}$         & 120 X 120 \\ 
			Alpha and Beta angles                          & 7.07 and 13.93 degree            \\
			Order range (Wavelength range)                   & 27-44 (5800$\AA$-9770$\AA$)              \\ 

		\hline
		\hline
	\end{tabular}
\end{table*}

%=====================================
\begin{table*} 
\caption{Details of various COTS components used in ProtoPol echelle spectrometer section}\label{proto_spectrometer_component}
\centering
\setlength{\tabcolsep}{5pt}
\renewcommand{\arraystretch}{1.2}
	\begin{tabular}{cc cc cc cc}
           \hline
		  \hline
		\textbf{Component} & \textbf{Component} & \textbf{Specification} & \textbf{Part} \\
		&\textbf{Type} &  & \textbf{Number} \\ 
        \hline
        \hline
		Collimator & Off-axis  & Focal length = 326.7 mm & 35-584$^a$ \\ 
                   & Parabolic Mirror  & Off-axis angle = 30 degrees & (M/S Edmund Optics)\\

        Echelle & Echelle & Groove density: 54.49 lp/mm & 53-*-416E\\
         Grating &        & Blaze angle: 46 degrees     & Richardson Gratings$^b$\\
                 &         &                            & (M/S Newport Corp.)\\

        Cross-     & Plane Ruled  & 1. Groove density: 150 lp/mm & 1. 53-*-426R\\
        Disperser  &  Reflection   &        Blazed at 8000$\AA$      & 2. 53-*-091R\\
        Gratings   &  Gratings  & 2. Groove density: 300 lp/mm     & Richardson Gratings\\
                 &              &    Blazed at 4220$\AA$        & (M/S Newport Corp.)\\

        Camera & Canon Camera  & Focal length:200mm, & Canon EF 200mm$^c$ \\ 
                   & Lens System  & $f/2$             & f/2L IS USM Lens \\

        Detector & ANDOR CCD  & 1K×1K with & ikon-M-934$^d$  \\ 
        System   & Camera     & 13 $\mu$m Pixel size   & \\

		\hline
		\hline
	\end{tabular}

    \begin{list}{}{}
        \item a:\href{https://www.edmundoptics.in/p/1016-x-3048mm-pfl-30-off-axis-parabolic-aluminum-mirror/33487/}{https://www.edmundoptics.in/p/1016-x-3048mm-pfl-30-off-axis-parabolic-aluminum-mirror/33487/}
        \item b: \href{https://www.newport.com/b/richardson-gratings}{https://www.newport.com/b/richardson-gratings}
		\item c: \href{https://www.canon-europe.com/lenses/ef-200mm-f-2l-is-usm-lens/}{https://www.canon-europe.com/lenses/ef-200mm-f-2l-is-usm-lens/}
		\item d: \href{https://andor.oxinst.com/products/ikon-large-ccd-series/ikon-m-934}{https://andor.oxinst.com/products/ikon-large-ccd-series/ikon-m-934}
    \end{list}
        
\end{table*}

%===================================

%==================================================

The tolerance analysis of the echelle spectrometer module of ProtoPol was conducted in a manner similar to that of the polarimeter module. The tolerance parameters were kept identical to the parameters assumed for the polarimeter module tolerance analysis as given in Table~\ref{ProtoPol_Polarimeter_tolerance_table}. The results (RMS diameters) for 10000 Monte-Carlo simulations are shown in Table~\ref{ProtoPol_Spectrometer_tolerance_table}. The RMS spot diameters for all the configurations were found to be $<$ 3 pixels ($\sim$ 1 arc-second) with a 98$\%$ confidence limit in worst-case scenario.

%=====================================
\begin{table*} 
\caption{Tolerance analysis of spectrometer unit in the wavelength range of 400-980 nm for on-axis field point for 10000 Monte Carlo runs. The distance of the CCD image plane from the ideal camera lens was used as a compensator ($\pm 1mm$).}\label{ProtoPol_Spectrometer_tolerance_table}
\centering
\setlength{\tabcolsep}{5pt}
\renewcommand{\arraystretch}{1.2}
	\begin{tabular}{cc cc cc cc cc cc cc}
           \hline
		  \hline

%\toprule
\multicolumn{1}{c}{\textbf{Order}} & \multicolumn{1}{c}{\textbf{Wavelength}} & \multicolumn{5}{c}{\textbf{RMS diameter ($\boldsymbol{\mu m}$)}}\\
\cmidrule(lr){3-7}

\textbf{Number} & \textbf{(nm)} & \textbf{Nominal} & \textbf{98$\%$} & \textbf{90$\%$} & \textbf{Mean} & \textbf{Sigma}\\

\hline
\hline
\textbf{Blue CD} &  \\
            64 & 399.0 & 15.44 & 27.10 & 23.36 & 16.64 & 4.86 \\ 
            64 & 405.6 & 12.60 & 24.80 & 20.78 & 14.38 & 4.50 \\
			64 & 412.0 & 10.52 & 22.60 & 18.54 & 12.86 & 3.92 \\
            
			54 & 472.9 & 9.88 & 20.36 & 16.76 & 12.32 & 3.14 \\ 
            54 & 480.7 & 10.42 & 21.20 & 17.56 & 12.88 & 3.30 \\
			54 & 488.3 & 10.90 & 22.20 & 18.40 & 13.44 & 3.52 \\

            44 & 580.3 & 9.96 & 20.26 & 16.62 & 12.46 & 3.02 \\ 
            44 & 590.0 & 9.82 & 20.60 & 16.90 & 12.60 & 3.10 \\
			44 & 599.3 & 9.70 & 20.82 & 17.02 & 12.74 & 3.12 \\
            
\textbf{Red CD} &  \\
			43 & 593.9 & 9.44 & 19.54 & 16.24 & 12.18 & 2.92 \\ 
            43 & 603.7 & 9.24 & 19.54 & 16.28 & 12.26 & 2.92 \\
			43 & 613.3 & 9.12 & 19.86 & 16.46 & 12.38 & 2.98 \\
            
			35 & 729.6 & 8.76 & 19.62 & 16.12 & 11.76 & 3.04 \\ 
            35 & 741.7 & 9.34 & 20.78 & 17.16 & 12.44 & 3.30 \\
			35 & 753.4 & 10.06 & 22.06 & 18.56 & 13.18 & 3.68 \\

            27 & 945.8 & 22.62 & 34.56 & 30.76 & 23.72 & 5.24 \\ 
            27 & 961.5 & 24.30 & 36.72 & 32.72 & 25.44 & 5.46 \\
			27 & 976.7 & 26.10 & 39.40 & 34.92 & 27.36 & 5.68 \\

		\hline
		\hline
	\end{tabular}

\end{table*}

%%%%%%%%%%%%%%%%%%%%%%%%%%%%%%%%%%%%%%%%%%%%%%%%%%

\subsection{Calibration Unit: Optical design}
\label{subsec-CalUnit}
\begin{figure}[]
	\centering
	\includegraphics[width=\textwidth]{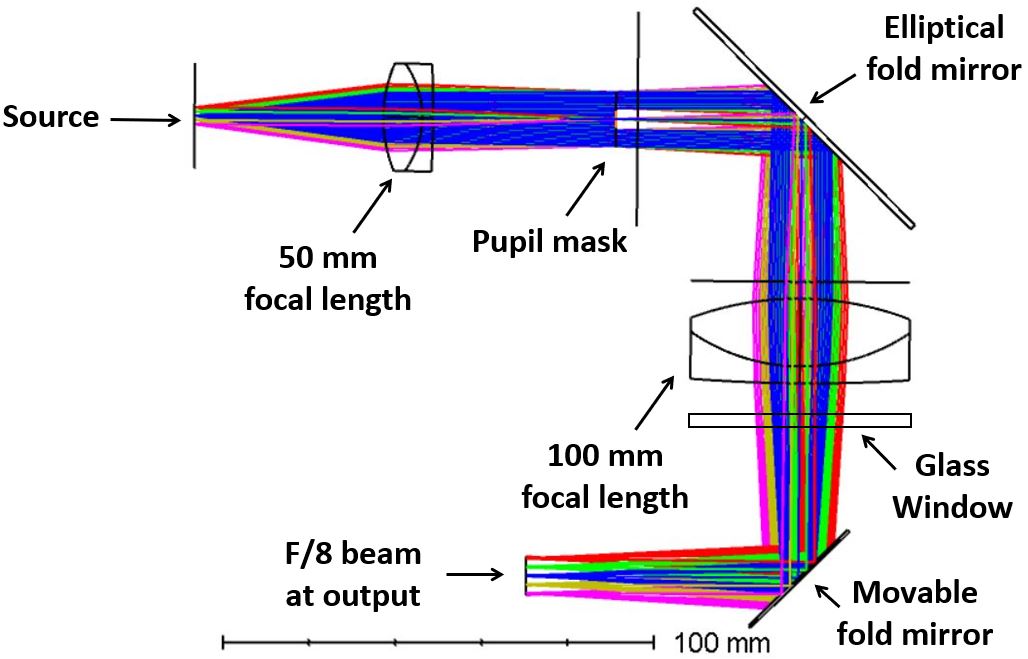}
	\caption{2-D optical design layout of Calibration unit of ProtoPol.}
         \label{calib_optics}
\end{figure}

The calibration unit is designed to perform wavelength calibration using Uranium-Argon (U-Ar) spectral lamp and determine the trace of echelle orders using halogen lamp. An oversize optical fiber (core diameter of 1 mm) is used to couple the light from these lamps to the slit-pinhole. A 30mm focal length lens is used to concentrate (overfill) the light of the U-Ar lamp onto the input of the optical fiber, whereas the light of the bright halogen lamp was coupled to the same optical fiber in a diffused manner.  The output end of the fiber  (acting as the source) is then mapped onto the pinhole using a relay system of COTS lenses having focal lengths of 50mm and 100mm, respectively, thereby providing a magnification of $\times$2 to the fiber core in a telecentric way.  The optical layout of the calibration optics is given in Figure \ref{calib_optics}. Two COTS fold mirrors are used in the optical design of the calibration unit (see Figure~\ref{calib_optics}). The second fold mirror is mounted on a linear translational stage, so it can be moved to direct the calibration beam into the ProtoPol's optics. In the configuration, this mirror also blocks the telescope beam. A pupil mask is inserted at the pupil plane such that the calibration beam mimics the actual telescope beam. The ratio of the aperture to obstruction of the pupil mask is the same as the ratio of the diameter of the primary mirror to the secondary mirror of the telescope. Two different interchangeable pupil masks are produced corresponding to PRL 1.2m and 2.5m telescopes, respectively, and the appropriate one can be placed at the pupil plane of the calibration unit optics.

%%%%%%%%%%%%%%%%%%%%%%%%%%%%%%%%%%%%%%%%
%%%%%%%%%%%%%%%%%%%%%%%%%%%%%%%%%%%%%%%%%%%%%

%%%%%%%%%%%%%%%%%%%%%%%%%%%%%%%%%%%%%%%%%%%%%%%%%%%%%%%%%%%
%%%%%%%%%%%%%%%%%%%%%%%%%%%%%%%%%%%%%%%%%%%%%%%%%%%%%%%%%%%

\section{Other Design Considerations}

Subtle design aspects of the echelle spectrometer play an important role in the successful design \& development of a spectro-polarimeter. Two such design considerations are discussed below, whose careful evaluation during the design phase and implementation during the development phase would ensure the desired resolution and minimization of cross-talk between adjacent spectral traces.

\subsection{Effect of slit-tilt and anamorphic magnification on slit projection}
\label{subsec-SlitTilt}

Due to the presence of an echelle and the CD grating arrangement in the optics chain as the dispersing elements, two effects come into play - the tilt in the projection of the slit on the CCD plane and the anamorphic magnification \cite{eversberg2014fundamentals} of the cross-dispersion grating. The first one is responsible for the rotation of the slit in the detector plane, thereby affecting the spectral resolution, while the other would cause a magnification/de-magnification effect along the cross-dispersion direction of the echelle orders. This would, thus, affect the inter-order and intra-order o- and e-ray separations. The contribution of these effects on the spot sizes on the image plane of the detector is, therefore, studied in detail and simulated as a part of the design process.
\par
While the geometrical magnifications of polarimeter and spectrograph optics can provide the first-order estimates of inter-order/intra-order separations between two consecutive o- and e- rays (of neighboring or same order), subtle effects may change their projected separation. The two such effects are (1.) the slit-tilt effect and (2.) the anamorphic magnification, which we discuss here. The out-of-plane angle ($\gamma$) of the echelle grating results in a slit-tilt effect \cite{eversberg2014fundamentals} which increases the projected slit width and lowers the resolution. If $\theta_{B}$ is the blaze angle and $\gamma$ is the out-of-plane angle of the echelle grating, then the tilt ($\chi$) in the slit with respect to the order normal is given by \cite{eversberg2014fundamentals},
$\tan \chi=2 \: \tan \theta_{B} \: \tan \gamma $
where $\theta_{B}=46^{\circ}$ and $\gamma=10.5^{\circ}$ are the blaze-angle and out-of-plane angle of the echelle grating. The above relation gives the projected slit-tilt $\chi=21^{\circ}$. This tilt in the projected slit increases the effective slit width on the CCD plane along the echelle dispersion direction by a factor of $1/\cos \chi$ while decreasing the corresponding parameter in the cross-dispersion direction by a factor $\cos \chi$. 
\par
On the other hand, anamorphic magnification of the system, caused by CD gratings, is given by,   
$m_{\textrm{anamorphic}} = \frac{\cos\alpha}{\cos \beta}$
where $\alpha$ and $\beta$ are the angles of incidence and dispersion, respectively, for the CD gratings. This anamorphic magnification further modifies the slit projection in the cross-dispersion direction, thereby affecting the separation between o-ray and e-rays in the cross-dispersion direction.

%----------------------------------

%%%%%%%%%%%%%%%%%%%%%%%%%%%%%%%%%%%%%%%%%%%%%%%%%%%%%%

\begin{figure}[]
	\centering
	\includegraphics[width=0.7\textwidth]{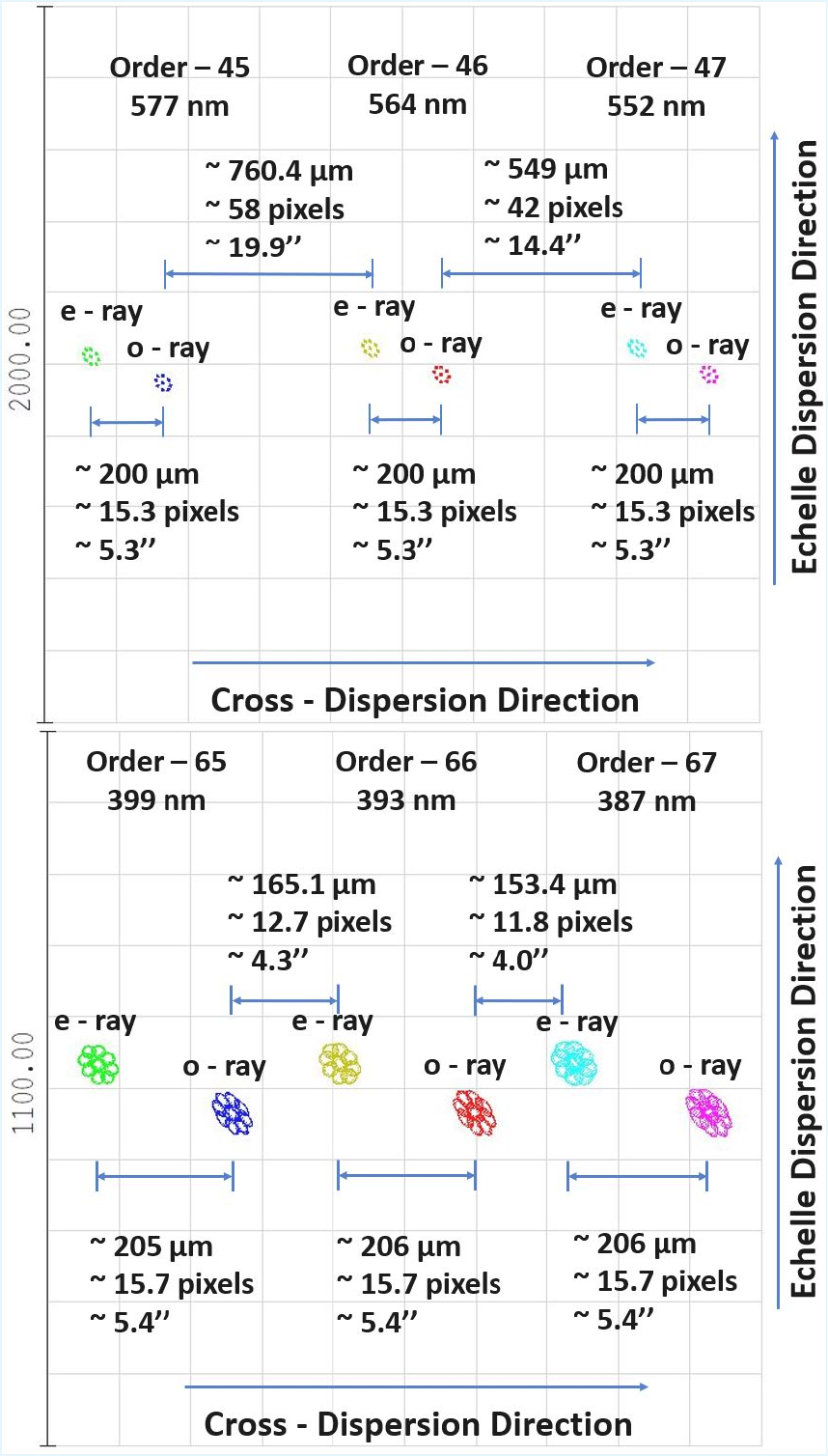}
    
    \caption{ZEMAX simulations to demonstrate the effect of slit-tilt and anamorphic magnification on the inter-order and intra-order separation of o-ray and e-ray for Blue CD. Seeing PSF (FWHM of 1 arc-second) is approximated by a circle of diameter 1 arc-second ($97\: \mu m$) at the telescope focal plane. The top panel shows the effect for the lower orders 45, 46, and 47, while the bottom panel shows the effect for the higher orders 65, 66, and 67, incident on the detector for the Blue CD. The central wavelengths corresponding to the orders are also mentioned. Box scales are mentioned in microns.} \label{Blue_6_orders}  
\end{figure}

%%%%%%%%%%%%%%%%%%%%%%%%%%%%%%%%%%%%%%%%%%%%%%%%%%%%%%%%%%%%%%%%%%%%%%%%%%

\par
To evaluate the impact of aforementioned effects on projected slit width (and therefore on spectral resolution), inter/intra order separations of spectra, etc., simulations were performed in ZEMAX using the entire optical design of ProtoPol. Here, we consider a source of circular diameter $97\mu$m at the object plane of the ProtoPol, which is taken as an approximation of seeing-dominated point spread function (PSF) having full-width-at-half-maximum (FWHM) of 1 arc-second ($\sim 97\: \mu m$ on 2.5m telescope's focal plane). Thereafter, 9 field points - one at the center and the remaining 8 points on the circumference of the circle - are selected, and their spot's positions on the CCD plane are determined by using the ProtoPol's optical design model in ZEMAX. Under the influence of the slit-tilt and anamorphic magnification effects, the circular field at the telescope focal plane will become elliptical at the CCD plane, and the ellipse will be rotated with respect to the dispersion direction. In the echelle dispersion direction, the circle gets elongated by a factor $1/\cos \chi$, and in the CD direction, the circle gets contracted by a factor $\cos \chi$. Such effects would also change the sampling scale of the PSF in both the echelle dispersion and cross-dispersion directions. Without the slit-tilt or anamorphic magnification, the simple geometric projection ($0.67 \times 0.61 \sim 0.41$) would cause 1 arc-second at the 2.5m telescope focal plane (97$\mu$m) to fall at 39.8$\mu$m (3.06 pixels) at the CCD plane.  As $\alpha$ and $\beta$ are equal for the echelle grating, it would not cause the echelle's anamorphic magnification in the echelle dispersion direction. So the dispersion direction would only be affected by the slit-tilt effect, and the pixel sampling in the dispersion direction would change to a higher value. On the other hand, the cross-dispersion direction would be affected by both the slit-tilt and the anamorphic magnification, and sampling in the CD direction would become less. Considering a circular PSF of diameter 1 arc-second (97 $\mu m$) at the telescope focal plane, the effect of slit-tilt and anamorphic magnification on the elongation and contraction of the circular PSF in the echelle dispersion direction and cross-dispersion direction, respectively, is analyzed for extreme orders (45 and 67) of Blue CD. The spot for o- and e-ray falls on approximately 3 pixels and 3.2 pixels along the dispersion and cross-dispersion directions, respectively. However, due to different pixel sampling along dispersion ($1^{''}\approx 3.2 \:$pixels) and cross-dispersion ($1^{''}\approx 2.88\:$pixels) directions under the influence of slit-tilt and anamorphic magnification, they sample $0.9^{''}$ and $1.1^{''}$ along the dispersion and cross-dispersion directions, respectively. Similar results are obtained for the extreme orders of Red CD (27 and 44) as expected, since the pixel sampling for Red CD ($1^{''}\approx 3.2\:$pixels along dispersion and $1^{''}\approx 2.86\:$pixels cross-dispersion) is very similar to the Blue CD values. The effect of slit-tilt and anamorphic magnification on the inter-order and intra-order separation of o-ray and e-ray for Blue CD is shown in Figure~\ref{Blue_6_orders} for the extreme orders. Similar results are also obtained for Red CD.

%%%%%%%%%%%%%%%%%%%%%%%%%%%%%%%%%%%%%%%%%%%%%%%%%%%%%%%%%%%%%%%%%%%%%%%%%%%%%

\subsection{Effect of varying seeing conditions on the cross-talk between o-ray \& e-ray seeing}

\begin{figure}[]

    \centering
	\includegraphics[width=\textwidth]{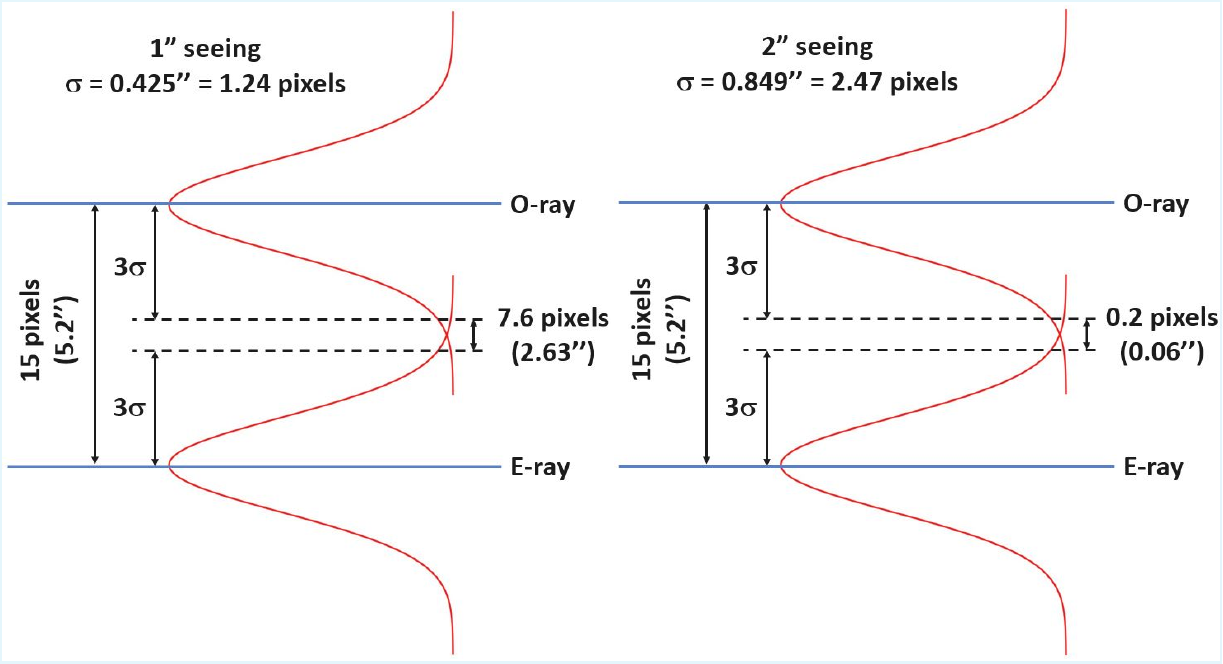}
  \caption{Minimum separation between o and e-ray trace for an order for 1 \textit{(left)} and 2 \textit{(right)} arc-second seeing conditions for Red CD. Even for 2 arc-second seeing conditions, the separation between o- and e-ray traces is determined to be 0.2 pixels at $3\sigma$ level. Similar results are also obtained for Blue CD.}\label{o-e_separation1}
\end{figure}

The spectral traces of o- and e- rays for various orders need to be well separated to minimize any cross-talk in the cross-dispersion direction between the traces. The extent of spectral traces along the cross-dispersion direction would depend on the pixel sampling of the PSF along the cross-dispersion direction and on the seeing profile. Figure~\ref{o-e_separation1} illustrates the situation for the seeing profiles of 1.0 and 2.0 arc-seconds for Red CD. The average separation between the traces for o and e-rays is 15.3 pixels and 15.0 pixels for Blue and Red CDs, respectively, close to the orders near the V-band, where the performance of the instrument is well optimized. As it can be seen from the plots, even for relatively poor seeing conditions (seeing FWHM$\sim$2 arc-seconds), the cross-dispersion intensity profiles of o- and e- rays are well separated at 3$\sigma$ level. There are a few common orders kept in both the Red and Blue CD data frames. These orders are used to stitch the spectra during data reduction. More details on the existing incident orders on the detector are discussed in the section~\ref{subsec-LabCharacterization-ProtoPol}.

%=================================================

%%%%%%%%%%%%%%%%%%%%%%%%%%%%%%%%%%%%%%%%%%%%%%%%%%%%%%%%%%%
%%%%%%%%%%%%%%%%%%%%%%%%%%%%%%%%%%%%%%%%%%%%%%%%%%%%%%%%%%%

\section{ProtoPol: The Design Performance \& Efficiency}
\label{sec-DesignPerformance}

\begin{figure}[htbp]
	\centering
	\fbox{\includegraphics[width=\textwidth]{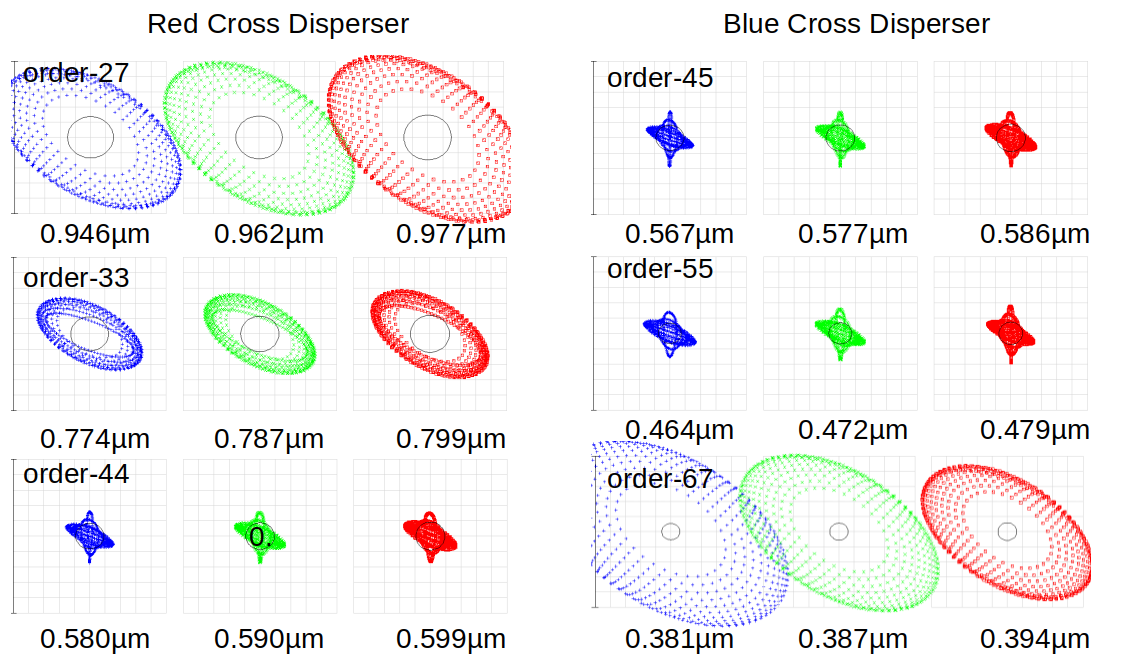}}
	\caption{\label{proto_ECHELLE-spt}Instrument performance with ideal lens as spectrometer camera (Box size is 26$\mu$m (2 pixels)).}
\end{figure}

\begin{figure}[htbp]

  \centering
  \includegraphics[width=\textwidth]{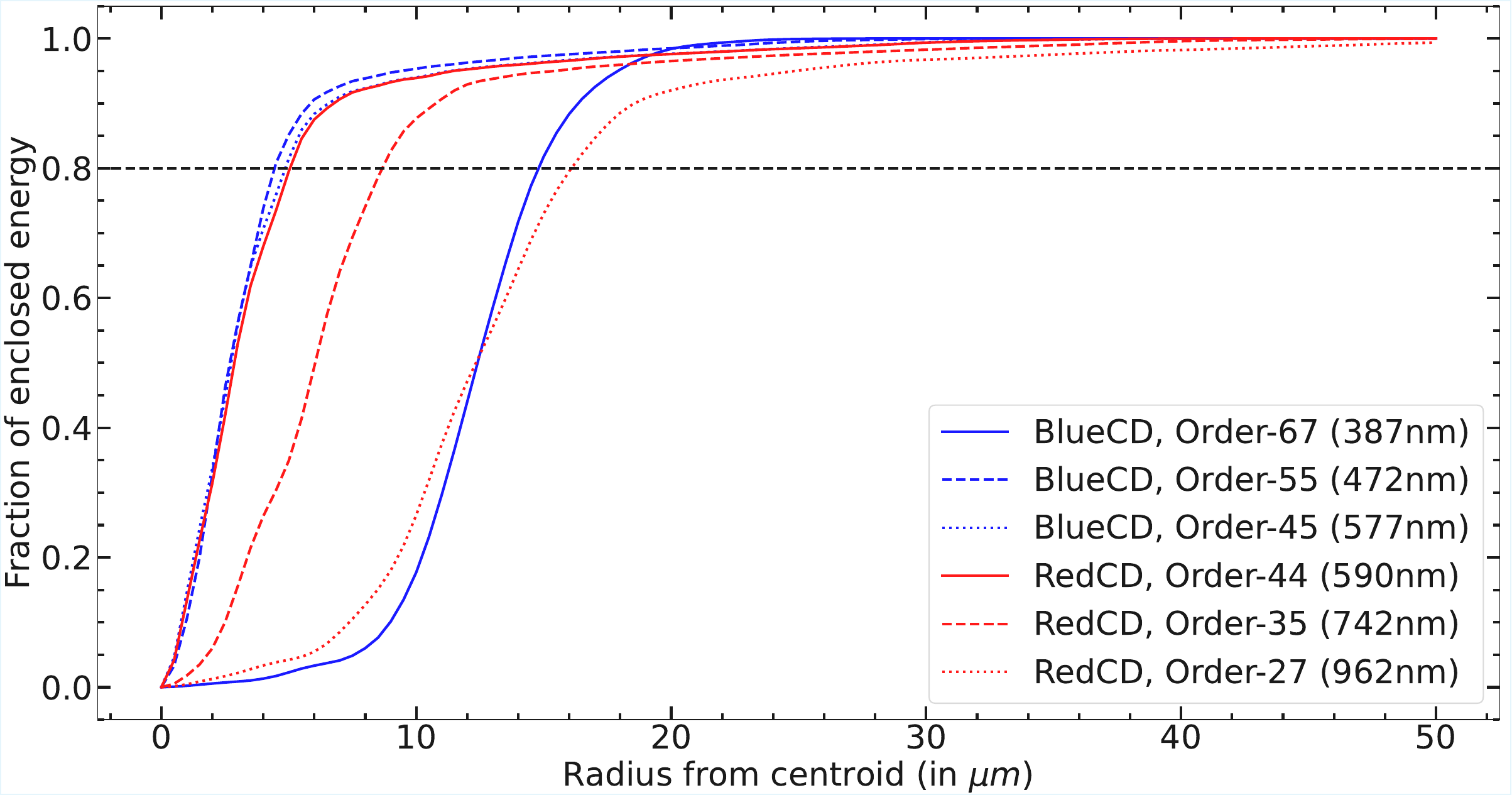}
    
  \caption{EE80 diagrams of Blue CD for orders 45 (dotted), 55 (dashed), and 67 (solid). The same for Red CD for orders 27 (dotted), 35 (dashed), and 44 (solid). The Blue and Red CD orders are color-coded by blue and red, respectively. The central wavelengths of the orders for which the EE80 plots were generated have also been mentioned in the legend. The 80$\%$ enclosed energy limit is shown by the black dashed line.}\label{ProtoPol_Blue_Red_EE80}
		
\end{figure}

\begin{figure}[]

    \centering
	\includegraphics[width=0.7\textwidth]{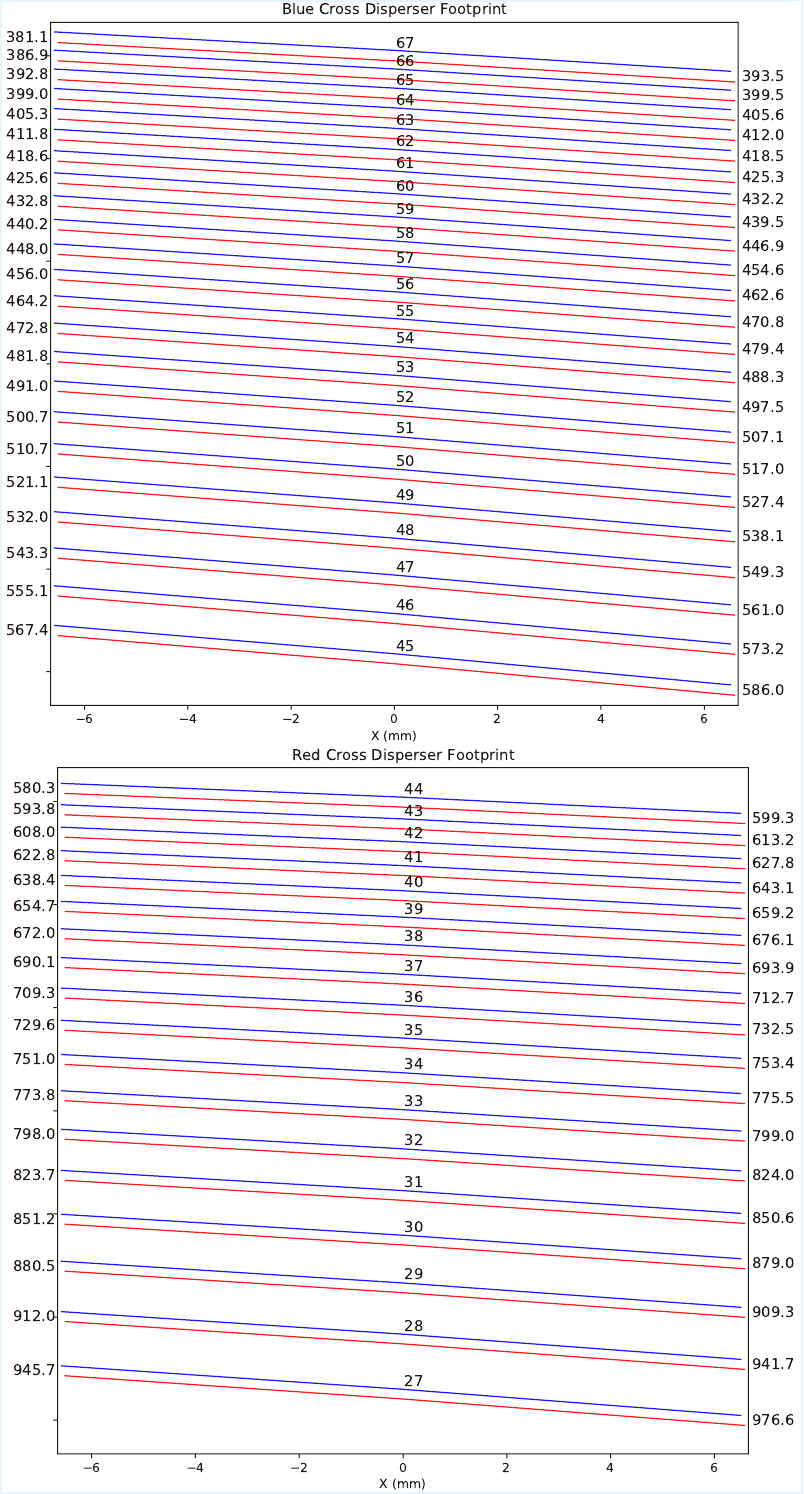}

  \caption{Echellogram footprint on  1K$\times$1K detector in the Blue CD \textit{(top)} and the Red CD \textit{(bottom)} mode. Red \& blue spectrum shows the same orders of o-ray and e-ray, respectively. Start and end wavelengths \& order numbers are mentioned on the left, right, and in the center of each of the orders, respectively.}\label{proto_blue}
\end{figure}

The optical design of the Canon camera lens system  (used in echelle spectrometer) is not available in the public domain, though unsuccessful efforts were made to obtain a design black-box from the manufacturers. Hence, to evaluate the performance of the instrument, an ideal lens with the same focal length as that of the Canon camera was incorporated as the camera optics in the ZEMAX design model of ProtoPol. Figure~\ref{proto_ECHELLE-spt} shows spot diagrams illustrating ProtoPol’s performance, with three cross-dispersed orders from the blue region and three from the red region shown. As can be seen, the spot diameters are well within 1.0-1.5 pixels for most of the orders. Only for the extreme wavelengths of Blue and Red CDs, the PSF tends to go poorer. The EE80 diagrams for o-rays for Blue and Red CD orders are shown in Figure~\ref{ProtoPol_Blue_Red_EE80}. The e-ray performance is almost identical to that of the o-ray, and hence only the o-ray EE80 diagrams are shown. Figure~\ref{proto_blue} depicts the o-ray and e-ray footprint of all orders on a 1K×1K CCD for the Blue CD and Red CD, respectively. Tables~\ref{tab:proto_per} provide root-mean-square (RMS) and EE80 diameters for some selected orders (spanning the entire wavelength range of the instrument over both CDs). For most configurations, the EE80 diameters are well within 1.5 pixels. As a conservative estimate, the optical performance of the  Canon lens system was approximated as Gaussian with a FWHM of 2 pixels. The convolution of this PSF with the PSFs of the rest of the optical system, as discussed above, may be approximated with a Gaussian of FWHM $\sim$ 2.5 pixels. This would broaden the effective slit width of 3 pixels (see section~\ref{subsec-SlitTilt}) to approximately 3.9 pixels. But this was the worst estimate; the later results during the characterization of the Canon camera and ProtoPol showed a much better performance, which is discussed in later sections.

%==========================================

%==========================================

%================================

%================================

%==========================================
\begin{table*} 
\caption{80\% EE, RMS, and Geometrical diameters for central field at detector plane (o-ray and e-ray have almost the same performance).}\label{tab:proto_per}
\centering
\setlength{\tabcolsep}{5pt}
\renewcommand{\arraystretch}{1.2}
	\begin{tabular}{cc cc cc cc}
           \hline
		  \hline
		\textbf{Order} & \textbf{Central} & \textbf{RMS} & \textbf{EE80} \\
		\textbf{No.} & \textbf{Wavelength ($\mu$m)} &\textbf{diameter ($\mu$m)} & \textbf{diameter ($\mu$m)} \\ 
        \hline
        \hline
		  27 & 0.962 & 27.1 & 32.6 \\ 
			33 & 0.787 & 14.7 & 20.5 \\ 
			44 & 0.590 & 5.7 & 10.1 \\ 
			45 & 0.577 & 5.7 & 10.0 \\ 
			55 & 0.472 & 5.7 & 9.1 \\ 
			67 & 0.387 & 27.8 & 30.2 \\ 
		\hline
		\hline
	\end{tabular}
\end{table*}

%===================================

%======================

%===============================

%==================================

\par
The following efficiencies are taken into consideration to calculate the throughput of ProtoPol : (1.) Atmospheric transmissivity was estimated from corresponding plot at 2 km altitude (Mt. Abu observatory altitude = 1.722 km) as simulated by \cite{smith2012lunar} using model MODTRAN 4.0; (2.) Reflectivities of the telescope’s primary and secondary mirrors were estimated from the aluminum reflectivity curves given by \cite{shanks2016optics}, scaled to peak reflectivity of $\sim 85 \%$  (for aluminum coatings of Mt. Abu telescope mirrors); (3). The CCD quantum efficiency curve is obtained from the specification sheet of the detector system. (4.) Effciency of ProtoPol optical chain was determined by considering transmissivity of each lens surface to be 98$\%$. Reflectivity of the mirrors and the off-axis parabola were considered from the reflectivity curves provided by the manufacturer. The grating reflectivities were estimated from the data sheets of the gratings as obtained from the manufacturer. As for the Canon camera lens system, since not much information was available in the public domain about its optical design, its expected throughput was taken as only 50\% as a conservative lower bound, as mentioned in \cite{harding2016chimera}.
\par
\begin{figure}[htbp]
	\centering
	\includegraphics[width=\textwidth]{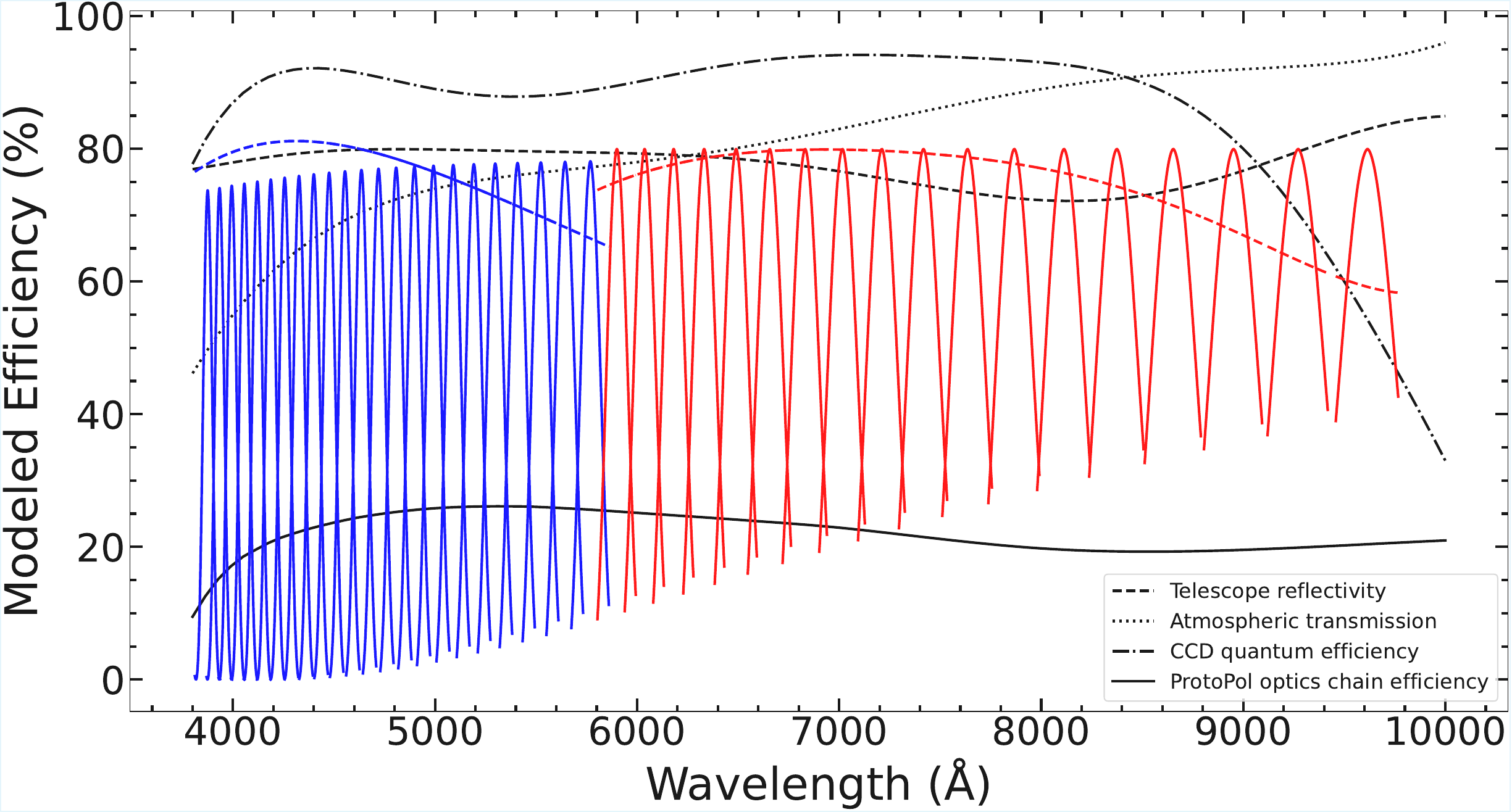}
	\caption{Figure shows the efficiencies of the atmospheric transmission (black dotted curve), telescope reflectivity (black dashed curve), the quantum efficiency of the CCD (black dash-dot curve), and the efficiency of ProtoPol's optical chain (black solid curve), which were considered to estimate the total throughput of the instrument as shown in Figure~\ref{theoretical_eff}. The blue and red dashed curves are the efficiencies of the blue and red CD, respectively, while the blue and red solid curves are the blaze efficiencies of the echelle grating for the blue and red CD orders, respectively.
    }\label{theoretical_eff_1}
\end{figure}
\begin{figure}[htbp]
	\centering
	\includegraphics[width=\textwidth]{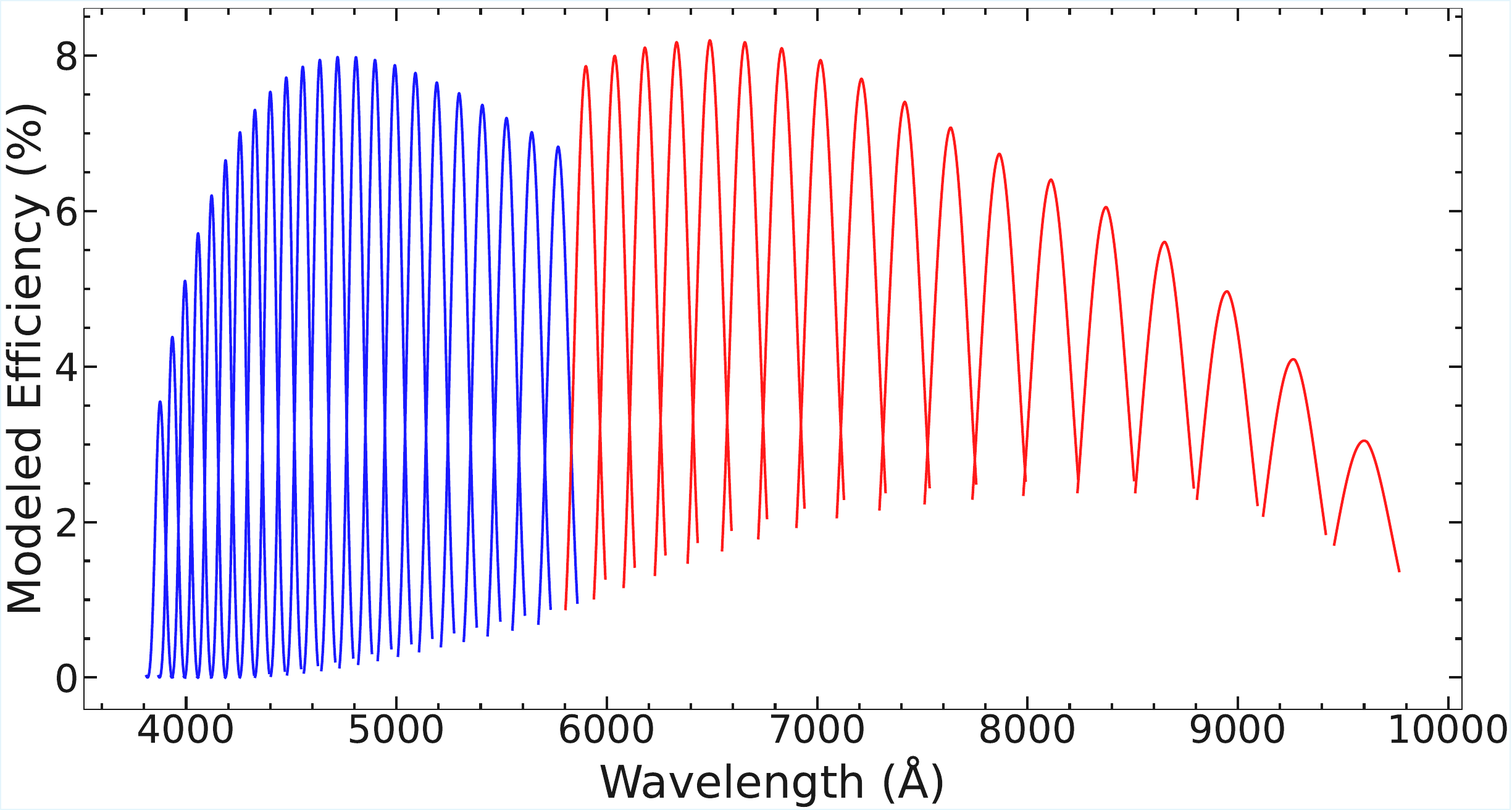}
	\caption{Cumulative expected efficiency curves of the echelle orders. In determining these efficiencies, the contribution of atmospheric transmission, reflectivity of telescope mirrors, the CCD quantum efficiency, the transmission of the ProtoPol optics chain, and the blaze efficiencies of the echelle and CD gratings are considered. The blue and red colored curves correspond to echelle orders dispersed by the Blue and Red CD gratings, respectively. The slit-losses are not considered.}
    \label{theoretical_eff}
\end{figure}

The efficiency curves of the Blue CD (\href{https://www.gratinglab.com/Products/Product_Tables/Efficiency/Efficiency.aspx?catalog=53-*-091R}{53-091R}) and Red CD
(\href{https://www.gratinglab.com/Products/Product_Tables/Efficiency/Efficiency.aspx?catalog=53-*-426R}{53-426R}) were obtained from the specification sheet of the respective gratings. The echelle order-wise grating efficiency was determined from the grating efficiency equations as given in \cite{bottema1981echelle}, with peak efficiency assumed from the reflectivity curve 
(\href{https://www.gratinglab.com/Products/Product_Tables/Efficiency/Efficiency.aspx?catalog=53-*-416E}{53-416E}) of the grating as given by the manufacturer. To estimate the combined efficiency of both gratings, the echelle grating efficiency for each order is determined, and then it is multiplied by the first-order efficiency of the corresponding CD grating. Figure \ref{theoretical_eff_1} shows the efficiency curves of the various components. The final cumulative efficiency of the various echelle orders (including atmospheric throughput, telescope optics, polarimeter optics chain, spectrometer optics chain, CCD quantum efficiency, and the respective efficiencies of the echelle and CDs) is shown in Figure \ref{theoretical_eff}. Peak efficiency of $\sim$6.8\% was thus estimated for the instrument incorporating all the factors. It is to be noted that this estimate does not include the slit-losses. More details on the finally achieved efficiency and on-sky performance of the instrument are discussed in Paper-II.

%==========================================

%==========================================

%==========================================

%==========================================

%%%%%%%%%%%%%%%%%%%%%%%%%%%%%%%%%%%%%%%%%%%%%%%%%%%%%%%%%%%%%%%%%%%%%
%%%%%%%%%%%%%%%%%%%%%%%%%%%%%%%%%%%%%%%%%%%%%%%%%%%%%%%%%%%%%%%%%%%%%

\section{Opto-mechanical Design of ProtoPol}
\label{sec-OptoMechDes}

\begin{figure}[]
\begin{center}
    \includegraphics[width=\textwidth]{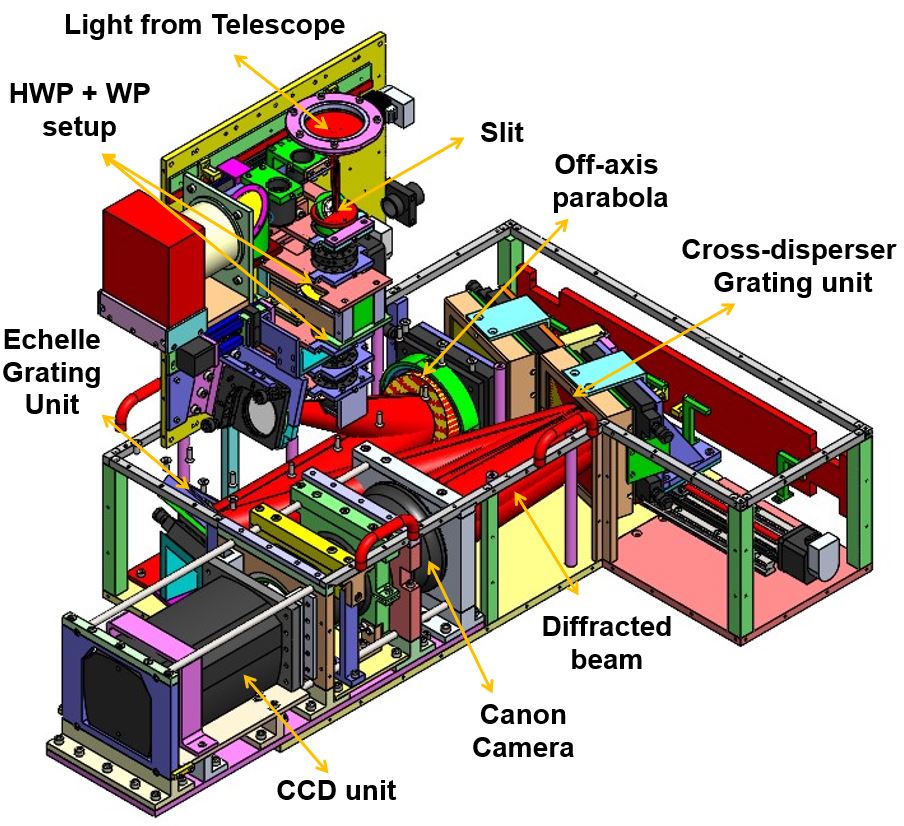}
    \caption{Complete CAD model of ProtoPol (without outside covers), including beam as imported from ZEMAX optical CAD design. The telescope beam enters the instrument through the pinhole slit and gets collimated by the collimator optics. The HWP+WP setup separates the beam into o- and e-rays, which subsequently get imaged by the camera optics and reflected by a set of three mirrors at the object plane of the spectrograph. An OAP acting as the collimator of the spectrograph section collimates the beam again, which then gets dispersed by echelle + CD gratings, and finally, the cross-dispersed orders get imaged on the detector by the Canon camera.}
    \label{complete_CAD}    
\end{center}    
\end{figure}

The opto-mechanical system of ProtoPol has been designed with a modular approach, wherein each of the sub-systems can be built and assembled independently before integrating into the complete instrument envelope. These sub-systems can be transported independently and reassembled on the telescope floor. This modular approach was in particular chosen as a test case for the development of the upcoming M-FOSC-EP instrument (see section~\ref{sec-Intro}), which would be bigger and heavier as compared to ProtoPol. ProtoPol's mechanical systems are divided into three modules: (1.) The polarimeter module, which interfaces with the telescope beam and at the output provides two orthogonal polarized o- and e- beams, which are to be fed to the echelle spectrometer module; (2.) The spectrometer module to disperse the polarized beams and to record the spectra onto the CCD detector; and (3.) a Calibration Unit hosting spectral lamps and additional optics to simulate the telescope beam and pupil, which is fed to the instrument for spectral calibration purposes. All these modules are discussed below. \href{https://www.solidworks.com/}{SOLIDWORKS} CAD software was used to develop the opto-mechanical design and CAD model of ProtoPol as shown in Figure~\ref{complete_CAD}.

%=========================================
%=========================================

%%%%%%%%%%%%%%%%%%%%%%%%%%%%%%%%%%%%%%%%%%%%%%%%%%%%%%%%%%%%%%%%%%%%%%%%%%%%%%%%%%%%%%%%%

\subsection{Polarimeter module and slit viewer optics}\label{sec:proto_polarimeter}

\begin{figure}[H]
	\centering
	\includegraphics[width=0.6\textwidth]{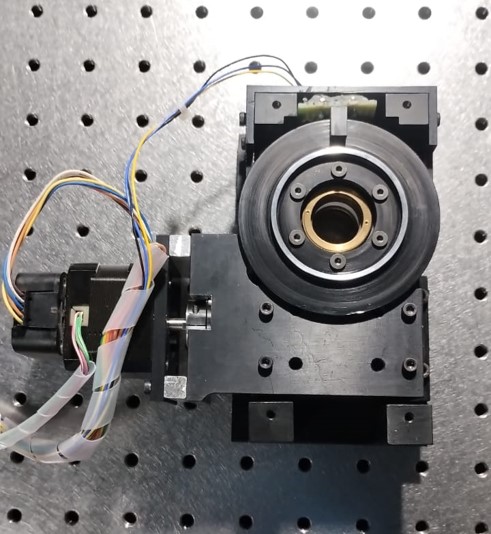}
	\caption{In-house developed stepper-motor driven rotational stage for HWP.}
    \label{HWP}
\end{figure}

Polarimeter module interfaces with the telescope mounting structure and couples the telescope beam to the instrument's optics. The telescope beams enter the instrument through a protective glass window and form the telescope's focal plane inside the instrument. A reflecting pinhole from ThorLabs (\href{https://www.thorlabs.com/thorproduct.cfm?partnumber=P150W}{P150W}) of diameter 150$\mu$m is fixed at the focal place at 45\degree. It would thus project an angle of $\sim$1.1 arc-second on the sky for PRL 2.5m telescope. While the pinhole would feed the on-axis target's beam into the instrument's optics, the reflecting surface would direct the rest of the field onto the slit-viewer guide camera. This slit-viewer tracks the source and provides an on-slit guiding of the telescope using the science target itself. The slit-viewer optics consists of a machine-vision camera lens (Model no.- \href{https://www.edmundoptics.in/p/16mm-f-18-1inch-HPi-Series-Fixed-Focal-Length-Lens/38837/}{33-814}, M/S Edmund Optics) and a guide camera (\href{https://www.astroshop.eu/astronomical-cameras/starlight-xpress-camera-lodestar-x2-autoguider-mono/p,44835}{Starlight Xpress Camera Lodestar X2}). The beam from the pinhole is collimated by the two achromat doublet lenses which are mounted and aligned within an off-the-shelf cage-rod system for 1-inch optics. The collimated beam is then passed through the polarization optics (Half-wave plate and Wollaston prism) mounted into the custom design mounts and a retractable long-pass cut-off filter (cutoff wavelength of 5250$\AA$; model no.\#84-744, M/S Edmund Optics) to prevent second-order contamination while observing at redder wavelengths. The long-pass filter is mounted on a linear translation stage and can be moved in or out of the path of the beam. The half-wave plate (HWP) is mounted on an in-house developed stepper-motor driven rotational stage (See Figure~\ref{HWP}), whose gear-ratio can provide angular resolution of 0.0016 \degree.

Thereafter, two achromatic doublet lenses will then image the separated ordinary (o-) and extraordinary (e-) rays onto the polarimeter focal plane, which coincides with the object plane of the spectrometer unit. These lenses are also mounted using a similar off-the-shelf cage-rod system as for the collimator lenses. Subsequently, a set of three fold-mirrors redirects the o- and e-rays towards the echelle spectrometer at $30\degree\:$ which is the requirements of the off-axis parabola (OAP) used within the spectrometer as collimator for the spectrometer unit.  In a normal mounting setup, with gravity vector pointing vertically downward, the polarimeter module is oriented vertically up while the spectrometer module is oriented horizontally, as depicted in Figure~\ref{complete_CAD}. Polarimeter module also consists of provisions for field viewer optics, Glan-Taylor prism for polarization calibration, and a fold mirror to direct calibration beam into the instrument while blocking the telescope beam. These are discussed in section~\ref{subsec-calUnit}.

%%%%%%%%%%%%%%%%%%%%%%%%%%%%%%%%%%%%%%%%%%%%%%%%%%%%%%%%%%%%%%%%%%%%%%%%%%%%%%%%%%

\subsection{Spectrometer Section}

The spectrograph section features four main components mounted on the horizontal base plate of the instrument (Figure~\ref{complete_CAD}): a parabolic collimator mirror, an echelle grating, CD gratings, and a Canon lens system coupled with a CCD camera. The incoming beam to the spectrometer section is first collimated by the off-axis parabolic (OAP) mirror (30$^\circ$ off-axis angle in the plane of the horizontal base plate). The collimated horizontal beam is then incident on the echelle grating, which directs it towards the CD gratings. For fine adjustments, the off-axis parabola, echelle grating, and CDs are mounted on tip-tilt stages (with locking provisions) for precise optical alignment. The two CD gratings are positioned on two separate linear motorized stages, allowing either to be selected and placed into the beam path as needed. After cross-dispersion, the light is directed towards the Canon lens system, which focuses the beam onto the ANDOR make CCD detector system. The Canon camera and ANDOR CCD form a separate sub-assembly which is firmly attached to the spectrometer base plate.

%%%%%%%%%%%%%%%%%%%%%%%%%%%%%%%%%%%%%%%%%%%%%%%%%%%%%%%%%%%%%%%%%%%%%%%%%%%%%%%%%%%

\subsection{Calibration Unit}
\label{subsec-calUnit}

The calibration unit is used for the wavelength calibration of the spectra as well as the characterization of instrumental polarization. A Glan-Taylor prism is used for polarization calibration, while Uranium-Argon (U-Ar) spectral calibration lamps are used for wavelength calibration purposes. A halogen lamp is also provided for tracing of echelle orders of the target stars. A 30mm focal length lens is placed in front of the U-Ar lamp to focus the light on the input of an optical fiber having a 1 mm core diameter. The output of the optical fiber is positioned at the object plane of a calibration re-imaging optical system, which consists of two commercial off-the-shelf achromat doublet lenses and two fold mirrors. This optical setup mimics the telescope’s beam, projecting the light from the lamps onto the telescope’s focal plane. One of the fold mirrors is mounted on a linear translational stage (along with the Glan-Taylor prism) to direct the calibration beam into the ProtoPol's optical chain while blocking the incoming beam from the telescope. This stage is mounted inside the polarimeter module above the slit/pin-hole. The calibration module is mounted in the same vertical orientation as the polarimeter module. 

\par
Additionally, the calibration module includes a field viewer with a separate field-viewing COTS lens and detector to map the sky around the target source for identification of the source. A movable fold mirror (mounted on the same linear translation stage holding calibration mirror and Glan-Taylor prism),  directs the telescope field towards the field viewer optics. The assembly features a CCD camera (Model no. \href{https://telescopes.net/sbig-stf-8300-monochrome-camera-complete-imaging-system-used.html}{SBIG STF-8300M}, 3326 $\times$ 2504 pixels) and a re-imaging lens system. The linear translational stage in this calibration setup holds the fold-mirrors for the calibration unit, and the field viewer, as well as the Glan-Taylor prism. Any of these can be moved in or out of the path of the telescope's beam, thus enabling various calibration, observation, and target acquisition modes of ProtoPol.
\par
An off-axis auto-guider unit is also developed and attached to the top plate of ProtoPol for 1.2m PRL telescope, considering the telescope does not have an in-built auto-guiding system. This auto-guider unit employs a pick-up mirror kept at 45\degree, which can be moved along a periphery with the help of a stepper motor. This mirror picks up an off-axis sky-field and directs it to a guide CCD - another Lodestar CCD camera - Which is coupled to the telescope's RA-DEC motion system for auto-guiding. This system is redundant for the latest 2.5m PRL telescope, which has an in-built off-axis guider system.

%%%%%%%%%%%%%%%%%%%%%%%%%%%%%%%%%%%%%%%%%%%%%%%%%%%%%%%%%%%%
%%%%%%%%%%%%%%%%%%%%%%%%%%%%%%%%%%%%%%%%%%%%%%%%%%%%%%%%%%%%

\section{Assembly-Integration-Testing (AIT) and Laboratory Characterization of ProtoPol}

As discussed above, ProtoPol is designed in a modular way; therefore, each of the sub-systems of ProtoPol was separately assembled and tested for their expected performances, then integrated and characterized in the laboratory for the final system performance. Before beginning the Assembly, Integration, and Testing (AIT) phase of the instrument, laboratory tests were conducted to verify the performance of the COTS Canon camera system. During the design phase of the instrument, image quality of the Canon-camera system was assumed to be of 2 pixels FWHM (see section~\ref{sec-DesignFull}). However, laboratory characterization of the camera revealed superior performance, with the point spread function (PSF) less than 1-pixel FWHM across visible wavelengths, though admittedly, only 50 mm beam diameter was used to test the camera performance, which is about 2/3rd the expected beam diameter ($\sim 78$mm). During transportation of the instrument from the laboratory to the observatory, the three modules viz. the polarimeter, calibration module, and spectrometer, were shipped separately and subsequently reassembled on-site. This helps in preserving the optical alignment achieved in the laboratory and prevents any performance degradation during the transportation process. The following sections detail the assembly and integration procedures for each module.

%%%%%%%%%%%%%%%%%%%%%%%%%%%%%%%%%%%%%%%%%%%

\subsection{Assembly and Integration of Polarimeter Unit}
\label{subsec-AITPolUnit}

\begin{figure}[]

    \centering
	\includegraphics[width=\textwidth]{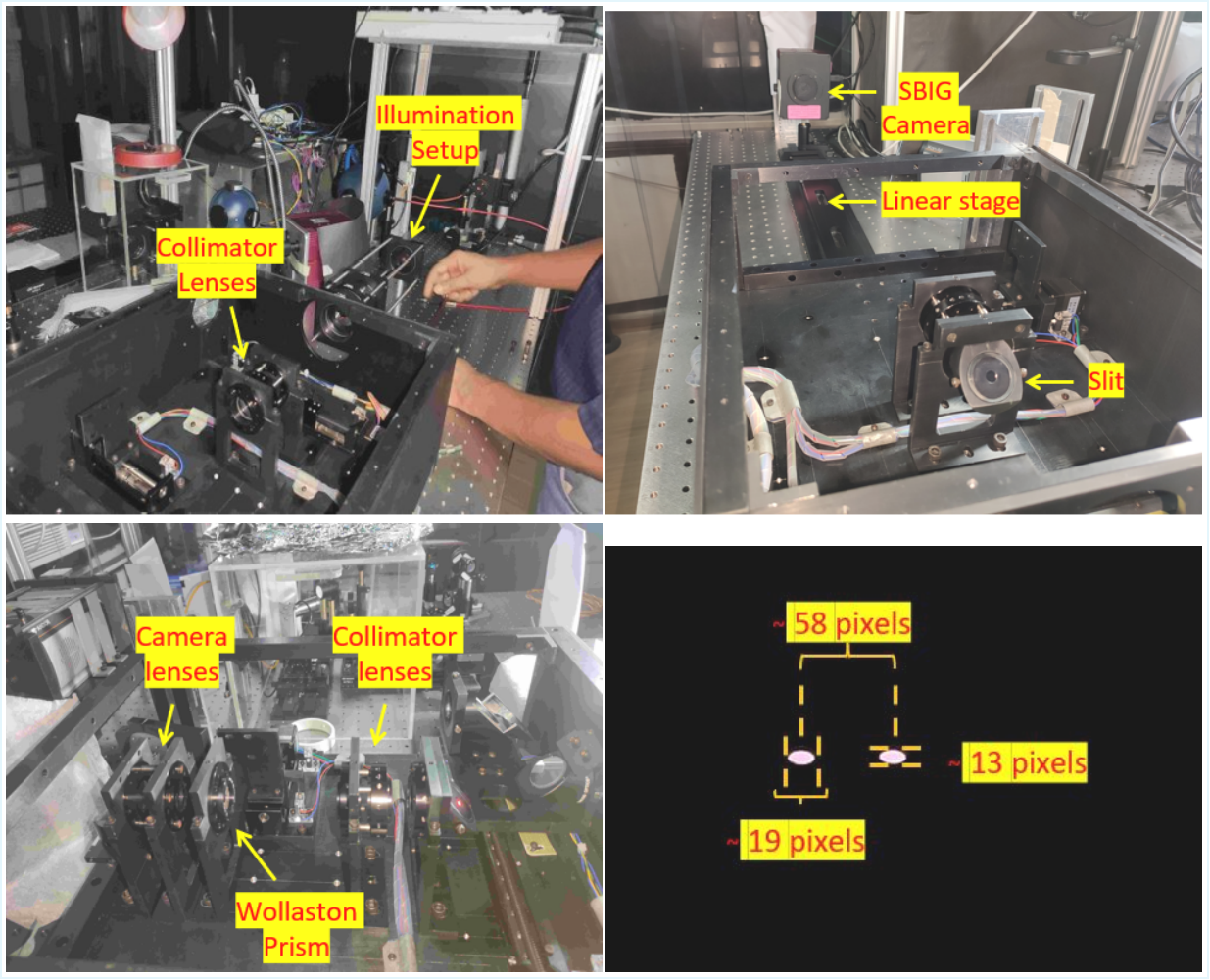}
    
  \caption{Figures show the assembly step of the ProtoPol's polarimeter unit. The top two panels show the setup used to assemble and align the polarimeter unit collimator lenses. The bottom-left panel shows the relative positioning of collimator optics, Wollaston prism, and camera optics inside the polarimeter unit. The image quality, dimension, and separation of o- and e-images of the pinhole on the polarimeter image plane are shown in the bottom-right panel.}\label{pol_col}
		
\end{figure}

The lenses forming the collimator and camera optics within the Polarimeter unit were aligned first using the laser retro-reflection technique. For this purpose, an illumination set-up was prepared. The illumination set-up consisted of two lenses (100 mm and 200 mm focal lengths) in $4-f$ configuration and an adjustable iris at their intermediate common focal plane to precisely control the f-number of the output beam. This illumination set-up was projecting the output of a 1-mm core-diameter optical fiber whose input in turn was uniformly fed by an integrating sphere. Thus, the output of this illuminating set-up was a uniform beam of the desired f-number. For the alignment of the collimator optics, its two achromat COTS lenses were first mounted within their lens mounts and fixed to the cage-rod system with adjustable separation. They were placed in front of a pinhole, which was illuminated by an $f/8$ beam coming out of the illumination setup discussed earlier. Their separation was adjusted to collimate the beam. The collimation was checked by placing a commercial \href{https://telescopes.net/sbig-stf-8300-monochrome-camera-complete-imaging-system-used.html}{SBIG-8300} camera mounted on a linear translation stage at a large distance. By moving the camera along the direction of collimation and by evaluating the beam diameter at multiple distances, the collimation was confirmed. The complete experimental setup for these alignment and verification procedures is shown in the top panels of Figure~\ref{pol_col}. 

%%%%%%%%%%%%%%%%%%%%%%

%%%%%%%%%%%%%%%%%%%%%%%%%%%%%%%%

\par 
Two other COTS achromatic lenses were used to form the camera optics of the polarimeter, which were aligned using the same laser retro-reflection method. It was done in two steps: 1.) first, the lenses were illuminated in reverse in the sense that they would provide a collimated beam once their proper separation is achieved; and 2.) they were again re-aligned along with the rest of the elements of the polarimeter optics. For this, the slit-pinhole and collimator optics were then mounted and fixed in the chassis of the polarimeter unit. Following the collimator optics, a Wollaston prism was mounted to split the collimated beam into o- and e-ray components. Thereafter, the assembled camera optics were mounted in the instrument setup. The o- and e- rays were focused by the polarimeter camera lenses onto the polarimeter focal plane. In the instrument setup as well, the separation between the lenses was again carefully adjusted by evaluating image quality at the polarimeter focal plane. As the focal plane formed very close to the last mechanical surface of the camera lens, direct imaging of the beam by the commercial SBIG camera (with a flange focal length of 17.5 mm) was not possible. Instead, a commercial webcam sensor was employed to assess image quality. The 150$\mu$m pinhole/slit, oriented at 45\degree, creates an elliptical projection of the circular pinhole with dimensions of 106$\mu$m and 150$\mu$m in orthogonal directions.  The Wollaston prism is appropriately rotated to align the shorter 106$\mu$m dimension  perpendicular to the direction of ray-separation forming the slit-width of the subsequent spectrometer optics. The greater dimension (150$\mu$m), thus, formed the slit-height. The demagnification ($\times$0.67) of the polarimeter unit would further reduce this elliptical projection to approximately 71$\mu$m $\times$ 100$\mu$m (equivalent to 13 $\times$ 19 webcam-sensor pixels) at the polarimeter focal plane. At the focal plane, the designed separation between o- and e-images of 324$\mu$m (about 58 webcam-sensor pixels) was successfully verified during the AIT process. The experimental configuration and resulting pinhole projection are displayed in the lower sections of Figure~\ref{pol_col}.
\par 
After the polarimeter camera optics, a set of three fold mirrors is used to direct the output beam of the polarimeter towards the input of the spectrometer section, wherein an off-axis parabolic (OAP) mirror serves as the collimator of the echelle spectrometer unit. In the polarimeter's collimated beam space, a rotatable half-wave plate (HWP) was positioned before the Wollaston prism. To prevent second-order spectral contamination in the longer wavelengths when operating in Red CD mode, a long-pass filter was installed immediately after the Wollaston prism. This filter is mounted on a linear motorized stage, allowing insertion into or retraction from the beam path as needed for observations in the Red CD and Blue CD configurations, respectively.
\par
\begin{figure}[]

    \centering
	\includegraphics[width=\textwidth]{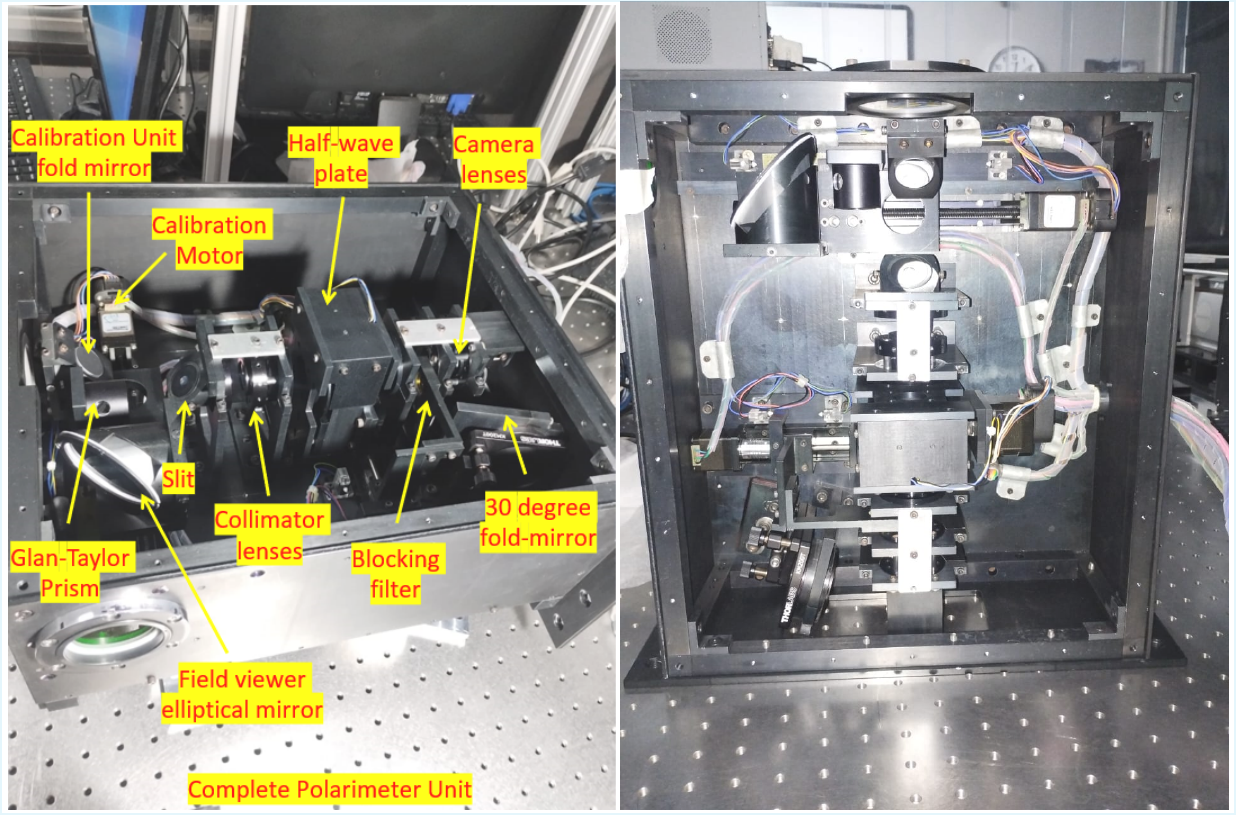}
    
  \caption{The completely assembled polarimeter unit of ProtoPol}\label{pol_complete}
			
\end{figure}

After completion of the optical alignment of the polarimeter optics, a motorized linear stage was installed before the pinhole, housing two fold-mirrors and a Glan-Taylor prism. This stage enables selection between different instrument operation modes by positioning the appropriate optical element in the beam path: (i) star observation mode - by removing all the elements from the path of telescope beam; (ii) polarization calibration mode where starlight passes through the Glan-Taylor prism to generate 100\% polarized output, (iii) wavelength-calibration mode by employing a fold-mirror to direct the beam from calibration unit (U-Ar or halogen lamp) into ProtoPol's main optical chain, and (iv) field-viewer mode where a fold mirror redirects the incoming telescope beam towards a field viewer camera for sky-field identification. The fully assembled polarimeter configuration is displayed in Figure~\ref{pol_complete}.

%%%%%%%%%%%%%%%%%%%%%%%%%%%%%%%%%%%%%%%%%%%%%%%%%%%%%

%%%%%%%%%%%%%%%%%%%%%%%%%%%%%%%%%%%%%%%%%%%%%%%%%%%%

%%%%%%%%%%%%%%%%%%%%%%%%%%%%%%%%%%%%%%%%%%%%%%%%%%%%%%%%%%%%
%%%%%%%%%%%%%%%%%%%%%%%%%%%%%%%%%%%%%%%%%%%%%%%%%%%%%%%%%%%%
%%%%%%%%%%%%%%%%%%%%%%%%%%%%%%%%%%%%%%%%%%%%%%%%%%%%%%%%%%%%

\par
\subsection{Assembly and Integration of	Calibration Unit}
\label{subsec-AITCalUnit}
\begin{figure}[]
\begin{center}
	\includegraphics[width=\textwidth]{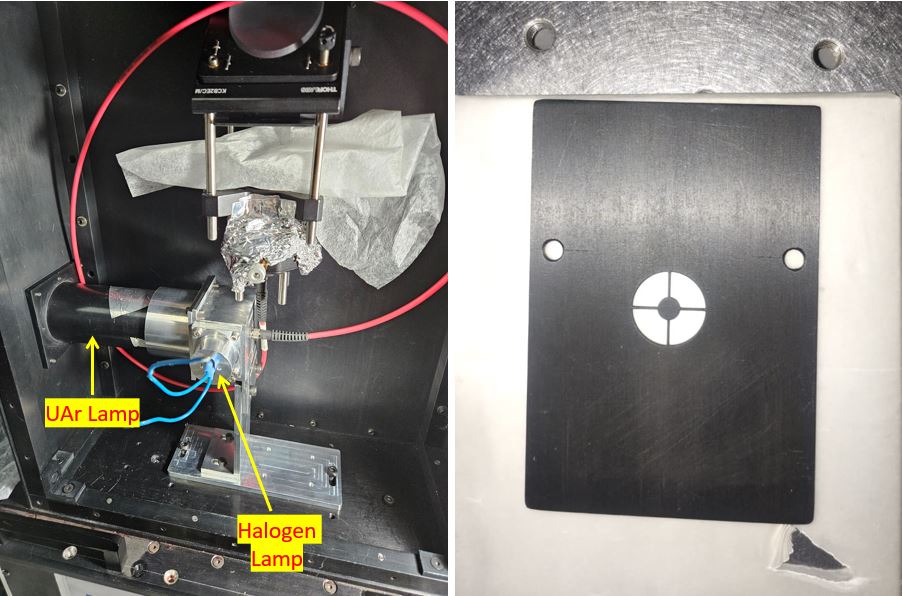}
    \caption{Calibration unit setup of ProtoPol (left) and a pupil mask (right) }\label{calib}    
\end{center}    
\end{figure}

As discussed in section~\ref{subsec-CalUnit}, the calibration unit of ProtoPol incorporates two light sources: a Uranium-Argon (U-Ar) hollow cathode lamp and a halogen lamp, and an optical re-imaging system to simulate an $f/8$ telescope beam (2.5m telescope) or an $f/13$ telescope beam (1.2m telescope) for wavelength calibration purposes of the stellar spectra. The optics (lenses and fold mirror) were assembled into the COTS mounts and cage-rod system and aligned for $4-f$ imaging configuration. A pupil mask is kept at an intermediate pupil plane in this opto-mechanical system to accurately replicate telescope beam characteristics.  The pupil mask features the same aperture-to-obstruction ratio as the ratio of the telescope primary to secondary mirror. The obstruction is suspended in the center of the aperture using four arms, mimicking the spider arms of the telescope's secondary mirror. Two distinct masks were manufactured, corresponding to the PRL 1.2m and 2.5m telescope configurations. The light from the lamps is coupled to this optics by an optical fiber of core-diameter 1 mm, whose input end is illuminated by the lamps. A 30mm focal length lens is employed to focus the light of the U-Ar lamp onto the fiber input, while the light of the halogen lamp is diffused in a small enclosure and coupled to the fiber directly.  The output of the optical fiber is mapped onto the slit-pinhole in the polarimeter unit by the calibration unit optics. The complete calibration assembly, including one of the pupil masks, is illustrated in Figure~\ref{calib}.

%%%%%%%%%%%%%%%%%%%%%%%%%%%%%%%%%%%
%%%%%%%%%%%%%%%%%%%%%%%%%%%%%%%%%%%%%%%%

\subsection{Assembly and Integration of	Spectrometer Unit}
\label{subsec-AITSpecUnit}

\begin{figure}[H]
	\begin{center}    
		
		\includegraphics[width=\textwidth]{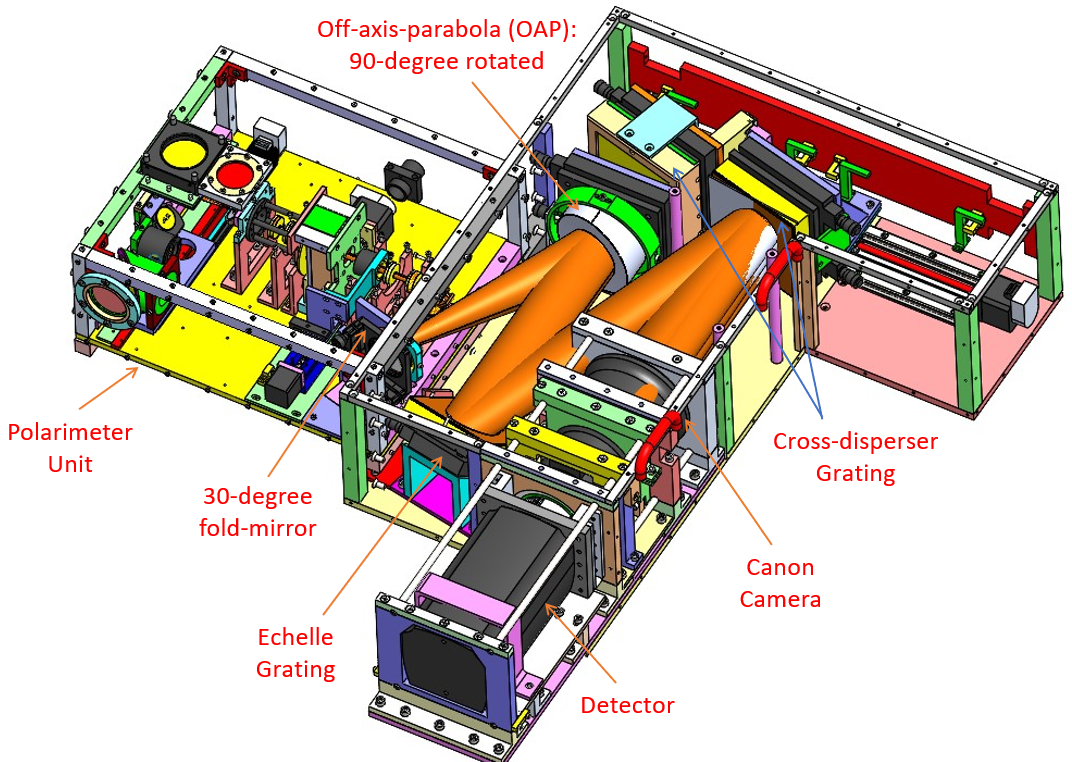}		
		\caption{The CAD model showing the horizontal set-up used for the alignment of the spectrometer unit with the polarimeter unit. The polarimeter unit was kept in the plane of the spectrometer unit for easier alignment of the spectrometer optics. The OAP was rotated 90 degrees to keep the horizontal setup symmetric to the original setup.}
		\label{protopol_horizontal_CAD}
		
	\end{center}    
\end{figure}

\begin{figure}[]

    \centering
	\includegraphics[width=\textwidth]{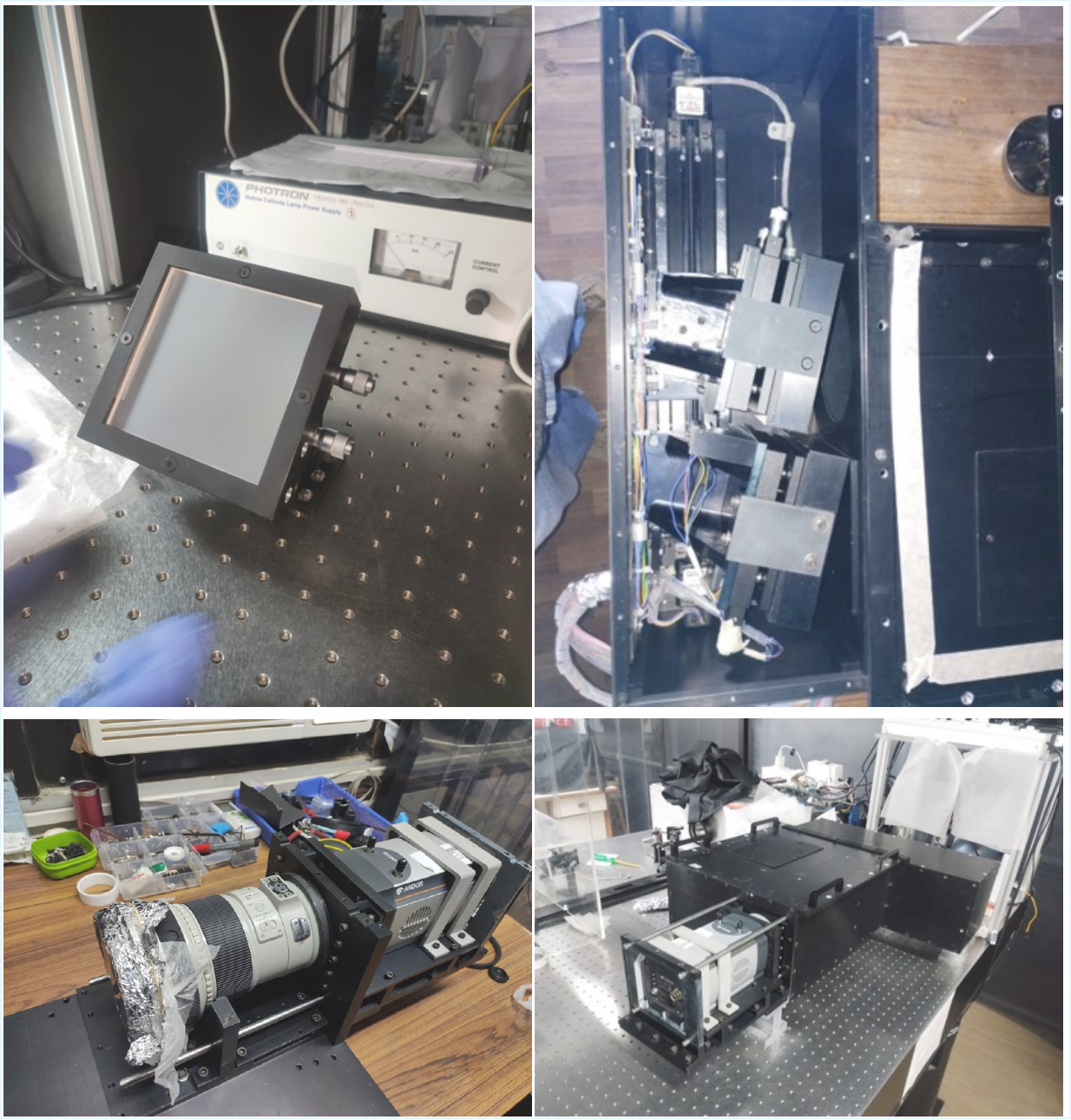}

   \caption{\textit{(Top left)} Assembled echelle grating on tip-tilt stage; \textit{(Top right)} CD gratings mounted on two linear stages to bring Blue and Red CD in and out of the beam path; \textit{(Bottom left)} Canon camera and detector assembly; \textit{(Bottom right)} Fully assembled spectrometer unit of ProtoPol}\label{spec_complete}
			
\end{figure}

\begin{figure}[]
	\begin{center}    
		
		\includegraphics[width=\textwidth]{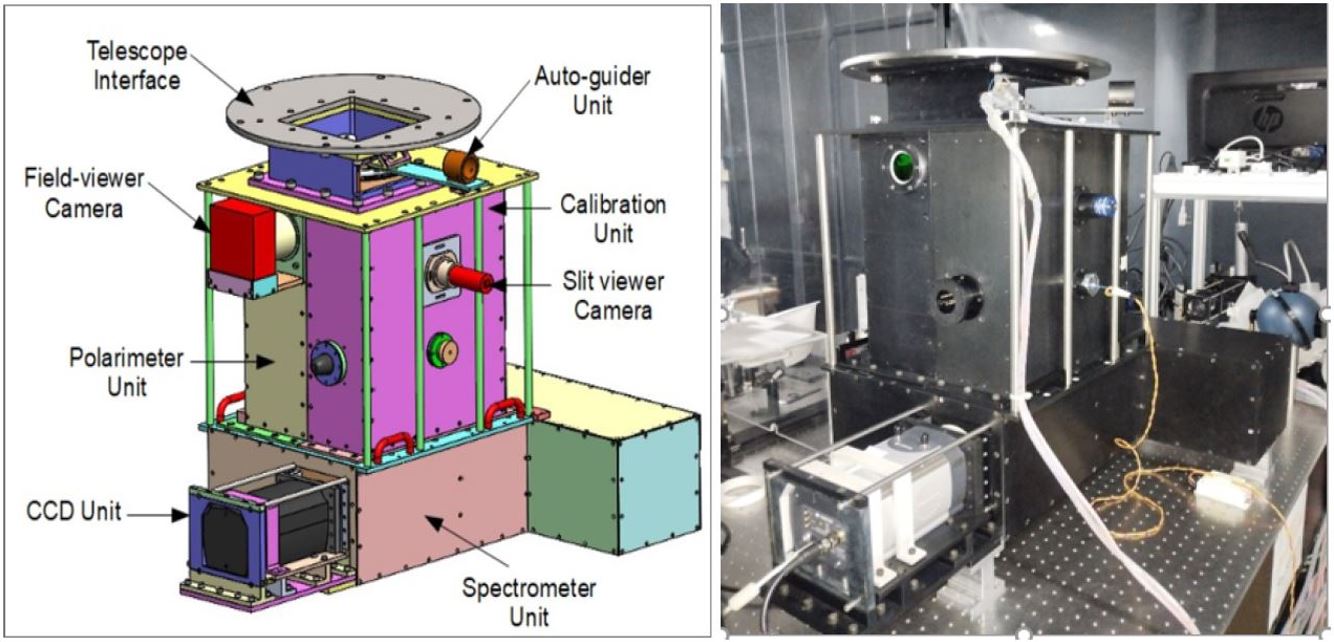}    
        \caption{Left: The CAD model of ProtoPol. Right: ProtoPol completed assembly in laboratory.}\label{protopol-complete}
		
	\end{center}    
\end{figure}

The spectrometer and polarimeter units were assembled in conjunction. To facilitate optical alignment between the polarimeter optics with the spectrometer optics, the entire setup was arranged in a horizontal configuration (Figure~\ref{protopol_horizontal_CAD}), achieved by rotating the off-axis parabolic collimator of the spectrometer by 90 degrees. The CCD detector system was first mounted with the Canon camera lens system and fixed into the spectrometer base plate. Two motorized linear translational stages for CD gratings were also fixed to the base plate. During the initial alignment process, temporary flat mirrors were placed in place of diffraction gratings to ensure precise optical adjustment using laser retro-reflection techniques. The slit-pinhole in the polarimeter unit was illuminated with a laser beam through the illumination set-up discussed in section~\ref{subsec-AITPolUnit}. This laser beam is then used to coarse-align all the optics through the retro-reflection method. Once successful alignment was achieved, the system was functionally verified by illuminating the slit-pinhole with a halogen lamp and confirming proper focus of both o- and e-ray images on the CCD detector. Thereafter, the mirrors were replaced by diffraction gratings. The mechanical mounts for all the gratings and OAP were designed and developed in-house and consist of the tip-tilt stages for fine alignment purposes. The gratings were then adjusted to properly orient the cross-dispersed spectral orders on the CCD detector. The tip-tilt stages, on which the gratings and off-axis parabolic mirror (OAP) were mounted, were carefully aligned for any fine adjustments and then locked in position. After completion of the alignment process, the OAP was returned to its original designed orientation by rotating back 90 degrees, and the mechanical enclosures of the polarimeter unit and spectrometer unit were assembled in their original positions. The calibration unit was also integrated into the system, and the spectral line sharpness of U-Ar lamp emissions lines in both dispersion and cross-dispersion directions was verified. Analysis of the emission lines of U-Ar lamp revealed the FWHM of the emission lines to be approximately 2.5 and 3.3 pixels for Blue and Red CD, respectively, across multiple echelle orders, corresponding to spectral resolutions between 0.3-0.4 $\AA$ and 0.6-0.75 $\AA$ for Blue and Red CD, across the 4000-9600 $\AA$ wavelength range - consistent with design specifications (see Paper-II for full discussion). These characterization aspects are discussed in the next section. Figure~\ref{spec_complete} illustrates the assembled grating and the spectrometer unit, while Figure~\ref{protopol-complete} shows the fully integrated ProtoPol instrument alongside its CAD model for reference.

%%%%%%%%%%%%%%%%%%%%%%%%%%%%%%%%%%%%%%%%%%%

%%%%%%%%%%%%%%%%%%%%%%%%%%%%%%%%%%%%%%%%%%%

%%%%%%%%%%%%%%%%%%%%%%%%%%%%%%%%%%%%%%%%%%%%%%%%%%%%%%%

%%%%%%%%%%%%%%%%%%%%%%%%%%%%%%%%%%%%%%%%%%%%%%%%%%%%%%%

\subsection{Laboratory Characterization tests of ProtoPol}
\label{subsec-LabCharacterization-ProtoPol}

\begin{figure}[]

    \begin{center}    
		
		\includegraphics[width=\textwidth]{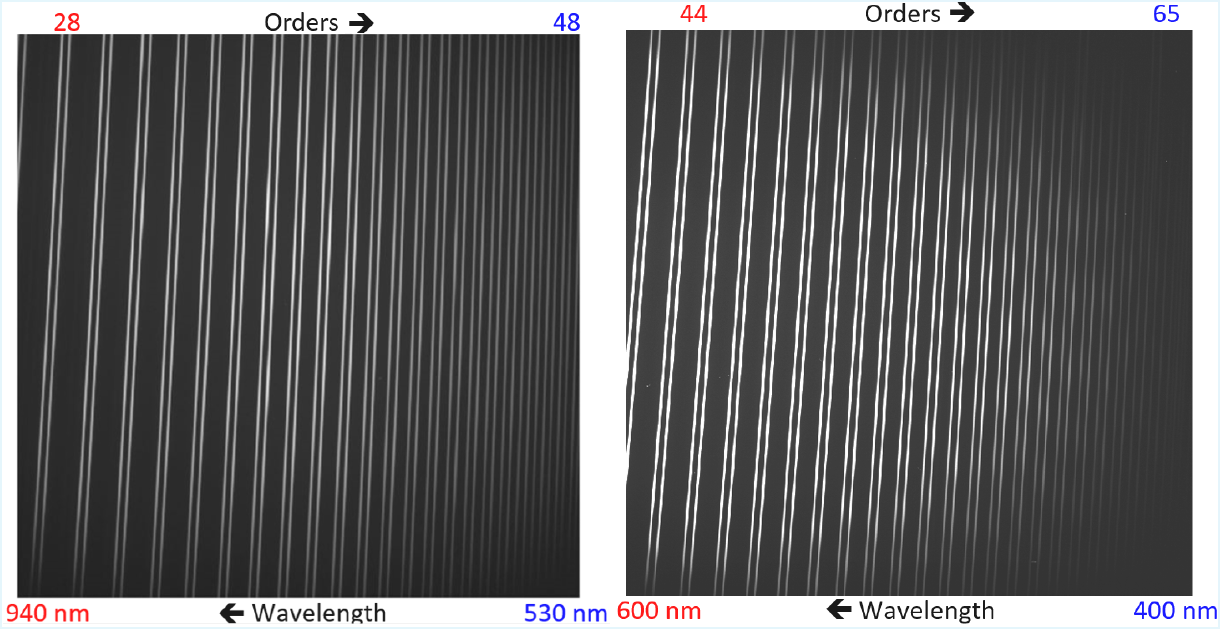}  
    
  \caption{Spectra of halogen lamp as obtained with Red (left) and Blue (right) CD gratings during laboratory characterization of ProtoPol. Order numbers are increasing from left to right with from 28 to 48 for the Red CD and 44 to 65 for the Blue CD. The wavelength range for each CD is also mentioned in the figure.}\label{hal_trace}

\end{center}
\end{figure}

\begin{figure}[]

    \centering
		\includegraphics[width=\textwidth]{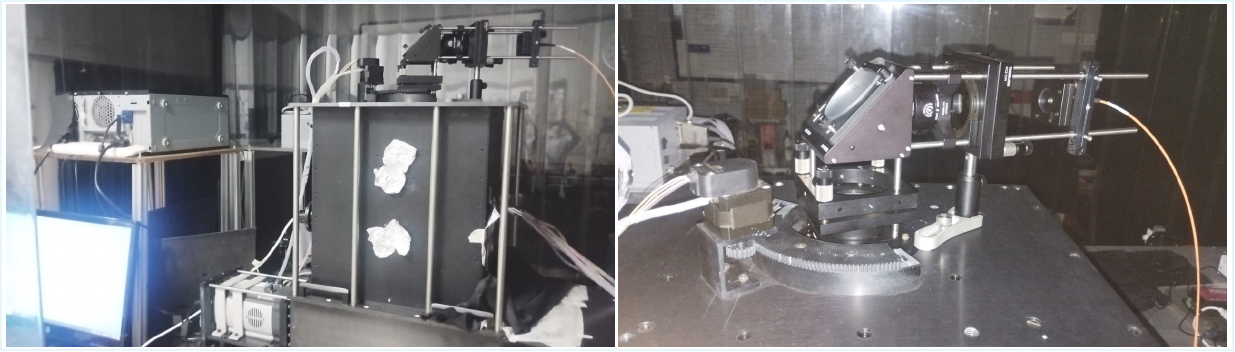}  
    
  \caption{The Telescope simulator assembly used to characterize ProtoPol in the laboratory.}\label{tel_sim_assembly}
		
\end{figure}

\begin{figure}[]
    \includegraphics[width=\textwidth]{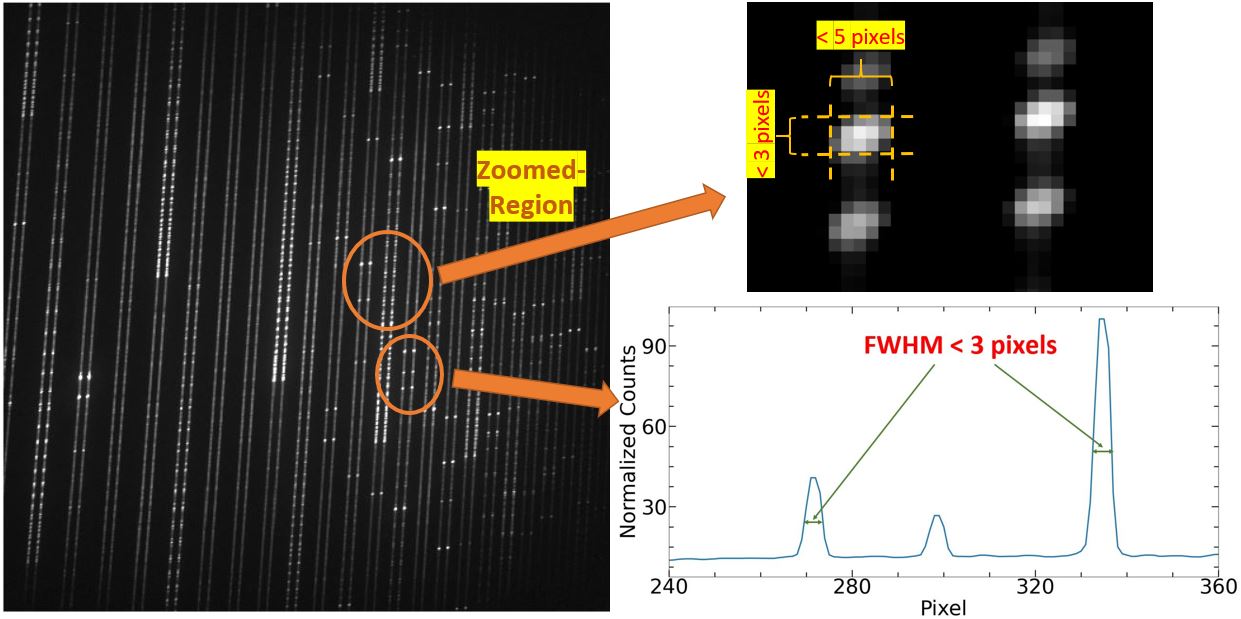}

  \caption{Raw data frame of ProtoPol showing spectra of the U-Ar lamp with Blue CD showing spectral line profile of emission lines and FWHM of emission lines along echelle dispersion direction. Typical dispersion values range from 0.12 to 0.17 $\AA$ per pixel from higher to lower orders with FWHMs of emission lines determined to be 2.6 - 2.8 pixels. This provides a spectral resolution $\sim$0.4 $\AA$ for Blue CD. See Paper-II for full discussion.}\label{blue_UAr}
		
\end{figure}

\begin{figure}[]
	\begin{center}    
		
		\includegraphics[width=\textwidth]{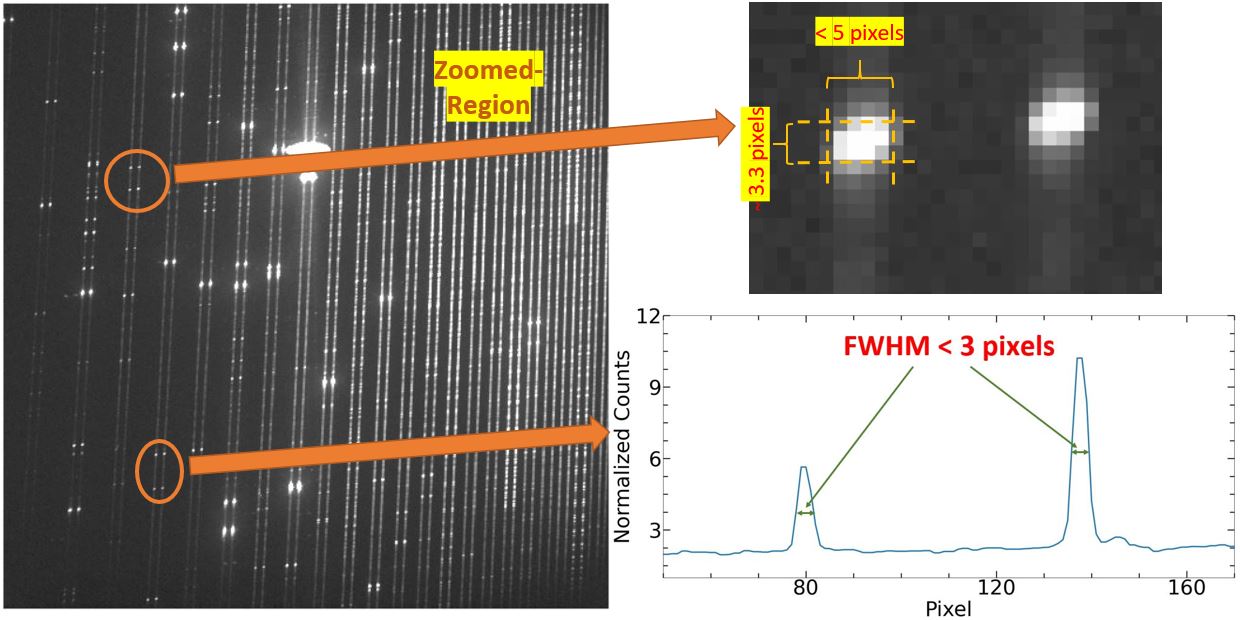}	
		
		\caption{Raw data frame of ProtoPol showing spectra of the U-Ar lamp with Red CD showing spectral line profile of emission lines and FWHM of emission lines along echelle dispersion direction. A very bright emission of argon can also be noticed. Typical dispersion values range from 0.17 to 0.25 $\AA$ per pixel from higher to lower orders with FWHMs of emission lines determined to be 3.0 - 3.6 pixels. This provides a spectral resolution in the range of 0.6 to 0.75 $\AA$ for Red CD. See Paper-II for full discussion.}	\label{red_UAr}	

	\end{center}    
\end{figure}

The laboratory characterization began with tests to characterize the detector system, such as measurements of background illumination, dark current, etc. Even though the CCD camera manufacturer provided some of these parameters, they were independently verified during this phase. Next, ProtoPol was characterized in the laboratory using various spectral and halogen lamps. Figure~\ref{hal_trace} displays the halogen lamp spectra taken with ProtoPol for both red and Blue CD gratings. Prior to deploying ProtoPol on the telescope, the instrument was tested in the laboratory using a telescope simulator, which was designed to mimic the telescope's optical beam. The setup consisted of two COTS achromatic lenses mounted on the instrument's top plate. A 550 $\mu$m core-diameter optical fiber, illuminated by different lamps (U-Ar or Halogen), served as the light source for the setup. The simulator set-up projected an image of the fiber core at the input onto the slit-pinhole, simulating the telescope's imaging performance. An iris was placed between the lenses, acting as the pupil mask, with its aperture set to produce an $f/8$ or $f/13$ beam, corresponding to the 2.5m and 1.2m telescopes, respectively. Additionally, a rotating polarizer was inserted between the lenses to modulate the input beam's polarization state, simulating the input polarization state of astronomical sources. Here, in the polarization characterization tests of the instrument, circular polarization effects and the modulator’s elliptical retardance are neglected. The telescope is assumed to be modeled using a fully linearly polarized input beam, even though the 5- layer PMMA modulators used in the polarimeter are in general known to have non-zero circular retardance \cite{harrington2020polarization}. The telescope simulator setup is illustrated in Figure~\ref{tel_sim_assembly}. The emission line profiles obtained using a U-Ar lamp are shown in Figures~\ref{blue_UAr} and ~\ref{red_UAr}. The dispersion ($\AA$ per pixel) and FWHMs of the emission line profiles in various orders, when considered together,  finally provide the spectral resolutions in the range of 0.40-0.75$\AA$ across various orders, which is consistent with the designed range. This is discussed in detail in Paper-II during the on-sky characterization of the instrument.

%%%%%%%%%%%%%%%%%%%%%%%%%%%%%%%%%%

%%%%%%%%%%%%%%%%%%%%%%%%%%%%%%%

%%%%%%%%%%%%%%%%%%%%%%%%%%%%%%

%%%%%%%%%%%%%%%%%%%%%%%%%%%%%%%%%%%%%%%

%%%%%%%%%%%%%%%%%%%%%%%%%%%%%%%%%%%%%%
%%%%%%%%%%%%%%%%%%%%%%%%%%%%%%%%%%%%%

\par
\begin{figure}[ht]
    \centering
		\includegraphics[width=\textwidth]{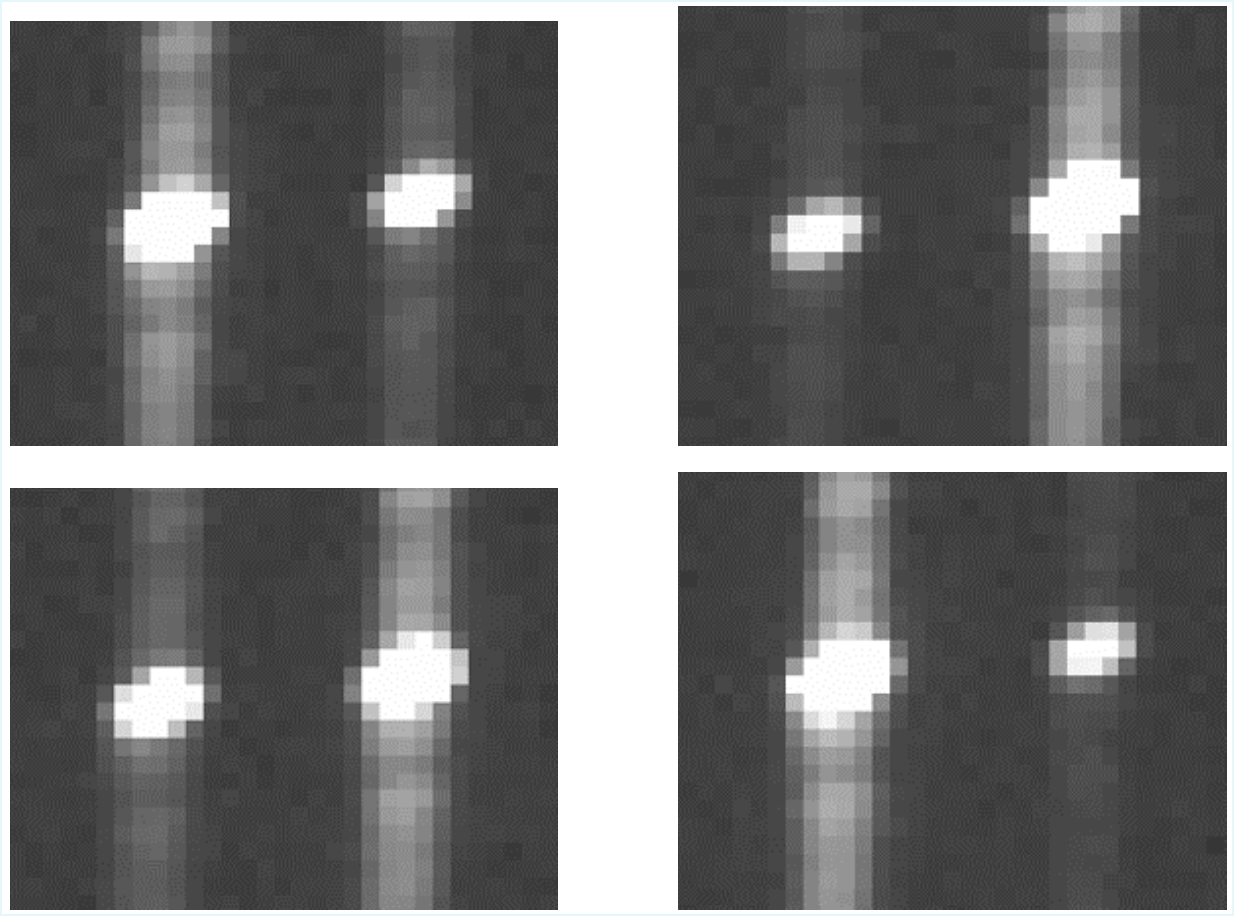}  

    % Main caption
    \caption{Variation in emission line intensity can be seen for four different HWP positions (0 \textit{(top-left)}, 22.5 \textit{(top-right)}, 45 \textit{(bottom-left)}, and 67.5 \textit{(bottom-right)}) but fixed input polarization (60 degrees) in the star-mode (i.e., without Glan-Taylor prism in beam path). Intensity flip can be noticed between (0, 45) and (22.5, 67.5) frames.}\label{tel_sim_test1}

\end{figure}

\begin{figure}[ht]
    \centering
		\includegraphics[width=\textwidth]{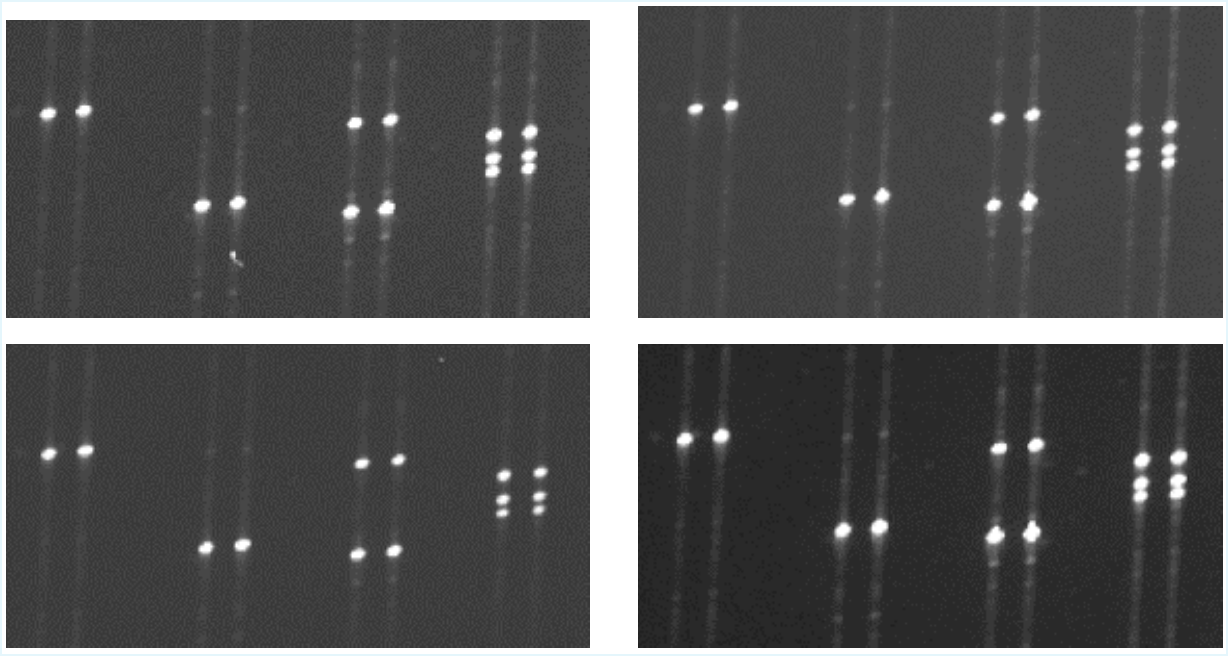}  

    % Main caption
    \caption{Figure shows that when a Glan-Taylor prism is used in the path of the incoming beam (in polarization calibration mode), no variation is seen in the intensities of emission lines for a given HWP position (22.5 degrees) by varying input polarization angle (0 \textit{(top-left)}, 60 \textit{(top-right)}, 120 \textit{(bottom-left)}, and 180 \textit{(bottom-right)}) of the input beam.}\label{tel_sim_test2}

\end{figure}

 
The telescope simulator was set up to gather multiple sets of polarization data by incrementally adjusting the input polarization from 0 degrees to 360 degrees in steps of 60 degrees. For each polarization setting, spectra of U-Ar and halogen lamps were recorded at four distinct HWP angles (0, 22.5, 45, and 67.5 degrees) in the star-mode (i.e, without the Glan-Taylor prism in the light path) and in the polarization calibration mode (i.e, with the Glan-Taylor prism in the light path). The same procedure was carried out for both the Blue and Red CD gratings. Taking measurements at four HWP positions not only helps mitigate biases from the optical paths followed by the o- and e-rays inside the instrument, but also eliminates variations in detector sensitivity where the o- and e-rays fall. Figure~\ref{tel_sim_test1} demonstrates the evident swapping of o- and e-rays between the (0, 45) degree and (22.5, 67.5) degree HWP configurations when operating in the star mode (without the Glan-Taylor prism in the light path). While in the polarization calibration mode (with the Glan-Taylor prism in the light path), the polarization state of the beam falling onto the slit-pinhole would be independent of the input polarization and is decided by the angle of the Glan-Taylor prism, as demonstrated in Figure~\ref{tel_sim_test2}.
\par
Subsequent to the above first-level laboratory characterization tests, ProtoPol was moved to Mt Abu observatory, where it was thoroughly characterized for its on-sky performance, and a data-reduction pipeline was developed. These aspects are discussed in detail in Paper-II.

%%%%%%%%%%%%%%%%%%%%%%%%%%%%%%%%%%%%%%%%%%%%%%%%%%%%%%%

\par 
\section{Automation and Control System}
\label{sec-ControlSys-ProtoPol}
\par 
ProtoPol's automation and control system is developed along the lines of a previously developed control system for an earlier instrument MFOSC-P (Mt. Abu Faint Object Spectrograph and Camera - Pathfinder) developed by the team \cite{srivastava2021design}. It consists of two main components: (1) the detector system controls and (2) the instrument system controls. The detector system includes integrated read-out electronics and is managed using SOLIS, a control software developed by the Original Equipment Manufacturer (OEM) \href{https://andor.oxinst.com/products/solis-software/; Accessed: 2020-01-23}{ M/S/ Andor Technology Ltd}. SOLIS allows for various acquisition settings for the detector, such as adjusting exposure times, shutter controls, and selecting different data acquisition modes (e.g., sub-region imaging, different read-out speeds, image spooling, kinetic acquisition, etc.). Additionally, the software has detector cooling and support for other safety features built-in, offering image processing tools for rapid data analysis. The remaining instrument operations of ProtoPol are handled by an in-house developed control system with a graphical user interface (GUI). The instrument incorporates five motion systems: a calibration mirror/GT prism/field viewer mirror positioning system, a rotational mechanism to rotate the HWP to different angles, a long-pass filter motion system, two separate grating motion mechanisms for the two CD gratings, and an off-axis auto-guider mirror (only needed for 1.2m telescope). These systems are powered by stepper motors or stepper-driven linear translational stages, with quadrature encoders fitted on all motors except for the off-axis auto-guider's stepper motor. Limit and home sensors are integrated into each motion system. The instrument control system also operates the calibration unit’s lamps alongside these motion mechanisms.
%====================================================
\begin{figure}[]
	\begin{center}    
		
		\includegraphics[width=\textwidth]{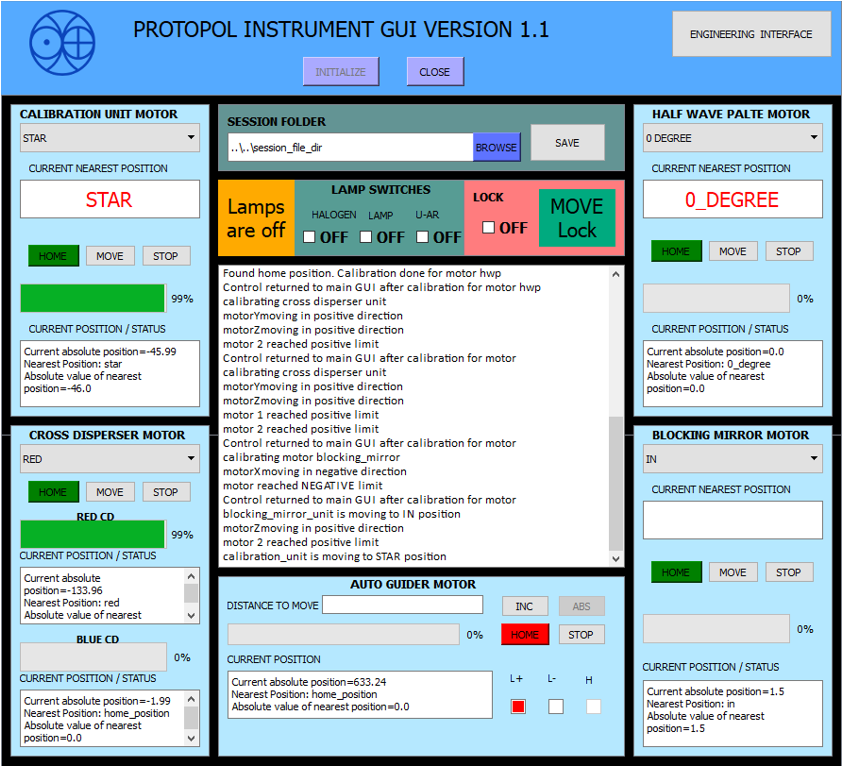}		
		
		\caption{The graphical user interface (GUI) for ProtoPol's operation.}\label{fig-GUI-ProtoPol}
		
	\end{center}    
\end{figure}
%================================================================

\par 
The GUI software has been developed with two primary functions: (1) enabling the control system to operate as a versatile 8-axis motion controller for various applications, and (2) providing a dedicated user interface for ProtoPol. The GUI and its underlying application software were built using Python 2.7 with PyQt4 framework. A block-level architecture of the application software can be found in \cite{srivastava2021design}. The front facade serves as the front-end GUI window, custom-designed for operations relating to ProtoPol instrument (see Figure~\ref{fig-GUI-ProtoPol}). It displays real-time status and visual feedback for all motion subsystems, allowing users to monitor movements during mode transitions on the GUI screen. This front-layer interface accepts inputs/tasks from the user and relays them to a general-purpose engineering interface via an instrument-specific configuration file and a subsystem-specific layer of user interface. These configuration files are stored in an editable ASCII text format, containing the default settings for various subsystems of ProtoPol, thus acting as a bridge between the instrument-specific main GUI screen and the controller interface engineering interface. For debugging and engineering tests, the GUI includes provisions for direct access to hardware components (e.g., motors, limit sensors, etc.) through a password-protected engineering interface. The engineering interface has another setup window to let users configure the parameters for motion profiles (start/top speeds, acceleration, deceleration, micro-stepping, pitch, etc.) and provides a command-line console interface to communicate with the controller via ASCII command sets (provided by the controller manufacturer via library functions). All commands are packaged and sent to the hardware system through a USB 2.0 communication layer.
\par
The software has been designed for cross-platform compatibility and easy scalability. While the front-end GUI is specifically tailored for ProtoPol operations, its design of the interface allows for straightforward adaptation for future instrumentation needs. Both the main GUI screen and the configuration file can be modified appropriately to enable the system to control different instruments without the necessity to change the underlying hardware infrastructure.

%%%%%%%%%%%%%%%%%%%%%%%%%%%%%%%%%%%%%%%%%%%%%%%%%%%%%%%%%%%%
%%%%%%%%%%%%%%%%%%%%%%%%%%%%%%%%%%%%%%%%%%%%%%%%%%%%%%%%%%%%

\section{On-sky Commissioning and Characterization Observations}

\begin{figure}[]
    \centering
		\includegraphics[width=\textwidth]{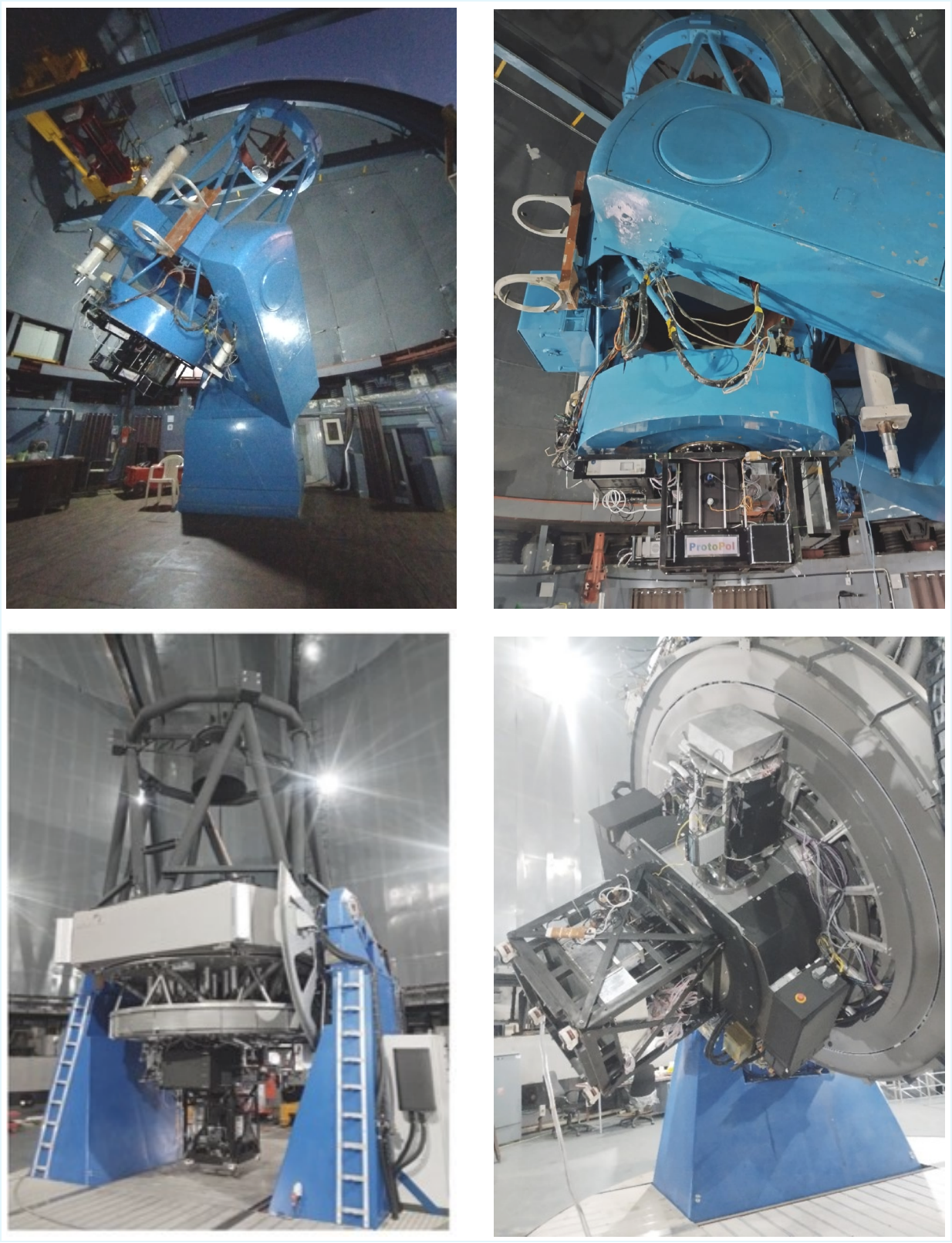}  

    \caption{Images showing ProtoPol mounted on PRL 1.2m (\textit{top}) and 2.5m telescopes at Mt.Abu (\textit{bottom}).}
		\label{ProtoPol_tel}

\end{figure}

\begin{figure}[]
\begin{center}
    \includegraphics[width=\textwidth]{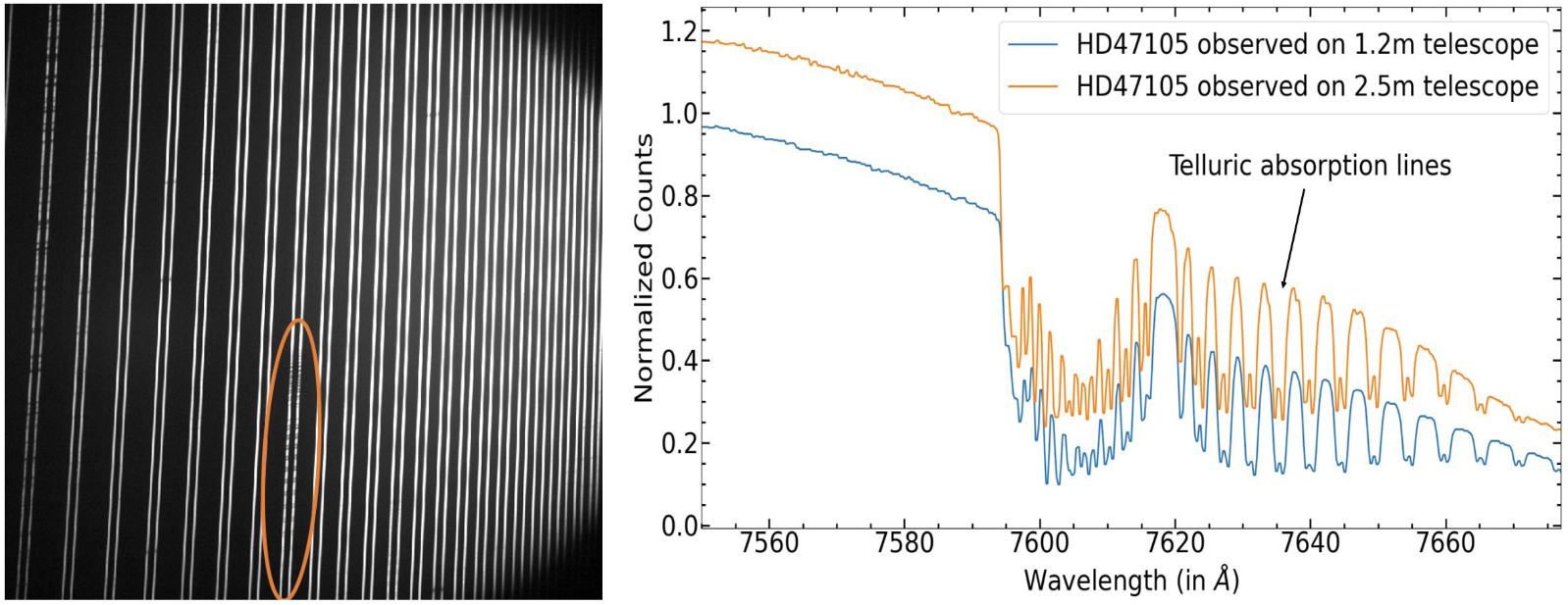}
    
    \caption{Left: The raw data frame of a star HD47105 containing telluric absorption features in one of the orders as marked by the \textit{orange} oval. Right: The reduced spectra of the echelle order. The \textit{orange} curve representing observation from 2.5m telescope is given a +0.2 offset (in arbitrary units) along the y-axis for better clarity.}\label{HD47105-spec}
    
\end{center}    
\end{figure}

After the completion of the laboratory tests of the instrument, the instrument was transported to Mt. Abu observatory. The instrument was initially mounted on the PRL 1.2m telescope between mid-December 2023 and mid-February 2024 to check for instrument compatibility with the 1.2m telescope and perform in-sky characterization tests. Since then, ProtoPol has been shifted to the PRL 2.5m telescope for science and characterization observations and has been fully operational. As mentioned earlier, ProtoPol was designed and developed in a modular way to facilitate ease of transportation. Therefore, the instrument was dismantled into three main units - polarimeter unit, calibration unit, and spectrometer unit- and then reassembled at the observatory site. Different mounting-cage structures were designed for mounting the instrument on PRL 1.2m and 2.5m telescopes. Some pictures of the first operations of ProtoPol on PRL 1.2m and 2.5m telescopes are shown in Figure~\ref{ProtoPol_tel}. The spectroscopic image quality of ProtoPol, both on the 1.2m and 2.5m telescopes, can be seen in Figure~\ref{HD47105-spec}, where the spectra of a standard unpolarized star HD47105 for an order covering telluric absorption lines are shown for observations from both the telescopes. The sharp absorption features are found to be well resolved, confirming very good on-sky performance of the instrument. Detailed on-sky characterization report, along with development of a fully-automated dedicated data-reduction pipeline for ProtoPol and the first science results obtained with the instrument are presented in Paper II.\\

%%%%%%%%%%%%%%%%%%%%%%%%%%%%%%%%

%%%%%%%%%%%%%%%%%%%%%%%%%%%%%

%%%%%%%%%%%%%%%%%%%%%%%%%%%%%%

%%%%%%%%%%%%%%%%%%%%%%%%%%%%%%%%%%%%%%

\section{Summary}

This is the first of the two-part paper series on the development and performance of ProtoPol - a medium resolution spectro-polarimeter for PRL 1.2m and 2.5m telescopes. The optical, opto-mechanical designs and assembly-integration-testing (AIT) of ProtoPol are discussed in this paper. The instrument has been developed completely in-house with commercially available off-the-shelf optical/opto-mechanical components, ensuring cost efficiency and a faster development schedule. ProtoPol was designed to provide a spectral resolution in the range of $\sim$0.4-0.75$\AA$ in the wavelength range of 4000-9600 $\AA$, though the final performance is determined to be better. The instrument has been successfully designed and developed and is currently fully operational at PRL 2.5 telescope. The development of a dedicated fully-automated echelle spectro-polarimetry data-reduction pipeline for ProtoPol, the on-sky characterization, and first science results obtained from the instrument are discussed in Paper-II of the series.

\appendix    % this command starts appendixes

\subsection*{Disclosures}
The authors declare that there are no financial interests, commercial affiliations, or other potential conflicts of interest that could have influenced the objectivity of this research or the writing of this paper

\subsection* {Code, Data, and Materials Availability} 
The data presented in this manuscript may be made available upon reasonable request. Corresponding authors may be contacted for that.

\subsection* {Acknowledgments}
We thank the anonymous reviewers for the review and their constructive comments and suggestion on the manuscript. Development of the ProtoPol instrument has been funded by the Department of Space, Government of India through Physical Research Laboratory (PRL), Ahmedabad. ProtoPol team is thankful to the Director, PRL, for supporting the ProtoPol and M-FOSC-EP development program. MKS expresses sincere thanks to Shyam N. Tandon (Inter-University Center for Astronomy and Astrophysics – IUCAA, Pune, India) for detailed discussions on several aspects of spectro-polarimeter design throughout the development process. AM is thankful to PRL for his Ph.D. research fellowship. ProtoPol team expresses sincere thanks to Mt. Abu observatory staff for their sustained help and support during ProtoPol commissioning and subsequent observations.

%%%%% References %%%%%

%\bibliography{report}   % bibliography data in report.bib

\begin{thebibliography}{10}

\bibitem{srivastava2021design}
M.~K. Srivastava, V.~Kumar, V.~Dixit, {\em et~al.}, ``Design and development of mt. abu faint object spectrograph and camera--pathfinder (mfosc-p) for prl 1.2 m mt. abu telescope,'' {\em Experimental Astronomy} {\bf 51}, 345--382  (2021).

\bibitem{Pirnay2018}
O.~{Pirnay}, G.~P. {Lousberg}, A.~{Lanotte}, {\em et~al.}, ``{Mt ABU 2.5m Telescope: design and fabrication},'' in {\em Proceedings of SPIE},  {\em Society of Photo-Optical Instrumentation Engineers (SPIE) Conference Series} {\bf 10700}, 107005S  (2018).

\bibitem{rajpurohit2018exploring}
A.~Rajpurohit, F.~Allard, S.~Rajpurohit, {\em et~al.}, ``Exploring the stellar properties of m dwarfs with high-resolution spectroscopy from the optical to the near-infrared,'' {\em Astronomy \& Astrophysics} {\bf 620}, A180  (2018).

\bibitem{Srivastava2016}
M.~K. Srivastava, D.~Banerjee, N.~Ashok, {\em et~al.}, ``Near-infrared studies of v2944 ophiuchi (nova ophiuchi 2015),'' {\em Monthly Notices of the Royal Astronomical Society} {\bf 462}(2), 2074--2084  (2016).

\bibitem{Kaur2017}
N.~Kaur, S.~Chandra, K.~S. Baliyan, {\em et~al.}, ``A multiwavelength study of flaring activity in the high-energy peaked bl lac object 1es 1959+ 650 during 2015--2016,'' {\em The Astrophysical Journal} {\bf 846}(2), 158  (2017).

\bibitem{Chakraborty2018}
A.~Chakraborty, A.~Roy, R.~Sharma, {\em et~al.}, ``Evidence of a sub-saturn around epic 211945201,'' {\em The Astronomical Journal} {\bf 156}(1), 3  (2018).

\bibitem{Joshi2017}
V.~Joshi, D.~P. Banerjee, and M.~Srivastava, ``Nova ophiuchus 2017 as a probe of 13c nucleosynthesis and carbon monoxide formation and destruction in classical novae,'' {\em The Astrophysical Journal Letters} {\bf 851}(2), L30  (2017).

\bibitem{Chakraborty2018b}
A.~{Chakraborty}, N.~{Thapa}, K.~{Kumar}, {\em et~al.}, ``{PARAS-2 precision radial velocimeter: optical and mechanical design of a fiber-fed high resolution spectrograph under vacuum and temperature control},'' in {\em Proceedings of SPIE},  {\em Society of Photo-Optical Instrumentation Engineers (SPIE) Conference Series} {\bf 10702}, 107026G  (2018).

\bibitem{rai2020optical}
A.~Rai, S.~Ganesh, S.~K. Paul, {\em et~al.}, ``Optical aspects of near-infrared imager spectrometer and polarimeter instrument (nisp),'' in {\em Ground-based and Airborne Instrumentation for Astronomy VIII},   {\bf 11447}, 1370--1377, SPIE  (2020).

\bibitem{sarkar2020electronics}
D.~R. Sarkar, A.~B. Shah, A.~Singh, {\em et~al.}, ``Electronics design and development of near-infrared imager, spectrometer, and polarimeter,'' in {\em Ground-based and Airborne Instrumentation for Astronomy VIII},   {\bf 11447}, 1605--1610, SPIE  (2020).

\bibitem{buzzoni1984eso}
B.~Buzzoni, B.~Delabre, H.~Dekker, {\em et~al.}, ``The eso faint object spectrograph and camera (efosc),'' {\em ESO Messenger (ISSN 0722-6691), Dec. 1984, p. 9-13.} {\bf 38}, 9--13  (1984).

\bibitem{andersen1995new}
J.~Andersen, M.~Andersen, J.~Klougart, {\em et~al.}, ``New power for the danish 1.54-m telescope.,'' {\em The Messenger} {\bf 79}, 12--14  (1995).

\bibitem{omar2017scientific}
A.~Omar, B.~Kumar, M.~Gopinathan, {\em et~al.}, ``Scientific capabilities and advantages of the 3.6 meter optical telescope at devasthal, uttarakhand,'' {\em Current Science} , 682--685  (2017).

\bibitem{kumar2022designs}
V.~Kumar, M.~K. Srivastava, V.~Dixit, {\em et~al.}, ``Designs of mt. abu faint object spectrograph and camera-echelle polarimeter (m-fosc-ep) and its prototype: spectro-polarimeters for prl 1.2 m and 2.5 m mt. abu telescopes, india,'' in {\em Ground-based and Airborne Instrumentation for Astronomy IX},   {\bf 12184}, 1696--1714, SPIE  (2022).

\bibitem{Rajpurohit2020}
A.~Rajpurohit, V.~Kumar, M.~K. Srivastava, {\em et~al.}, ``First results from mfosc-p: low-resolution optical spectroscopy of a sample of m dwarfs within 100 parsecs,'' {\em Monthly Notices of the Royal Astronomical Society} {\bf 492}(4), 5844--5852  (2020).

\bibitem{kumar2022optical}
V.~Kumar, M.~K. Srivastava, D.~P. Banerjee, {\em et~al.}, ``Optical and near-infrared spectroscopy of nova v2891 cygni: evidence for shock-induced dust formation,'' {\em Monthly Notices of the Royal Astronomical Society} {\bf 510}(3), 4265--4283  (2022).

\bibitem{ganesh2020empol}
S.~Ganesh, A.~Rai, A.~Singh, {\em et~al.}, ``Empol: an emccd based optical imaging polarimeter,'' in {\em Ground-based and Airborne Instrumentation for Astronomy VIII},   {\bf 11447}, 2032--2038, SPIE  (2020).

\bibitem{aarthy2019nicspol}
E.~Aarthy, A.~Rai, S.~Ganesh, {\em et~al.}, ``Nicspol: a near-infrared polarimeter for the 1.2-m telescope at mount abu infrared observatory,'' {\em Journal of Astronomical Telescopes, Instruments, and Systems} {\bf 5}(3), 035006--035006  (2019).

\bibitem{piskunov2011harpspol}
N.~Piskunov, F.~Snik, A.~Dolgopolov, {\em et~al.}, ``Harpspol—the new polarimetric mode for harps,'' {\em The Messenger} {\bf 143}(7)  (2011).

\bibitem{donati2003espadons}
J.-F. Donati, ``Espadons: An echelle spectropolarimetric device for the observation of stars at cfht,'' in {\em Solar Polarization},   {\bf 307}, 41  (2003).

\bibitem{gratton2001sarg}
R.~Gratton, G.~Bonanno, P.~Bruno, {\em et~al.}, ``Sarg: the high resolution spectrograph of tng,'' {\em Experimental Astronomy} {\bf 12}(2), 107--143  (2001).

\bibitem{bramall2010salt}
D.~Bramall, R.~Sharples, L.~Tyas, {\em et~al.}, ``The salt hrs spectrograph: final design, instrument capabilities, and operational modes,'' in {\em Ground-based and Airborne Instrumentation for Astronomy III},   {\bf 7735}, 1692--1701, SPIE  (2010).

\bibitem{strassmeier2008pepsi}
K.~Strassmeier, M.~Woche, I.~Ilyin, {\em et~al.}, ``Pepsi: the potsdam echelle polarimetric and spectroscopic instrument for the lbt,'' in {\em Ground-based and Airborne Instrumentation for Astronomy II},   {\bf 7014}, 277--288, SPIE  (2008).

\bibitem{donati2020spirou}
J.~Donati, D.~Kouach, C.~Moutou, {\em et~al.}, ``Spirou: Nir velocimetry and spectropolarimetry at the cfht,'' {\em Monthly Notices of the Royal Astronomical Society} {\bf 498}(4), 5684--5703  (2020).

\bibitem{kim2007boes}
K.-M. Kim, I.~Han, G.~G. Valyavin, {\em et~al.}, ``The boes spectropolarimeter for zeeman measurements of stellar magnetic fields,'' {\em Publications of the Astronomical Society of the Pacific} {\bf 119}(859), 1052  (2007).

\bibitem{baudrand1992musicos}
J.~Baudrand and T.~Bohm, ``Musicos-a fiber-fed spectrograph for multi-site observations,'' {\em Astronomy and Astrophysics (ISSN 0004-6361), vol. 259, no. 2, p. 711-719.} {\bf 259}, 711--719  (1992).

\bibitem{arasaki2015very}
T.~Arasaki, Y.~Ikeda, Y.~Shinnaka, {\em et~al.}, ``The very precise echelle spectropolarimeter on the araki telescope (vespola),'' {\em Publications of the Astronomical Society of Japan} {\bf 67}(3), 35  (2015).

\bibitem{ikeda2003development}
Y.~Ikeda, H.~Akitaya, K.~Matsuda, {\em et~al.}, ``Development of the high-resolution spectropolarimeter: Lips,'' in {\em Polarimetry in Astronomy},   {\bf 4843}, 437--447, SPIE  (2003).

\bibitem{srivastava2024development}
M.~K. Srivastava, A.~Maiti, V.~Kumar, {\em et~al.}, ``Development of protopol: a medium resolution echelle spectro-polarimeter for prl 1.2 m and 2.5 m telescopes, mt abu, india,'' in {\em Ground-based and Airborne Instrumentation for Astronomy X},   {\bf 13096}, 577--592, SPIE  (2024).

\bibitem{schmid1994raman}
H.~Schmid and H.~Schild, ``Raman scattered emission lines in symbiotic stars: a spectropolarimetric survey,'' {\em Astronomy and Astrophysics (ISSN 0004-6361), vol. 281, no. 1, p. 145-160} {\bf 281}, 145--160  (1994).

\bibitem{harries1996raman}
T.~Harries and I.~Howarth, ``Raman scattering in symbiotic stars. i. spectropolarimetric observations,'' {\em Astronomy and Astrophysics Supplement Series} {\bf 119}(1), 61--90  (1996).

\bibitem{kenyon1986symbiotic}
S.~J. Kenyon, ``Symbiotic stars,'' in {\em Interacting Binaries},  179--203, Springer  (1986).

\bibitem{munari2019symbiotic}
U.~Munari, ``The symbiotic stars,'' {\em The Impact of Binary Stars on Stellar Evolution} {\bf 54}, 77  (2019).

\bibitem{nussbaumer1989raman}
H.~Nussbaumer, H.~Schmid, and M.~Vogel, ``Raman scattering as a diagnostic possibility in astrophysics,'' {\em Astronomy and Astrophysics (ISSN 0004-6361), vol. 211, no. 2, March 1989, p. L27-L30. Research supported by SNSF.} {\bf 211}, L27--L30  (1989).

\bibitem{schmid1989identification}
H.~Schmid, ``Identification of the emission bands at 6830, 7088 a,'' {\em Astronomy and Astrophysics (ISSN 0004-6361), vol. 211, no. 2, March 1989, p. L31-L34. Research supported by SNSF.} {\bf 211}, L31--L34  (1989).

\bibitem{lee2000raman}
H.-W. Lee, ``Raman-scattering wings of h$\alpha$ in symbioticstars,'' {\em The Astrophysical Journal} {\bf 541}(1), L25  (2000).

\bibitem{yoo2002polarization}
J.~J. Yoo, J.-Y. Bak, and H.-W. Lee, ``Polarization of the broad h$\alpha$ wing in symbiotic stars,'' {\em Monthly Notices of the Royal Astronomical Society} {\bf 336}(2), 467--476  (2002).

\bibitem{lee2018h}
S.-J. Lee and S.~Hyung, ``H $\alpha$ and h $\beta$ raman scattering line profiles of the symbiotic star ag pegasi,'' {\em Monthly Notices of the Royal Astronomical Society} {\bf 475}(4), 5558--5569  (2018).

\bibitem{ikeda2004polarized}
Y.~Ikeda, H.~Akitaya, K.~Matsuda, {\em et~al.}, ``Polarized h$\alpha$ wings in the symbiotic stars ag draconis and z andromedae,'' {\em The Astrophysical Journal} {\bf 604}(1), 357  (2004).

\bibitem{de1994new}
D.~De~Winter, M.~Perez, {\em et~al.}, ``A new catalogue of members and candidate members of the herbig ae/be (haebe) stellar group,'' {\em Astronomy and Astrophysics Suppl., Vol. 104, p. 315-339 (1994)} {\bf 104}, 315--339  (1994).

\bibitem{mottram2007difference}
J.~C. Mottram, J.~Vink, R.~Oudmaijer, {\em et~al.}, ``On the difference between herbig ae and herbig be stars,'' {\em Monthly Notices of the Royal Astronomical Society} {\bf 377}(3), 1363--1374  (2007).

\bibitem{waelkens1997comet}
C.~Waelkens, K.~Malfait, and L.~Waters, ``Comet hale-bopp, circumstellar dust, and the interstellar medium,'' {\em Earth, Moon, and Planets} {\bf 79}(1), 265--274  (1997).

\bibitem{vink2002probing}
J.~S. Vink, J.~E. Drew, T.~J. Harries, {\em et~al.}, ``Probing the circumstellar structure of herbig ae/be stars,'' {\em Monthly Notices of the Royal Astronomical Society} {\bf 337}(1), 356--368  (2002).

\bibitem{henry2024character}
T.~J. Henry and W.-C. Jao, ``The character of m dwarfs,'' {\em Annual Review of Astronomy and Astrophysics} {\bf 62}  (2024).

\bibitem{Deshpande1995}
M.~R. {Deshpande}, ``{A brief report on the Infrared Telescope at Gurushikhar, MT Abu},'' {\em Bulletin of the Astronomical Society of India} {\bf 23}, 13  (1995).

\bibitem{Banerjee1997}
D.~P.~K. {Banerjee}, A.~D. {Bobra}, and D.~V. {Subhedar}, ``{The effect of aberrations on image quality - A study for the 1.2m Gurusikhar Infrared Telescope},'' {\em Bulletin of the Astronomical Society of India} {\bf 25}, 555  (1997).

\bibitem{eversberg2014fundamentals}
T.~Eversberg and K.~Vollmann, ``Fundamentals of echelle spectroscopy,'' in {\em Spectroscopic Instrumentation: Fundamentals and Guidelines for Astronomers},  193--227, Springer  (2014).

\bibitem{smith2012lunar}
A.~W. Smith, S.~R. Lorentz, T.~C. Stone, {\em et~al.}, ``Lunar spectral irradiance and radiance (lusi): New instrumentation to characterize the moon as a space-based radiometric standard,'' {\em Journal of Research of the National Institute of Standards and Technology} {\bf 117}, 185  (2012).

\bibitem{shanks2016optics}
K.~Shanks, S.~Senthilarasu, and T.~K. Mallick, ``Optics for concentrating photovoltaics: Trends, limits and opportunities for materials and design,'' {\em Renewable and Sustainable Energy Reviews} {\bf 60}, 394--407  (2016).

\bibitem{harding2016chimera}
L.~K. Harding, G.~Hallinan, J.~Milburn, {\em et~al.}, ``Chimera: a wide-field, multi-colour, high-speed photometer at the prime focus of the hale telescope,'' {\em Monthly Notices of the Royal Astronomical Society} {\bf 457}(3), 3036--3049  (2016).

\bibitem{bottema1981echelle}
M.~Bottema, ``Echelle efficiency and blaze characteristics,'' in {\em Periodic Structures, Gratings, Moire Patterns, and Diffraction Phenomena I},   {\bf 240}, 171--177, SPIE  (1981).

\bibitem{harrington2020polarization}
D.~M. Harrington, S.~A. Jaeggli, T.~A. Schad, {\em et~al.}, ``Polarization modeling and predictions for daniel k. inouye solar telescope, part 6: fringe mitigation with polycarbonate modulators and optical contact calibration retarders,'' {\em Journal of Astronomical Telescopes, Instruments, and Systems} {\bf 6}(3), 038001--038001  (2020).

\end{thebibliography}
%\bibliographystyle{spiejour}   % makes bibtex use spiejour.bst

%%%%% Biographies of authors %%%%%

\vspace{2ex}\noindent\textbf{First Author} is currently an associate professor in the astronomy and astrophysics division at Physical Research Laboratory (PRL), Ahmedabad, India. Dr. Mudit K. Srivastava is the Principal Investigator (PI) of ProtoPol and M-FOSC-EP instruments. He is an astronomer \& an instrumentation scientist and have been involved in the development of optical instrumentation for the last two decades. 

\vspace{2ex}\noindent\textbf{Second Author} is currently a Senior Research Fellow (SRF) - Ph.D. scholar -  at Physical Research Laboratory (PRL), Ahmedabad, India. Mr Arijit Maiti is holding a Master of Science Degree in Physics from National Institute of technology (NIT) Rourkela, India. He is responsible for the assembly-integration-test (AIT), subsequent on-sky characterization, and science verification of ProtoPol. His current research interests include optical instrumentation for ground-based telescopes, spectro-polarimetric and spectroscopic studies of symbiotic stars and novae.

\vspace{2ex}\noindent\textbf{Third Author} is currently a post-doctoral fellow at I. Physikalisches Institut, Universit\"at zu K\"oln. Dr. Vipin Kumar completed his doctoral thesis from the Physical Research Laboratory, Ahmedabad, India, wherein, as a part of his Ph.D. thesis, he developed the optical design of ProtoPol. He has also contributed towards the on-sky characterization of the instrument. His research focuses on developing Optical/IR instrumentation for ground-based telescopes, with a scientific emphasis on studying cool, low-mass stars.

\vspace{2ex}\noindent\textbf{Fourth Author} is a Technical Assistant (Mechanical) in the Physical Research Laboratory (PRL), Ahmedabad, India. Mr Bhaveshkumar Mistry received his Bachelor of Engineering (B.E.) degree in mechanical engineering. His expertise lies in the mechanical design and development of ground-based astronomical instrumentation. He has been responsible for the opto-mechanical design and subsequent fabrication of ProtoPol.

\vspace{2ex}\noindent\textbf{Fifth Author} is a Scientist/Engineer in the Astronomy and Astrophysics Division at the Physical Research Laboratory, Ahmedabad, India. Mrs. Ankita Patel received her Bachelor of Technology (B.Tech.) degree from Visvesvaraya Technological University, Bengaluru, India, in Electronics and Communication Engineering. She has been working on the control system and software aspects of instrumentation for ground-based astronomy. She has been responsible for the development of the control system and the instrument operation software of ProtoPol.

\vspace{2ex}\noindent\textbf{Sixth Author} is currently an engineer at Advanced Engineering Group, Azista Industries, Ahmedabad, India. Prior to this, Mr Vaibhav Dixit was a scientist/Engineer at Physical Research Laboratory, Ahmedabad, India. He has got his Bachelor's degree in Physical Sciences from Indian Institute of Space Science and Technology (IIST), Thiruvananthapuram, India and Master in Computational Sciences from University of Heidelberg, Germany. He contributed towards the optics and system design of the ProtoPol

\vspace{2ex}\noindent\textbf{Seventh Author} is a Scientist/Engineer at Physical Research Laboratory (PRL), Ahmedabad. Mr Kevikumar A. Lad completed his Bachelor's degree at the Indian Institute of Space Science and Technology (IIST), Thiruvananthapuram, India in Aerospace Engineering. Mr Lad had prepared the preliminary opto-mechanical design of ProtoPol and later contributed to other design aspects of the instrument.

\vspace{1ex}
\noindent Biographies and photographs of the other authors are not available.

\listoffigures
\listoftables

\end{spacing}
\end{document}